\RequirePackage{xcolor}
\documentclass[journal,twoside,web]{ieeecolor}          

\usepackage{etoolbox}
\makeatletter
\@ifundefined{color@begingroup}%
  {\let\color@begingroup\relax
   \let\color@endgroup\relax}{}%
\def\fix@ieeecolor@hbox#1{%
  \hbox{\color@begingroup#1\color@endgroup}}
\patchcmd\@makecaption{\hbox}{\fix@ieeecolor@hbox}{}{\FAILED}
\patchcmd\@makecaption{\hbox}{\fix@ieeecolor@hbox}{}{\FAILED}

\usepackage{tmi}

\usepackage{array}

\usepackage{amsthm}

\usepackage{amssymb}
\usepackage{amsmath}
\usepackage{amsfonts}
\usepackage{float}
\usepackage{graphicx}
\usepackage{mwe}
\usepackage{graphbox}
\usepackage{setspace}
\usepackage{epstopdf}
\usepackage{epsf}
\usepackage{url}
\usepackage[scr=esstix,cal=boondox]{mathalfa} 

\usepackage{textcomp}

\usepackage{bm}
\usepackage{acronym}

\usepackage{algorithmic}
\usepackage{graphicx}

\theoremstyle{remark}	
\theoremstyle{remark}	
\theoremstyle{remark}	

\usepackage{xcolor}

\makeatletter
\let\MYcaption\@makecaption
\makeatother
\usepackage[font=footnotesize]{subcaption}
\makeatletter
\let\@makecaption\MYcaption
\makeatother

\begin{document}

\title{Domain-Aware Few-Shot Learning for Optical Coherence Tomography Noise Reduction}

\author{Deborah Pereg
\thanks{D. Pereg is with Wellman Center for Photomedicine MGH, Harvard Medical School, and MIT CSAIL. 
\texttt{deborahp@mit.edu}}}


\maketitle

\begin{abstract}
Speckle noise has long been an extensively studied problem in medical imaging. 
In recent years, there have been significant advances in leveraging deep learning methods for noise reduction. Nevertheless, adaptation of supervised learning models to unseen domains remains a challenging problem. Specifically, deep neural networks (DNNs) trained for computational imaging tasks are vulnerable to changes in the acquisition system's physical parameters, such as: sampling space, resolution, and contrast. Even within the same acquisition system, performance degrades across datasets of different biological tissues.
In this work, we propose a few-shot supervised learning framework for optical coherence tomography (OCT) noise reduction, that offers a dramatic increase in training speed and requires only a single image, or part of an image, and a corresponding speckle suppressed ground truth, for training. Furthermore, we formulate the domain shift problem for OCT diverse imaging systems, and prove that the output resolution of a despeckling trained model is determined by the source domain resolution. We also provide possible remedies. We propose different practical implementations of our approach, verify and compare their applicability, robustness, and computational efficiency. Our results demonstrate significant potential for generally improving sample complexity, generalization, and time efficiency, for coherent and non-coherent noise reduction via supervised learning models, that can also be leveraged for other real-time computer vision applications. 

\end{abstract}
%



\section{Introduction}
OCT employs low coherence interferometry to produce cross-sectional tomographic images of internal structure of biological tissue. It is routinely used for diagnostic imaging, primarily of the retina and coronary arteries \cite{Villiger:2020}. The resolutions obtainable are in the range 1 to 15 $\mu$m, with depth range of millimeters. Unfortunately, OCT images are often degraded by speckle noise \cite{Schmitt:1999,Goodman:2007}, creating apparent grain-like structures in the image, with size as large as the spatial resolution of the OCT system. Speckle noise significantly degrades images and complicates interpretation and medical diagnosis by confounding tissue anatomy and masking changes in tissue scattering properties. 

Speckle suppression is often achieved by incoherent averaging of images with different speckle realizations \cite{Pircher:2003}, e.g., through angular compounding \cite{Desjardins:2007, Zhao:2020}. Averaging methods attempt to preserve the resolution while suppressing speckle arising from non-resolved tissue structure, yet some methods produce blurred images. Moreover, although effective at suppressing speckle in ex vivo tissues or in preclinical animal research, the additional time and data throughput required to obtain multiple speckle realizations can often make this approach incompatible with clinical in vivo imaging.

Consequently, many numerical algorithms attempt to computationally suppress speckle. To name a few: non-linear filtering \cite{ozcan:2007}, non-local means (NLM) \cite{yu:2016,cuartas:2018}, and block matching and 3D filtering (BM3D) \cite{chong:2013}. The majority of these algorithms employ an image denoiser treating speckle as independent and identically distributed (i.i.d) Guassian noise. The solution can sometimes be sensitive to parameters' fine-tuning. Some algorithms also rely on accurately registered volumetric data, which is challenging to obtain in clinical settings. 

Recently, the speckle reduction task has been extensively investigated from a supervised learning perspective \cite{shi:2019,devalla:2019,gour:2020}. As known, most supervised learning data-driven methods require a large training dataset. 
In OCT, Dong et al.(2020) \cite{Dong:2020} trained a super-resolution generative adversarial network (SRGAN) \cite{Ledig:2017,Goodfellow:2014} with hardware-based speckle suppressed ex vivo samples defining the ground-truth. Namely, they used 200,000 speckle-modulating OCT images of size $800 \times 600$ for training. Chintada et al. (2023) \cite{Chintada:2023} use a conditional-GAN (cGan) \cite{isola:2017} trained with hundreds of retinal data B-scans, with NLM \cite{cuartas:2018} as ground truth.
Ma et al. \cite{ma:2018} (2018) also use a cGAN to perform speckle reduction and contrast enhancement for retinal OCT images by adding an edge loss function to the final objective. The clean images for training are obtained by averaging of B-scans from multiple OCT volumes.

That said, there has been a growing amount of evidence demonstrating that supervised learning methods, specifically in the context of computational imaging and inverse problems, may require significantly smaller datasets. 
For example, it was observed that for image restoration for florescence microscopy \cite{weigert:2018}, even a small number of training images led to 
an acceptable image restoration quality (e.g., 200 patches of size $64 \times 64 \times  16$). 
Pereg et al. (2020) \cite{Pereg:2020} used a single synthetic example for seismic inversion.

In learning theory, domain shift is a change of data distribution between the source domain (training dataset) and target domain (the test dataset). Despite advances in data augmentation and transfer learning, neural networks often fail to adapt to unseen domains. For example, convolutional neural networks (CNNs) trained for segmentation task can be highly sensitive to changes in resolution and contrast. Performance often degrades even within the same imaging modality. A general review of domain adaptation (DA) for medical image analysis can be found in \cite{Guan:2021}. The different approaches are separated into shallow and deep DA models, further divided into supervised, semi-supervised and unsupervised DA, depending on the availability of labeled data in the target domain. Generally speaking, the appropriate DA approach depends on the background and the properties of the specific problem. Many DA methods suggest ways to map the source and target domains to a shared latent space. Whereas, generative DA methods attempt to translate the source to the target or vice versa. 
In our study, we focus on a simple yet efficient physics-aware unsupervised DA approach for the case of a change in the OCT imaging system. Namely, only unlabeled data is available for the target domain. This problem is also referred to in the literature as domain generalization \cite{wang:2022}, and it has been hardly explored in medical imaging so far \cite{zhang:2020}.


In this work, we investigate few-shot learning as a powerful platform for OCT speckle reduction with limited ground truth training data. 
We prove that the output resolution of a supervised learning speckle suppression system is determined by the sampling space and the resolution of the source acquisition system. We also mathematically define the effects of the domain shift on the target output image. 
In light of the theoretical analysis, we promote the use of a patch-based recurrent neural net (RNN) framework to demonstrate the applicability and efficiency for few-shot learning for OCT speckle suppression. We demonstrate the use of a single image training dataset, that generalizes well. The proposed approach introduces a dramatic decrease in training time and required computational resources. Training takes about \textit{2-25 seconds} on a GPU workstation and a few minutes on a CPU workstation (2-4 minutes). 
We further propose novel upgrades for the original RNN framework and compare their performance. Namely, we introduce a one-shot patch-based RNN-mini-GAN architecture. We further demonstrate increased SNR achieved via averaging overlapping patches. Furthermore, we recast the speckle suppression network to a deblurring system. We further compare the 3 different RNN models' results with a patch-based one-shot learning U-Net \cite{ronneberger:2015}.  
We illuminate the speckle reduction dependence of the acquisition system, via known lateral and axial sampling space and resolution, and offer strategies for training and testing under different acquisition systems.  
Finally, our approach can be applicable to other learning architectures as well as other applications where the signal can be processed locally, such as speech and audio, video, seismic imaging, MRI, ultrasound, natural language processing and more. The results in this paper are a substantial extension that replaces our non-published previous work (\cite{Pereg:2022}, Section 6.2).

\vspace{-5pt}
\section{Preliminaries}\label{sec2}

\subsection{RNN Encoder-Decoder Framework} \label{sec2.1}
Assume an observation data sequence $\mathbf{y}=[\mathbf{y}_{0}, \mathbf{y}_{1}, ..., \mathbf{y}_{L_{\mathrm{t}-1}}]$, $\mathbf{y}_t \in \mathbb{R}^{L_{\mathrm{t}} \times 1}, \ t \in [0,{L_{\mathrm{t}}-1}] $, and a corresponding output sequence $\mathbf{x}=[\mathbf{x}_{0}, \mathbf{x}_{1}, ..., \mathbf{x}_{L_{\mathrm{t}}-1}]$, $\mathbf{x}_t \in \mathbb{R}^{P \times 1}$. The RNN forms a map $f : \mathbf{y}\rightarrow \mathbf{z}$, from the input data to the latent space variables. 
That is, for input $\mathbf{y}_t$ and state $\mathbf{z}_t$ at time step $t$, the RNN output is generally formulated as 
$\mathbf{z}_t =  f(\mathbf{z}_{t-1},\mathbf{y}_t)$ \cite{Bengio:2013}.
Hereafter, we focus on the specific parametrization:
\begin{equation}\label{2.2}
\mathbf{z}_t=\sigma(\mathbf{W}^T_{zy} \mathbf{y}_t + \mathbf{W}^T_{zz} \mathbf{z}_{t-1} + \textbf{b}),
\end{equation}
where $\sigma$ is an activation function, $\mathbf{W}_{zy} \in \mathbb{R}^{N \times n_{\mathrm{n}}}$ and $\mathbf{W}_{zz} \in \mathbb{R}^{n_{\mathrm{n}} \times n_{\mathrm{n}}}$ are weight matrices and $\mathbf{b}\in \mathbb{R}^{n_{\mathrm{n}} \times 1}$ is the bias vector. At $t=0$ previous outputs are zero. Here, we use the ReLU activation function, $\mathrm{ReLU}(z)=\max\{0,z\}$. We wrap each cell with a fully connected layer with the desired final output $\mathbf{x}_t \in \mathbb{R}^{P \times 1}$, such that
$\mathbf{x}_t = \mathrm{FC}(\mathbf{z}_t)$.

Traditionally, RNNs are used for processing of time related signals, to predict future outcomes, and for natural language processing tasks such as handwriting recognition \cite{Graves:2009} and speech recognition \cite{Graves:2013}. In computer vision, recurrent convolutional networks (RCNNs) were proposed for object recognition \cite{Liang:2015}. Pixel-RNN \cite{Van:2016} sequentially predicts pixels in an image along the two spatial dimensions.

\vspace{-10pt}
\subsection{Speckle Statistics}
OCT tomograms display the intensity of the scattered light, as the log-valued squared norm of the complex-valued tomogram. It is assumed that the contributions from structural features beyond the imaging resolution of OCT add up coherently, and generate a random speckle pattern \cite{Schmitt:1999,Goodman:2007}. Speckle is not an additive statistically independent noise, but rather unresolved spatial information originating in the interference of many sub-resolution spaced scatterers \cite{Curatolo:2013}.
Speckle also plays an important role in other fields, e.g., synthetic-aperture radar, and ultrasound medical imaging. 
Exact analogs of the speckle phenomenon appear in many other fields and
applications. The squared magnitude of the finite-time Fourier
transform (FFT) (the periodogram) of a sample function of almost any random process shows
fluctuations in the frequency domain that have the same single-point (pixel) statistics
as speckle \cite{Porat:1996}.
Generally speaking, speckle appears in a signal when the signal is a linear combination of
independently random phased additive complex elements. 
The resulting sum is a random
walk, that may exhibit constructive or destructive interference depending on the
relative phases. The intensity of the
observed wave is the squared norm this sum .

As mention above, speckle in OCT arises from sub-resolution reflectors. 
In addition optical frequency domain imaging (OFDI) OCT is the FFT of its measured spectral components, and therefore exhibits noise typical to any periodogram. 
When a pixel's value is a result of a a sum of large enough number of reflectors, the sum, 
according to the central limit theorem, has a Gaussian distribution. 
In this case, assuming a uniform phase, for fully developed speckle,
the intensity is distributed according to an exponential probability density,
\vspace{-5pt}
\begin{equation}
p_{Y|X}(y|x)=\frac{1}{x}\exp \Big(-\frac{y}{x}\Big), \quad y>0
\label{2.4}
\end{equation}
where $y$ is the measured intensity pixel value, and $x$ is the mean intensity, defining the ground truth. 
In other words, the fluctuations of fully developed speckle are of the same order as the
the ground truth pixel value, which renders this type of noise to be particularly destructive and visually disturbing.
Speckle that is not fully developed would have more complicated distributions formulation, depending on the number of phasors, and their amplitudes and phases distribution. 

\section{Domain Aware Speckle Suppression} \label{sec3} 
Let us denote $f(\mathscr{z,x}) \in \mathbb{C}$ as the ground truth ideal tomogram perfectly describing the depth sample reflectivity.
Here $\big\{(\mathscr{z,x}): \mathscr{z,x} \geq 0,  (\mathscr{z,x})\in \mathbb{R}^2 \big\}$ are continuous axial and lateral spatial axes. A measured tomogram can be formulated as 
\begin{equation}\label{3.1}
Y(\mathscr{z,x})=10\log_{10}\big(|f(\mathscr{z,x})*\alpha(\mathscr{z,x})|^2\big).
\end{equation}
where $*$ denotes the convolution operation and $\alpha(z,x)$ is a point spread function (PSF).
In the discrete setting, assuming $F^{z}_{\mathrm{s}}$, $F^{x}_{\mathrm{s}}$ axial and lateral sampling rates respectively, and that the set of measured values at $\{\mathscr{z}_m,\mathscr{x}_n\}$ lie on the grid $m/F^{z}_{\mathrm{s}}$ and $n/F^{x}_{\mathrm{s}}$, $m,n \in \mathbb{N}$, 
\begin{equation}\label{3.2}
Y[m,n]=10\log_{10}\big(\big|f[m,n]*\alpha[m,n]\big|^2\big)
\end{equation}
A speckle suppressed tomogram can be viewed as the incoherent mean of coherent tomograms with different speckle realizations \cite{Goodman:2007,mehta:2010},
\begin{equation}\label{3.3}
X[m,n]=10\log_{10}\big(\big|f[m,n]\big|^2*\big|\alpha[m,n]\big|^2\big)
\end{equation}

In OFDI OCT (using a wavelength-swept source) \cite{Villiger:2015,Drexler:2015}, the axial range $\Delta_\mathrm{z}$ is given by the central wavelength and the wavelength sampling. The axial sampling space is $\delta z = \frac{\Delta_\mathrm{z}}{n_\mathrm{z}}$, where $n_{\mathrm{z}}$ the total A-line number of pixels. 
In the axial direction, the PSF effective width $\omega_\mathrm{z}$ is determined by the FFT of a zero-padded Hanning window. 
In the lateral direction, the PSF has a Gaussian shape proportional to {\small$\exp(-2x^2/w_\mathrm{x}^2)$}, where the $w_\mathrm{x}$ is referred to as the waist. $\delta x$ is the lateral sampling space. 
Therefore, $\alpha[m,n]$ is separable and can be expressed as 
$\alpha[m,n]=\alpha_{\mathrm{x}}[m]*\alpha_{\mathrm{z}}[n]$.
Note that, the resolution and sampling rate are known parameters of an OCT imaging system. 


In matrix-vector form, we denote an input (log-scaled) image $\mathbf{Y}\in\mathbb{R}^{L_\mathrm{r} \times J}$ that is a corrupted version of $\mathbf{X}\in\mathbb{R}^{L_\mathrm{r} \times J}$, such that
$\mathbf{Y}=\mathbf{X}+\mathbf{N}$,
where $\mathbf{N}\in\mathbb{R}^{L_\mathrm{r} \times J}$ is an additional noise term.
Note that for the case of image despeckling we do not assume that the entries of $\mathbf{N}$ are neither i.i.d nor that it is uncorrelated with $\mathbf{X}$ \footnote{The speckled noise setting can also be described as $y=f(x)+n(x)$, where $f(\cdot)$ and $n(\cdot)$ are some functions.}.
Our task is to recover $\mathbf{X}$. That is, we attempt to find an estimate $\hat{\mathbf{X}}$ of the unknown ground truth $\mathbf{X}$. 

Let us assume a source training set $\mathcal{D}_{\mathcal{S}}=\big\{\{\mathbf{y}_{i},\mathbf{x}_{i}\}_{i=1}^m : {\mathbf{y}}_{i} \in \mathbb{R}^{L_\mathrm{t} \times N_{\mathrm{x}}}, {\mathbf{x}}_{i} \in \mathbb{R}^{L_\mathrm{t} \times N_{\mathrm{x}}} \big\}$, where $\{\mathbf{y}_{i},\mathbf{x}_{i}\} \sim P_{\mathcal{S}}$ are image patches sampled from a source domain $\mathcal{S}$ as ground truth. The learning system is trained to output a prediction rule $\mathcal{F}_\mathcal{S}: \mathcal{Y} \rightarrow \mathcal{X}$. We assume an algorithm that trains the predictor by minimizing the training error (empirical error or empirical risk). The domain shift problem assumes a target domain $\mathcal{T}$ with samples from a different distribution $\{\mathbf{y}_{i},\mathbf{x}_{i}\} \sim P_{\mathcal{T}}$.

\textit{Assumption 1.} (Speckle Local Ergodicity).
Denote $\mathbf{y}_i$ a patch centered around pixel $i$ of the image $\mathbf{Y}$.
$P_Y(\mathbf{y}_i)$ is the probability density  of a patch $\mathbf{y}_i$.
Under the assumption that pixels in close proximity are a result of shared similar sub-resolution scatterers we assume ergodicity of the Markov random field (MRF, e.g. \cite{li:2009}) of patches $\mathbf{y}_i$ consisting of pixels in close proximity. In other words, the probability distribution of a group of pixels' values in close spatial proximity is defined by the same density across the entire image. This assumption takes into account that some of these patches correspond to fully developed speckle, non-fully developed speckle, and a combination of both. Note that the measured pixels' values are correlated. 
That said, this assumption could be somewhat controversial, particularly in surrounding of abrupt changes in the signal intensity. But since our images tend to have a layered structure, and the PSF visible range is about 7-9 pixels in each direction, we will make this assumption. 

\textit{Defionition 2.} (Sampling-Resolution Ratio).\label{Def2} We define the lateral sampling-resolution ratio $p_\mathrm{x} \triangleq \big[\frac{\omega_\mathrm{x}}{\delta x}\big]$ in \textit{pixels}, and the axial sampling-resolution ratio as $p_\mathrm{z}\triangleq\big[\frac{\omega_\mathrm{z}}{\delta z}\big]$ where $[\cdot]$ denotes rounding to the closest integer. That is, in a discrete setting, $p_\mathrm{z}$ and $p_\mathrm{x}$ are the number of pixels capturing the effective area of the PSF in each direction. The superscripts $^t$ and $^s$ denote the target and source respectively.

\textit{Theorem 3.} (Domain-Shift Speckle Suppression Theorem). \label{Theorem3}
A (learned) speckle suppression mapping $\mathcal{F}_\mathcal{S}: y^s \rightarrow x^s $ does not require domain adaptation. But, the output $\hat{x}^t=\mathcal{F}_{\mathcal{S}}\big\{y^t\big\}$ resolution will be determined by the source domain resolution.  
Mathematically, denote $\alpha^s[m,n]$, $\alpha^t[m,n]$ as the discrete PSF in the source and target domain, respectively, such that
\begin{equation}\label{3.5}
\alpha^s[m,n]*\alpha^{s \rightarrow t}[m,n]=\alpha^t[m,n]*\alpha^{t \rightarrow s}[m,n],
\end{equation}
where $\alpha^{s \rightarrow t}[m,n]$ and $\alpha^{t \rightarrow s}[m,n]$ are complementary impulse responses leading from one domain to the other
\footnote{(\ref{3.5}) does not apply to any general pair of PSFs but in our case study it is safe to assume that there exist $\alpha^{s \rightarrow t}[m,n]$ and $\alpha^{t \rightarrow s}[m,n]$ that approximately satisfy (\ref{3.5}).}. 
When applying the trained system to the target input we have,
{\footnotesize
\begin{flalign}\label{3.6}
\nonumber
& \hat{x}^t[m,n] =
\mathcal{F}_\mathcal{S}\big(y^t[m,n]\big) 
 = x^s[m,n]\Big|_{y^s[m,n]=y^t[m,n]}\\
& = \big|f^s[m,n]\big|^2*\big|\alpha^s[m,n]\big|^2
\Big|_{f^t[m,n]*\alpha^{s \rightarrow t}[m,n]=f^s[m,n]*\alpha^{t \rightarrow s}[m,n]}.
\end{flalign}
}We refer the reader to Appendix A for the proof and a detailed explanation.
For example, if $p_\mathrm{z}^s=p_\mathrm{z}^t$, and $p_\mathrm{x}^s<p_\mathrm{x}^t$, there exist $\alpha^{s \rightarrow t}_{\mathrm{x}}[n]$ such that $\alpha_{\mathrm{x}}^t[n] = \alpha_{\mathrm{x}}^s[n]* \alpha_{\mathrm{x}}^{s \rightarrow t}[n]$.
Then, we have
{\footnotesize
\begin{flalign*}
\hat{x}^t[m,n] &= \mathcal{F}_\mathcal{S}\big(y^t[m,n]\big) = \\
& = \big|f^t[m,n]*\alpha^{s \rightarrow t}[m,n]\big|^2*\big|\alpha^s[m,n]\big|^2.
\end{flalign*}
}In other words, the output resolution is determined by the source resolution $\alpha^s[m,n]$.
The tomogram component $|f^t[m,n]*\alpha^{s \rightarrow t}[m,n]|$ is a low resolution version of the target tomogram $|f^t[m,n]|$. 
\begin{enumerate} 
\item If $p_\mathrm{x}^{s}<p_\mathrm{x}^{t}$ or $p_\mathrm{z}^{s}<p_\mathrm{z}^{t}$ then the system's prediction for an input in the target domain may have additional details or an artificially enhanced resolution details, that would not naturally occur with other denoising mechanisms. Examples illustrating this phenomena are illustrated in Fig.~\ref{fig1}(e) and Fig.~\ref{fig3}(e)-(f). 
Possible remedies: train with larger analysis patch size, longer training, upsampling (interpolation) of source images (or decimation of target images).
\item If $p_\mathrm{x}^{s}>p_\mathrm{x}^{t}$ or $p_\mathrm{z}^{s}>p_\mathrm{z}^{t}$ then the network's output is blurred in the corresponding direction (e.g., Fig.~\ref{fig6}(e) and Fig.\ref{fig3}(h)). Possible remedies: train with smaller analysis patch size 
, downsample (decimate) training image (or upsample target images).
In this case the target has details that are smaller (in pixels) than the minimal speckle size of the source, which could be interpreted by the trained predictor as noise, thus the trained predictor may simply smear them out.
\end{enumerate}
Any combination of relations $p_{\mathrm{z},\mathrm{x}}^s \lessgtr p_{\mathrm{z},\mathrm{x}}^t$ along the different image axes is possible. For our OCT data, the resolution ratio mostly differs in the lateral direction (see Table 1). Note that for some OCT systems sampling space is below Nyquist rate. 
Preprocessing domain adaptation stage can be applied either to the source data or the target data, interchangeably, depending on the desired target resolution.

\section{Few Shot Learning via RNN}\label{sec4}
%

The initial RNN setting described in this subsection has been previously employed for seismic imaging \cite{Biswas:2018,Pereg:2019B,Pereg:2020}. 
Hereafter, the mathematical formulation focuses on the settings of the OCT despeckling task. Nonetheless, the model can be applied to a wide range of applications. 
We emphasize the potential of this framework, expand and elaborate its application, while connecting it to the theoretical intuition in Theorem 3. We also propose possible upgrades to further enhance the results in our case study. 

Most OCT images have a layered structure, and exhibit strong relations along the axial and lateral axes. RNNs can efficiently capture those relations and exploit them. 
That said, as demonstrated below, the proposed framework is not restricted to images that exhibit a layered structure, nor to the specific RNN-based encoder-decoder architecture. 

\textit{Definition 4 (Analysis Patch)\cite{Pereg:2019B}}.\label{Def4} We define an \textit{analysis patch} as a 2D patch of size $L_t\times N_\mathrm{x}$ enclosing $L_\mathrm{t}$ time (depth) samples of $N_\mathrm{x}$ consecutive neighboring columns of the observed image $\mathbf{Y}\in\mathbb{R}^{L_\mathrm{r} \times J}$. Assume $\{n_\mathrm{L},n_\mathrm{R} \in \mathbb{N} :n_\mathrm{L}+n_\mathrm{R}=N_\mathrm{x}-1\}$. Then the analysis patch associated with a point at location $(i,j)$ is
{\small
\begin{equation*}
\mathbf{A}^{(i,j)}_{k,l} = \{ Y[i+k,j+l]: k,l \in \mathbb{Z}, 1-L_t \leq k \leq 0, \ -n_L \leq l \leq n_R \}.
\end{equation*}
}An analysis patch $\mathbf{A}^{(i,j)}\in\mathbb{R}^{L_\mathrm{t} \times N_\mathrm{x}}$ is associated with a pixel $X[i,j]$ in the output image. To produce a point in the estimated $\hat{X}[i,j]$, we set an input to the RNN as an analysis patch, i.e., $\mathbf{y} = \mathbf{A}^{(i,j)}$. 
Each time step input is a group of $N$ neighboring pixels of the same corresponding time (depth). In other words, in our application $n_\mathrm{i}=N_\mathrm{x}$ and
$\mathbf{y}_t = \big[ Y[t,j-n_L], ...,Y[t,j+n_R]\big] ^T$.
We set the size of the output vector $\mathbf{z}_t$ to one expected pixel ($P=1$), such that $\mathbf{x}$ is expected to be the corresponding reflectivity segment, 
$\mathbf{x} = \big[ X[i-(L_t-1),j], ... , X[i,j] \big] ^T$.
Lastly, we ignore the first $L_t-1$ values of the output $\mathbf{x}$ and set the predicted reflectivity pixel $\hat{X}[i,j]$ as the last one, i.e, $\mathbf{x}_{L_t}$. 
The analysis patch moves across the image and produces all predicted points in the same manner. Each analysis patch and a corresponding output segment (or patch) are an instance for the net. The size and shape of the analysis patch defines the geometrical distribution of data samples for inference.

\paragraph*{Despeckling Reformulated as Image Deblurring}
Despite low-frequency bias of over-parametrized DNNs \cite{basri:2020}, previous works \cite{Pereg:2019B} demonstrate the ability of the proposed framework in promoting high frequencies and super-resolution.
To explore this possibility, we recast the framework described above to a deblurring task. This is achieved simply by applying a low-pass filter to the input speckled image and then training the system to deblur the image. Namely, given a noisy image $\mathbf{Y}$, the analysis patches will be extracted from the input image $\hat{\mathbf{Y}}=\mathbf{HY}$, where $\mathbf{H}$ is a convolution matrix of a 2D low pass filter. We will refer to this denoiser as deblurring RNN (DRNN).

\paragraph*{Averaging patches} 
Given a noisy image $\mathbf{Y}$, an alternative approach is to decompose it into overlapping patches, denoise
every patch separately and ﬁnally combine the results by simple averaging. This approach of averaging of
overlapping patch estimates is common in patch-based algorithms \cite{lebrun:2012,lebrun:2013}, such as Expected Patch Log-Likelihood (EPLL) \cite{hurault:2018}. It also improves SNR since we are averaging for any pixel a set of diﬀerent estimates. Mathematically speaking, the input analysis patch is still  $\mathbf{y} = \mathbf{A}^{(i,j)} \in\mathbb{R}^{L_\mathrm{t} \times N_\mathrm{x}}$. But in this configuration the output in no longer a 1D segment, but a corresponding output 2D patch. In other words, $n_\mathrm{i}=N_\mathrm{x}$, 
$\mathbf{x}_t = \big[X[t,j-n_L], ...,X[t,j+n_R]\big]^T$ such that $P=N_\mathrm{x}$.

\paragraph*{Incremental Generative Adversarial Network}
Image restoration algorithms are typically evaluated by some distortion measure (e.g. PSNR, SSIM) or
by human opinion scores that quantify perceived perceptual quality. It has long been established that distortion and perceptual quality are at odds with each other \cite{Blau:2018}. 
As mentioned above, previous works adopt a two-stages training \cite{Dong:2020,Ledig:2017}. The first stage trains the generator with a content loss. While, in the second stage, initialized by the generator’s pre-trained weights, we train both a generator $G$ and a discrimnator $D$. 
Therefore, we propose adding a second stage of training with a combined MSE and adversarial loss,
\begin{equation}
\mathcal{L}_\mathrm{G}  = \mathcal{L}_\mathrm{MSE} + \lambda \mathcal{L}_\mathrm{ADV},
\end{equation}
where $\lambda$ is a constant balancing the losses. 
The generator $G$ remains a patch-to-patch RNN-based predictor (with or without averaging patches).
To this end, we design and showcase a patch-discriminator of extremely low complexity, that consists simply of 2 fully-connected layers. We will refer to this approach as RNN-GAN.



\section{Experimental Results} \label{sec5}

Here, we show examples of our proposed few-shot domain-aware supervised learning despeckling approach with OCT experimental data, for demonstration. 
We investigated three one-shot learning challenging cases: (1) Matching tissue and matching acquisition systems, where we use one image or part of an image for training, and other images of the same tissue acquired by the same system for testing; (2) Tissue type mismatch; (3) Tissue type and acquisition system mismatch. 
Table \ref{Table1} presents the acquisition parameters, namely, axial and lateral sampling spaces in tissue, $N_{\mathrm{H}}$ - the effective number of measured spectral points vs $N_{\mathrm{FFT}}$ - total number of FFT points after zero padding, $\omega_\mathrm{x}$ - waist in $\mu$m, axial and lateral sampling-resolution ratios in pixels, and cropped image sizes.

For all experiments, we set the number of neurons as $n_n=1000$. Increasing the number of neurons did not improve the results significantly, but increases training time. The analysis patch size is $[15,15]$. Patch size can affect the results' higher frequencies. Larger patches create frequency bias in favor of lower frequencies. 
For the DRNN we used a Gaussian filter of size [7,7] and standard deviation $\sigma=1$.
For the RNN-GAN we employed overlapping patches averaging to promote additional noise reduction. 
\vspace{-10pt}
\begin{table}[h] 
\caption{Acquisition System Parameters} 
\label{Table1}
\vspace{-18pt}
\begin{center}
{\scriptsize
\begin{tabular}{|m{4.5em}|m{3.5em}|m{3.1em}|m{3.4em}|m{3.4em}|m{3.1em}|m{3.1em}|}
    \hline
	  & Chicken $\&$ Blueberry &Chicken Skin & Cucumber    
		& Retina  & Cardio-\newline vascular-I \cite{Villiger:2015}  & Cardio-\newline vascular-II \cite{otsuka:2020}
										\\ \hline \hline
		$\delta z (\mu m)$		& 6 &4.78 & 4.78  & 3.75 &4.84 & 4.43  \\ \hline
		$N_{\mathrm{H}}/N_{\mathrm{FFT}}$ &1600/2048 &844/1024 & 844/1024  &1024/2048 & 768/1024 & 800/1024 \\ \hline
		$p_\mathrm{z}$  		& 3    & 3    & 3    & 3  & 3    & 3     \\ \hline
		$\delta x (\mu m)$	& 3.06			& 2.5  & 8     & 9  & $\sim$12.2 & $\sim$24.4  \\ \hline
		$ \omega_\mathrm{x} (\mu m)$ & 8.28 & 4.14 & 8.28  & 18 & 30   & 30    \\ \hline
		$ p_\mathrm{x}$ 		& 3 & 2    & 1				 & 2	& 2    & 1     \\ \hline
		image size 		  & $350 \times 700$, $380 \times 944$ & $320 \times 650$   & $420 \times 401$ 	& $448 \times 832$		 & $1024 \times 1024$	& $1024 \times 1024$     \\ \hline
\end{tabular}  
}
\vspace{-10pt}
\end{center}
\end{table}

\paragraph*{Ex vivo OCT samples}
As ground truth for training and testing, we used hardware-based speckle mitigation obtained by dense angular compounding, in a method similar to \cite{Desjardins:2007}.
That is, ground truth images for chicken muscle, blueberry, chicken skin and cucumber sample tissues, as presented in Figs.~\ref{fig3}-\ref{fig6}(b), were acquired by an angular compounding (AC) system using sample tilting in combination with a model-based affine transformation to generate speckle suppressed ground truth data \cite{Keahey:2023}. Note that AC via sample tilting is not possible for in vivo samples.

\paragraph*{Retinal Data}
We used retinal data acquired by a retinal imaging system similar to \cite{braaf:2018}.
As ground truth for training and testing we used NLM-based speckle suppressed images \cite{cuartas:2018}. 
Note that NLM is considered relatively slow (about 23 seconds for a B-scan of size $1024 \times 1024$). Images were cropped to size $448 \times 832$.

\paragraph*{Cardiovascular OCT}
Finally, we tested our trained systems with OCT data of coronary arteries from two imaging systems. For this data we have no ground truth available. The first data referred to as Cardiovascular-I \cite{Villiger:2015} was acquire with in-house built catheters, for human cadaver imaging. The human heart coronary second data, Cardiovascular-II \cite{otsuka:2020}, was acquired with a second clinical system, where there is usually a guidewire in place. Since imaging time is critical, only 1024 A-lines per rotation were acquired.



\begin{figure*}[t]
    \begin{subfigure}[t]{0.25\textwidth}
                \includegraphics[width=\linewidth]{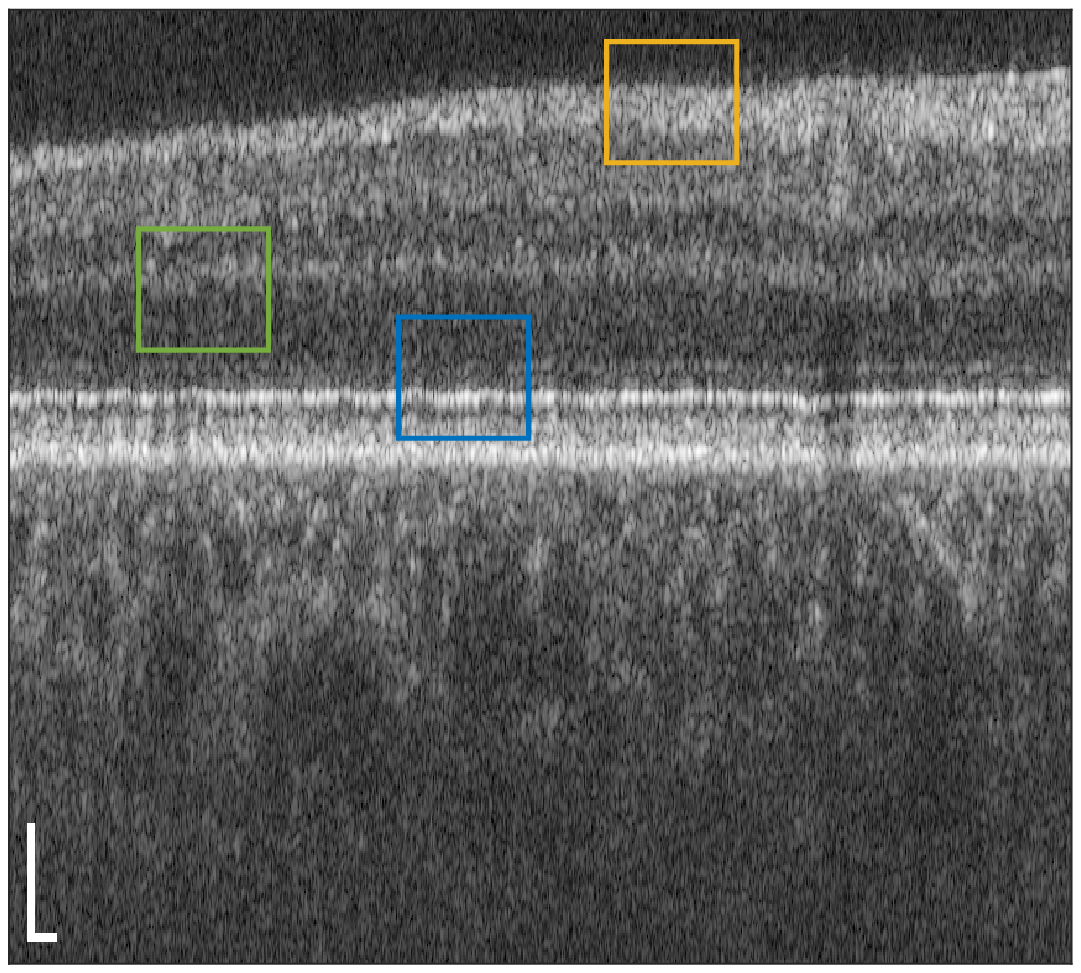} \\
				\includegraphics[width=0.33\linewidth]{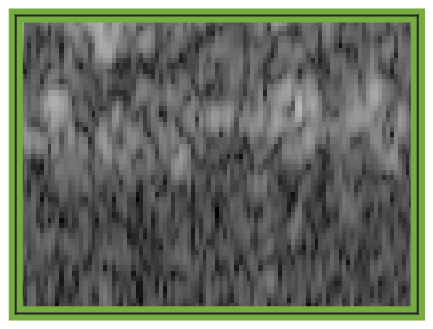}%
				\includegraphics[width=0.33\linewidth]{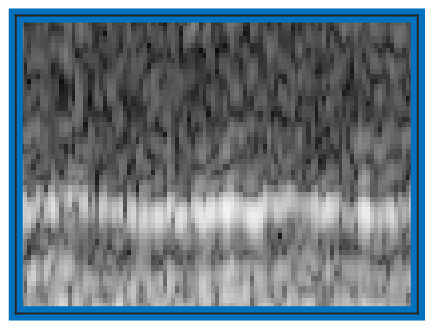}%
				\includegraphics[width=0.33\linewidth]{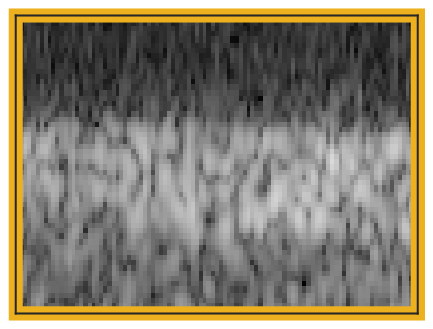}%

				\caption{{24.94dB, SSIM 0.47}}
    \end{subfigure}%
		\hfill 
		\begin{subfigure}[t]{0.25\textwidth}
        \includegraphics[width=\linewidth]{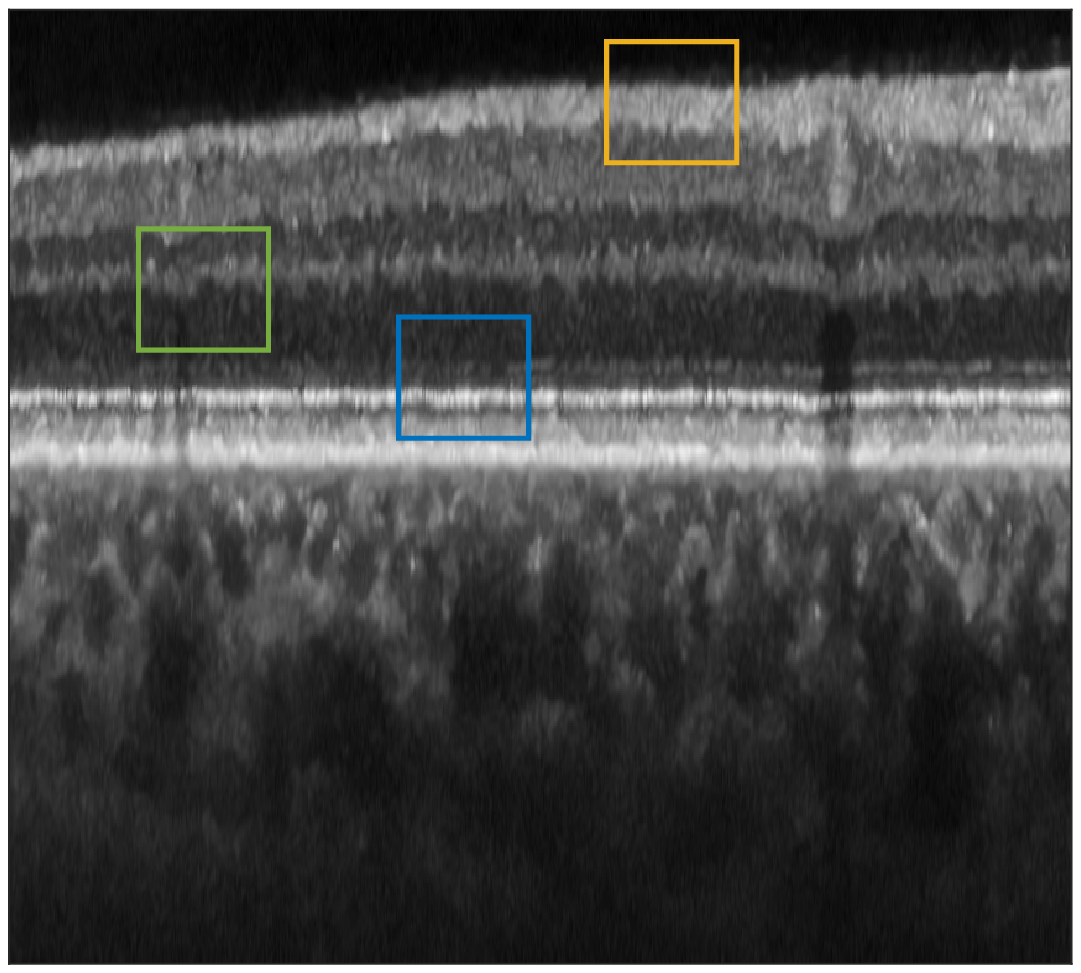} \\
				\includegraphics[width=0.33\linewidth]{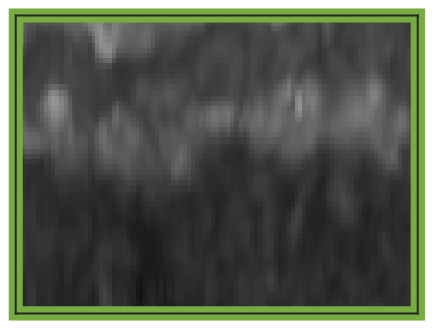}%
				\includegraphics[width=0.33\linewidth]{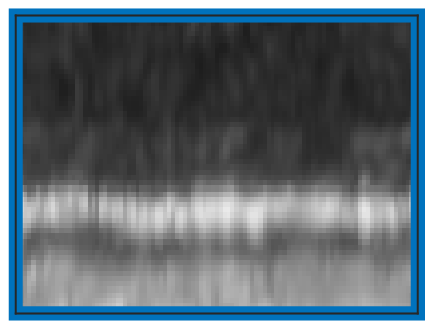}%
				\includegraphics[width=0.33\linewidth]{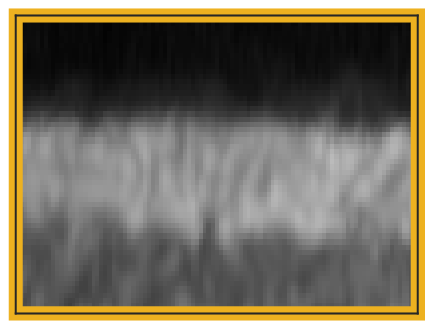}%
				\caption{}
    \end{subfigure}%
		\hfill 
    \begin{subfigure}[t]{0.25\textwidth}
        \includegraphics[width=\linewidth]{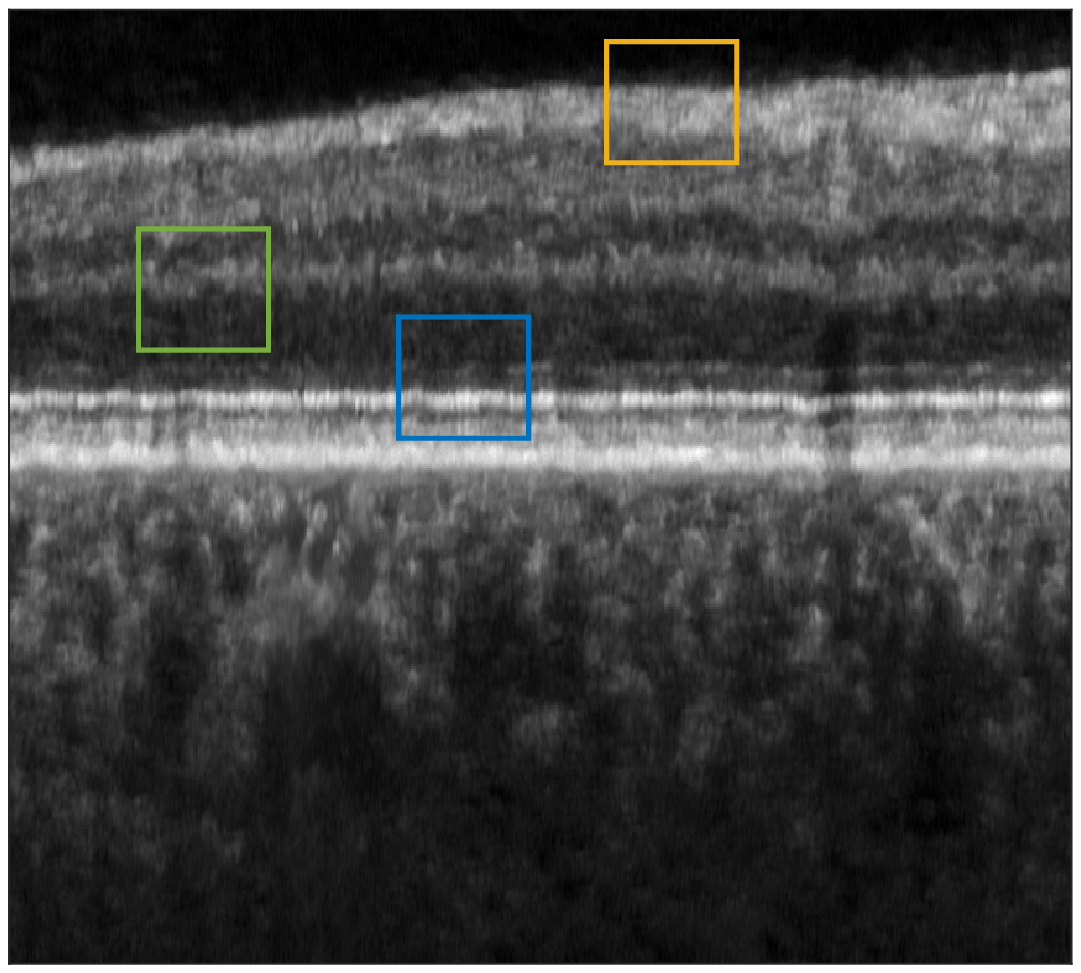} \\
				\includegraphics[width=0.33\linewidth]{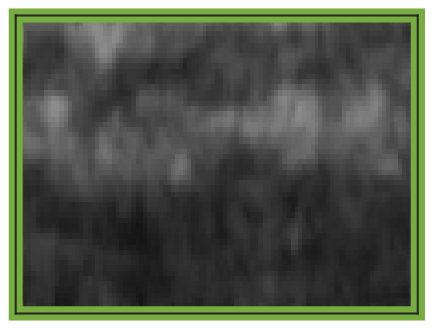}%
				\includegraphics[width=0.33\linewidth]{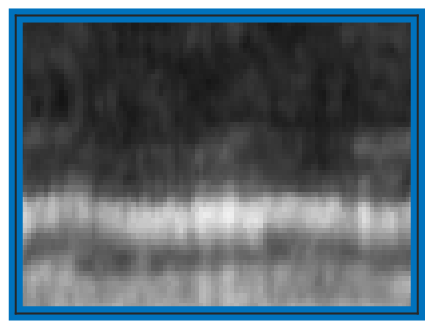}%
				\includegraphics[width=0.33\linewidth]{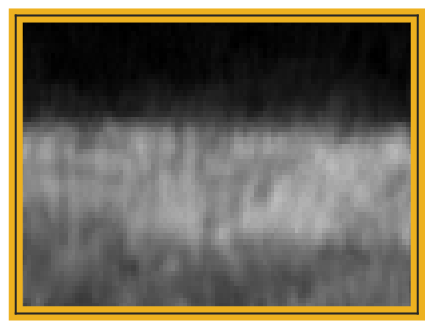}%
				\caption{31.68dB, SSIM 0.85}
    \end{subfigure}%
		\hfill 
		\begin{subfigure}[t]{0.25\textwidth}
        \includegraphics[width=\linewidth]{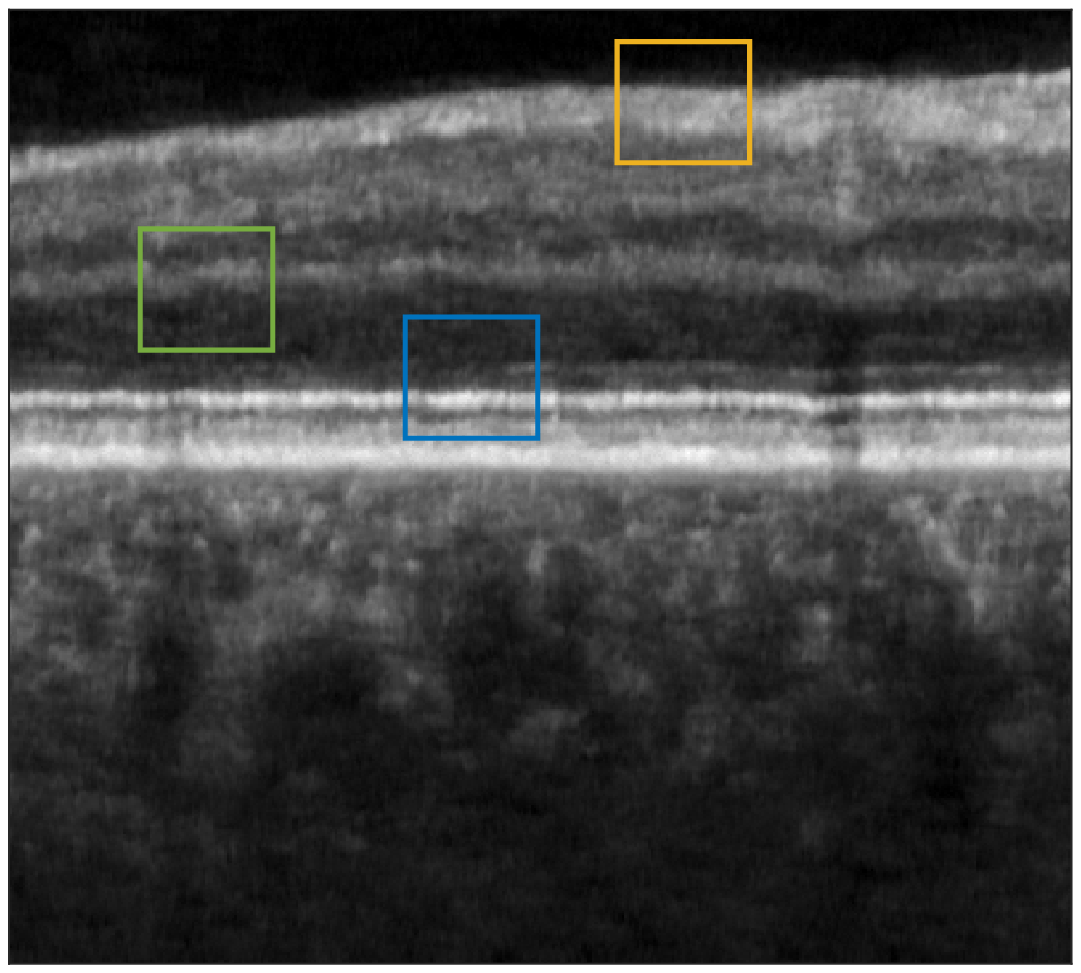}\\
				\includegraphics[width=0.33\linewidth]{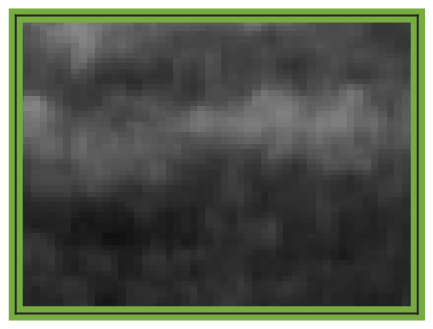}%
				\includegraphics[width=0.33\linewidth]{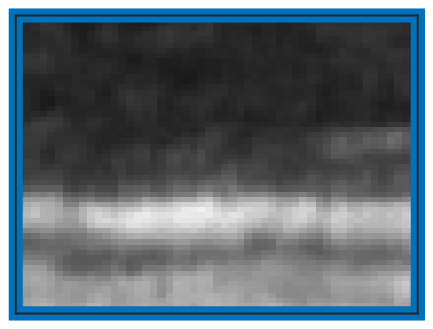}%
				\includegraphics[width=0.33\linewidth]{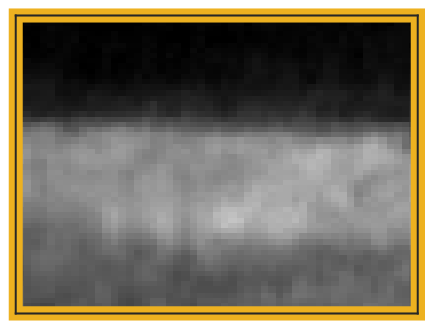}%
				\caption{31.21dB, SSIM 0.82}
    \end{subfigure}%
		\hfill
		\begin{subfigure}[t]{0.25\textwidth}
        \includegraphics[width=\linewidth]{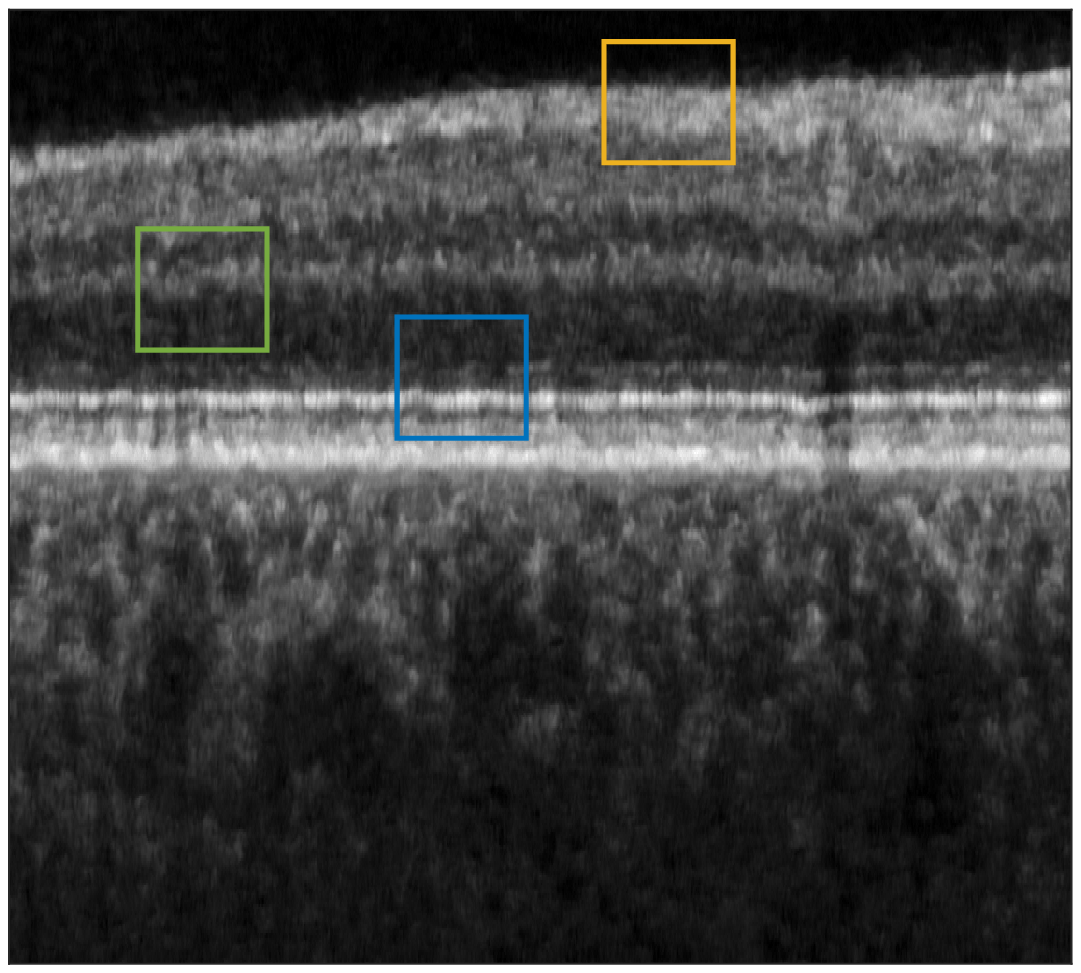}
				\\
				\includegraphics[width=0.33\linewidth]{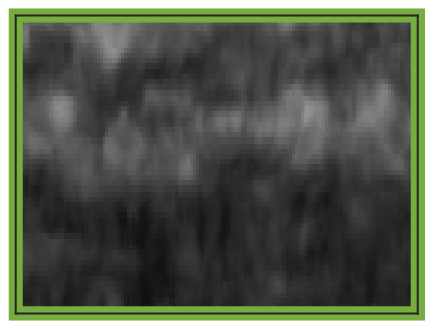}%
				\includegraphics[width=0.33\linewidth]{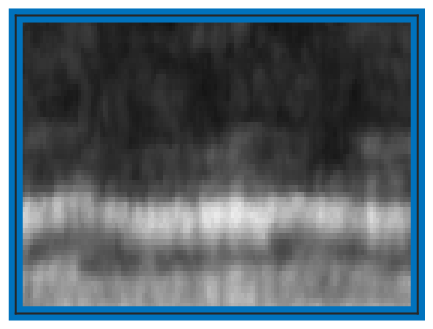}%
				\includegraphics[width=0.33\linewidth]{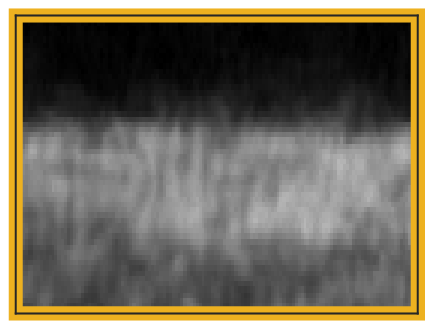}%
				\caption{31.26dB, SSIM 0.84}
    \end{subfigure}%
		\begin{subfigure}[t]{0.25\textwidth}
        \includegraphics[width=\linewidth]{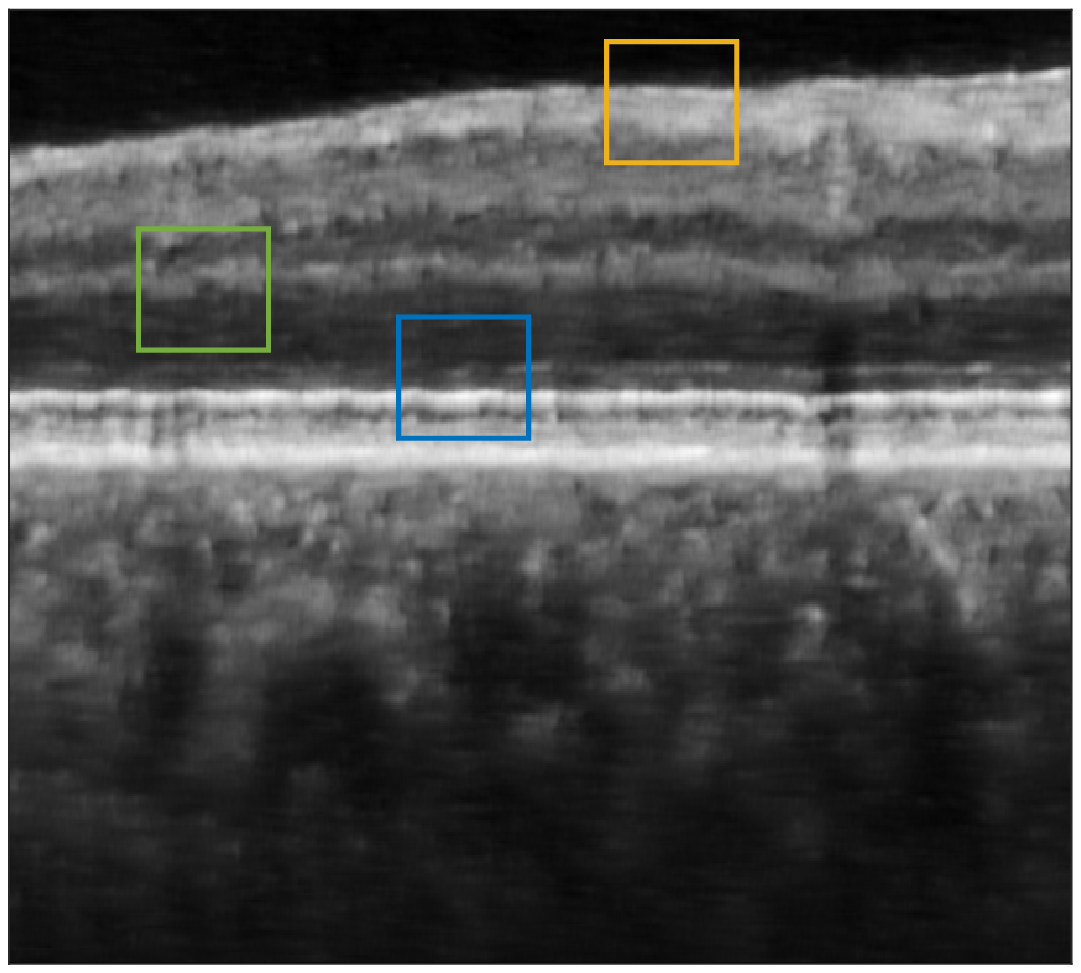}\\
				\includegraphics[width=0.33\linewidth]{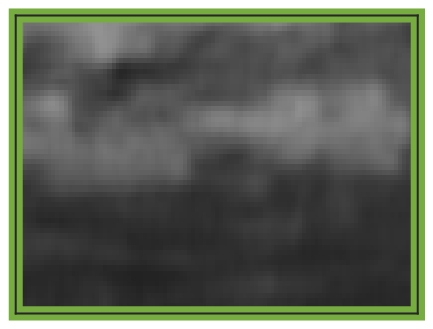}%
				\includegraphics[width=0.33\linewidth]{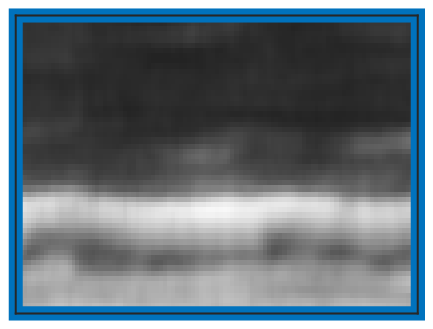}%
				\includegraphics[width=0.33\linewidth]{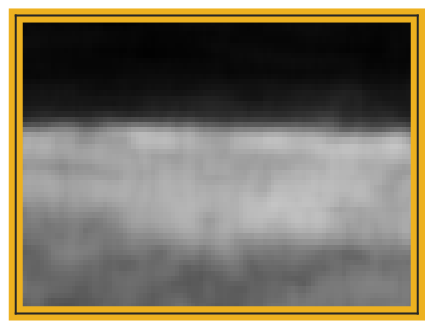}%
				\caption{30.72dB, SSIM 0.86}
    \end{subfigure}%
		\begin{subfigure}[t]{0.25\textwidth}
        \includegraphics[width=\linewidth]{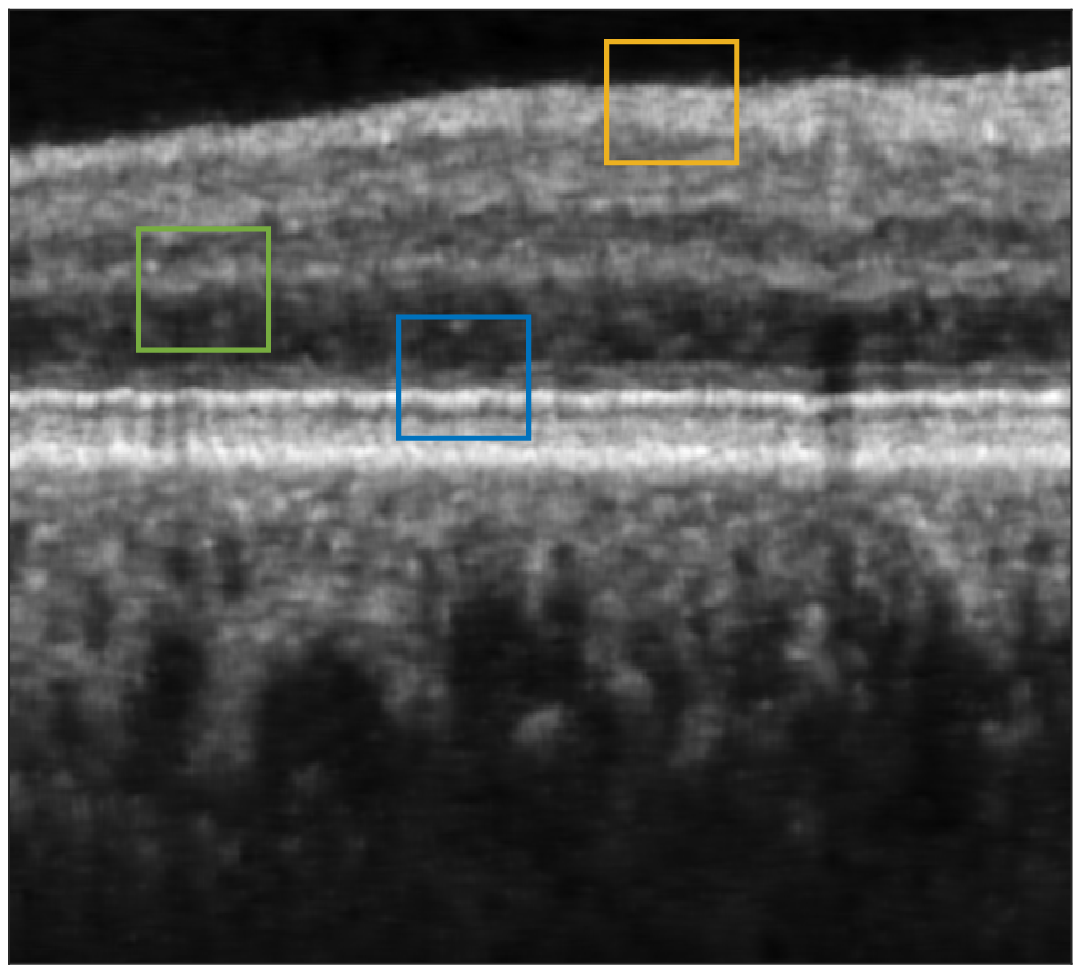}\\
				\includegraphics[width=0.33\linewidth]{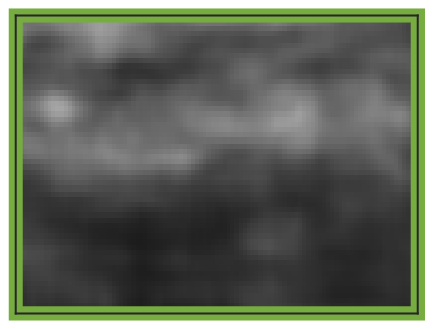}%
				\includegraphics[width=0.33\linewidth]{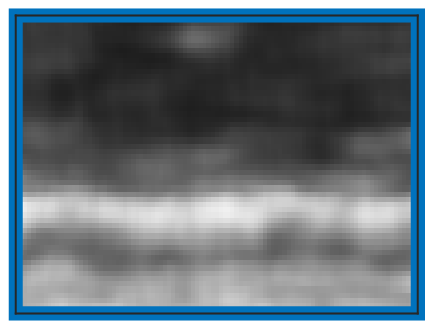}%
				\includegraphics[width=0.33\linewidth]{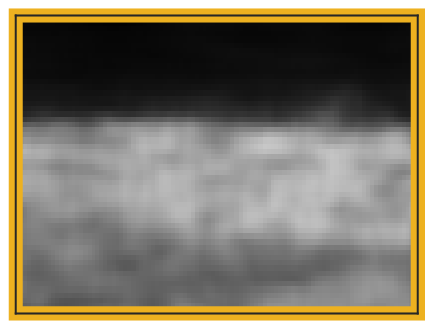}%
				\caption{29.90dB, SSIM 0.83}
    \end{subfigure}%
		\begin{subfigure}[t]{0.25\textwidth}
        \includegraphics[width=\linewidth]{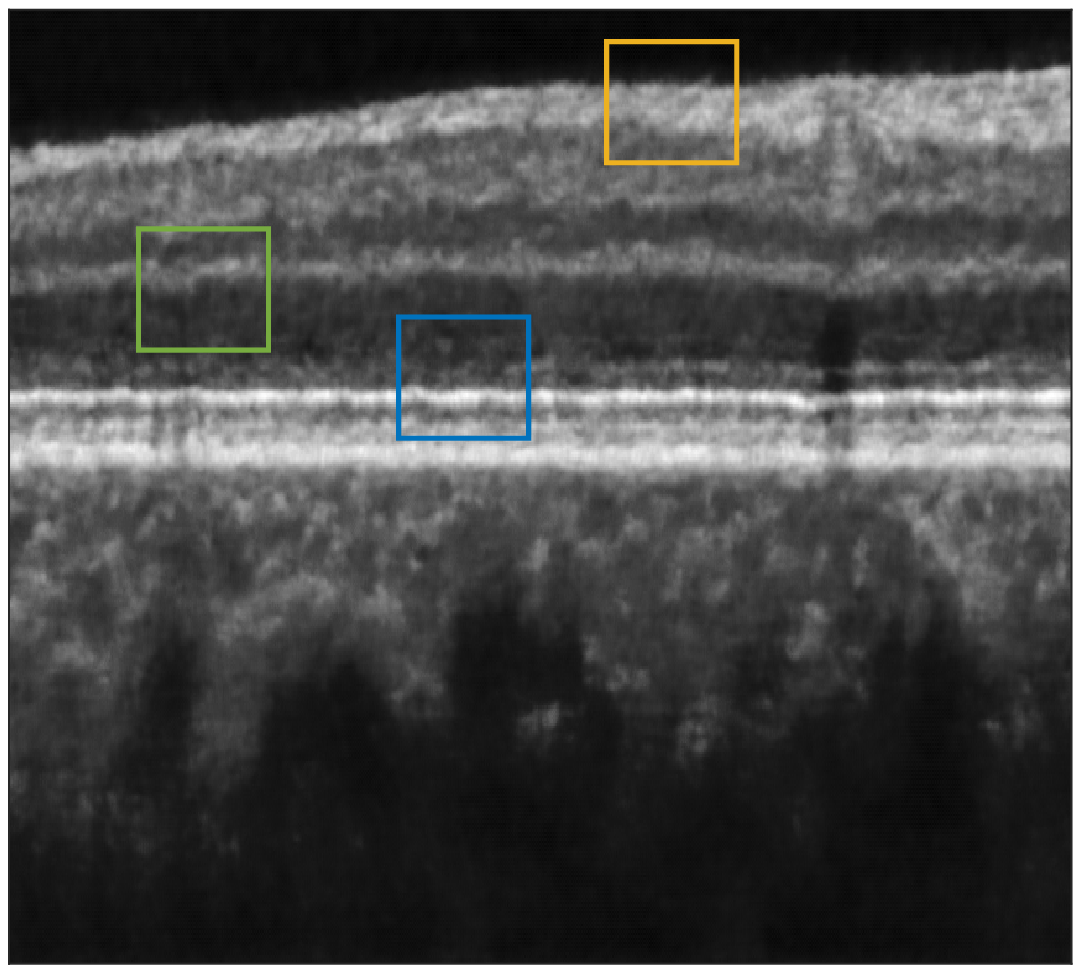}\\
				\includegraphics[width=0.33\linewidth]{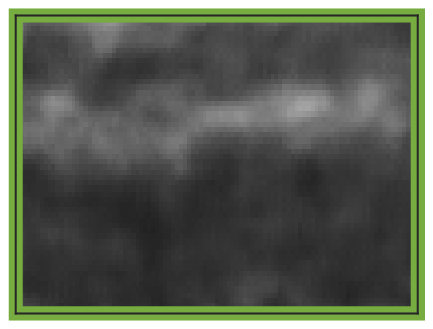}%
				\includegraphics[width=0.33\linewidth]{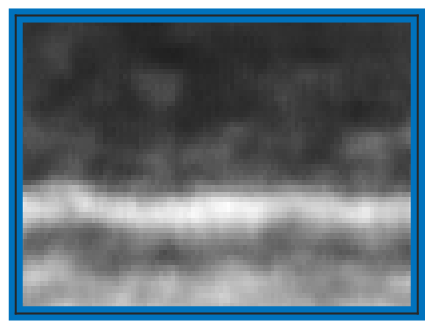}%
				\includegraphics[width=0.33\linewidth]{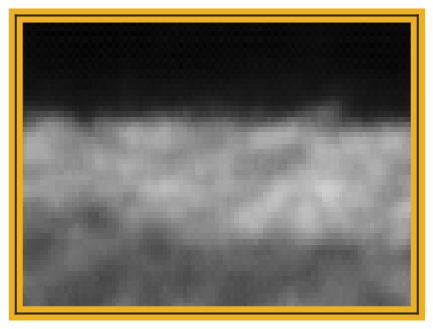}%
				\caption{30.20dB, SSIM 0.80}
    \end{subfigure}%
\caption{Retinal data speckle suppression: (a) Cross sectional human retina in vivo $p^s_{\mathrm{x}}=2$; (b) Despeckled (NLM) image used ground truth; (c) DRNN trained with 100 columns of retinal image $p^s_{\mathrm{x}}=p^t_{\mathrm{x}}=2$; System mismatch: (d) DRNN following lateral decimation of the target input by a factor of 2, $p^t_{\mathrm{x}}=1$; (e) DRNN following lateral interpolating of the input, $p^t_{\mathrm{x}}=3$; System and tissue mismatch:(f) RNN-GAN trained with 100 first columns of muscle chicken and blueberry, $p^s_{\mathrm{x}}=3$; (g) RNN-GAN trained with 200 last columns of blueberry; (h) U-Net trained with blueberry image of size $256 \times 256$, $p^s_{\mathrm{x}}=3$. Scale bars, 200 $\mu$m.}
\label{fig1}
\end{figure*}

\begin{figure*}[t]
    \begin{subfigure}[t]{0.197\textwidth}
        \includegraphics[width=\linewidth]{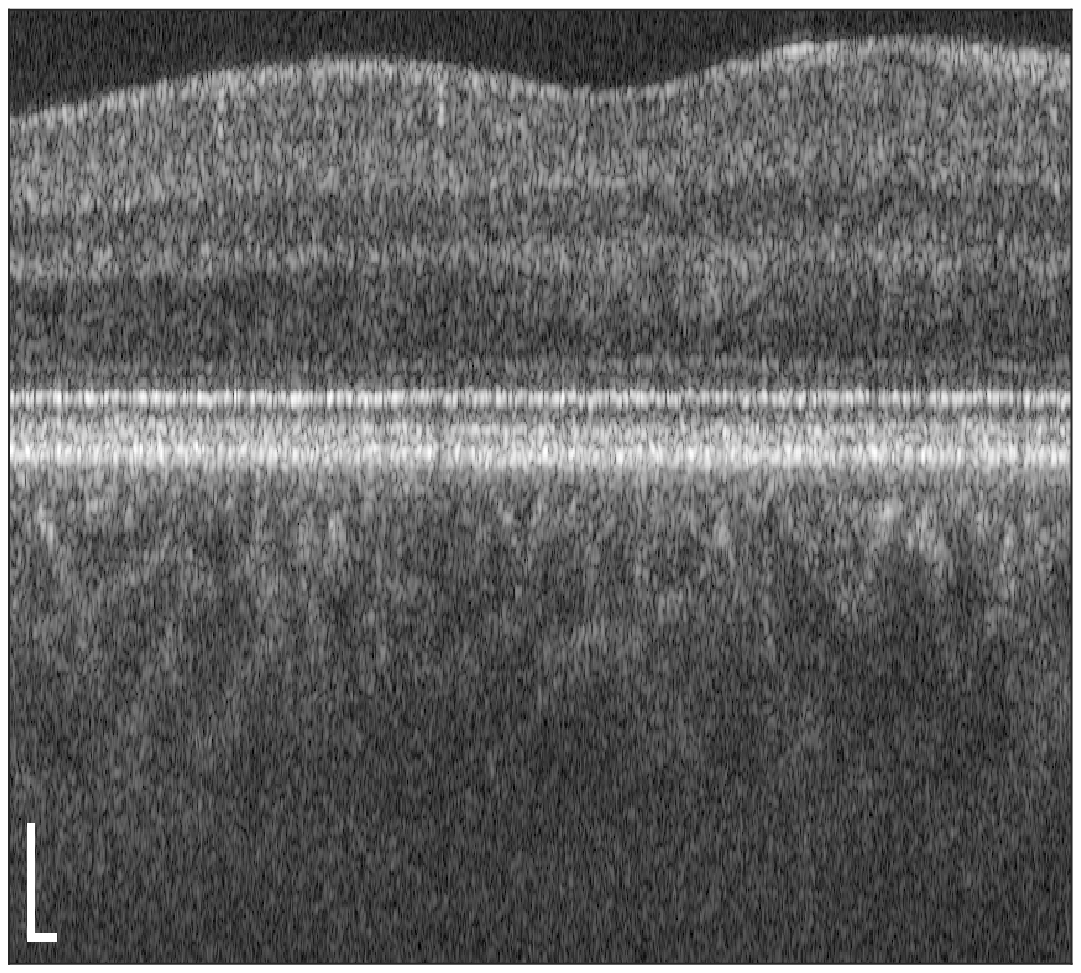}
				\caption{24.60dB, SSIM 0.46}
    \end{subfigure}%
		\hfill 
		\begin{subfigure}[t]{0.197\textwidth}
        \includegraphics[width=\linewidth]{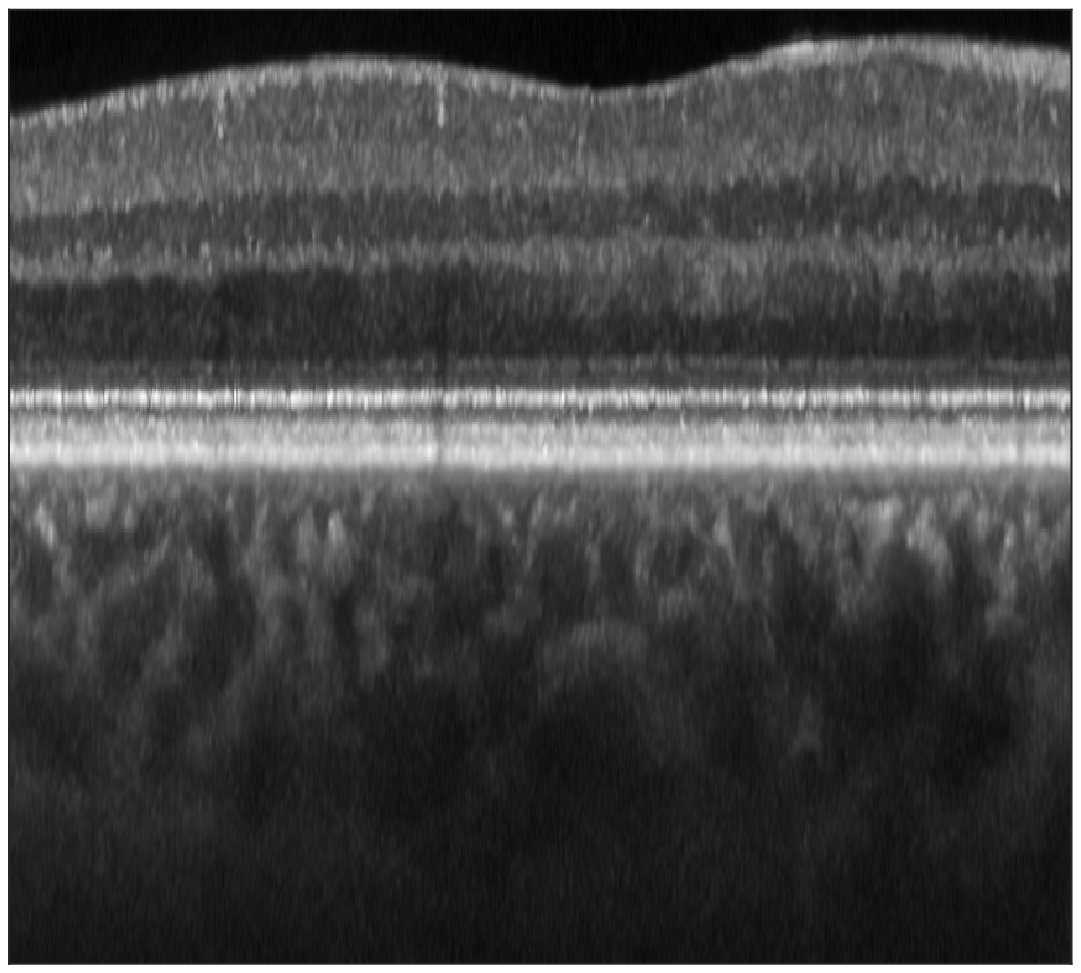}
				\caption{}
    \end{subfigure}%
		\hfill 
    \begin{subfigure}[t]{0.197\textwidth}
        \includegraphics[width=\linewidth]{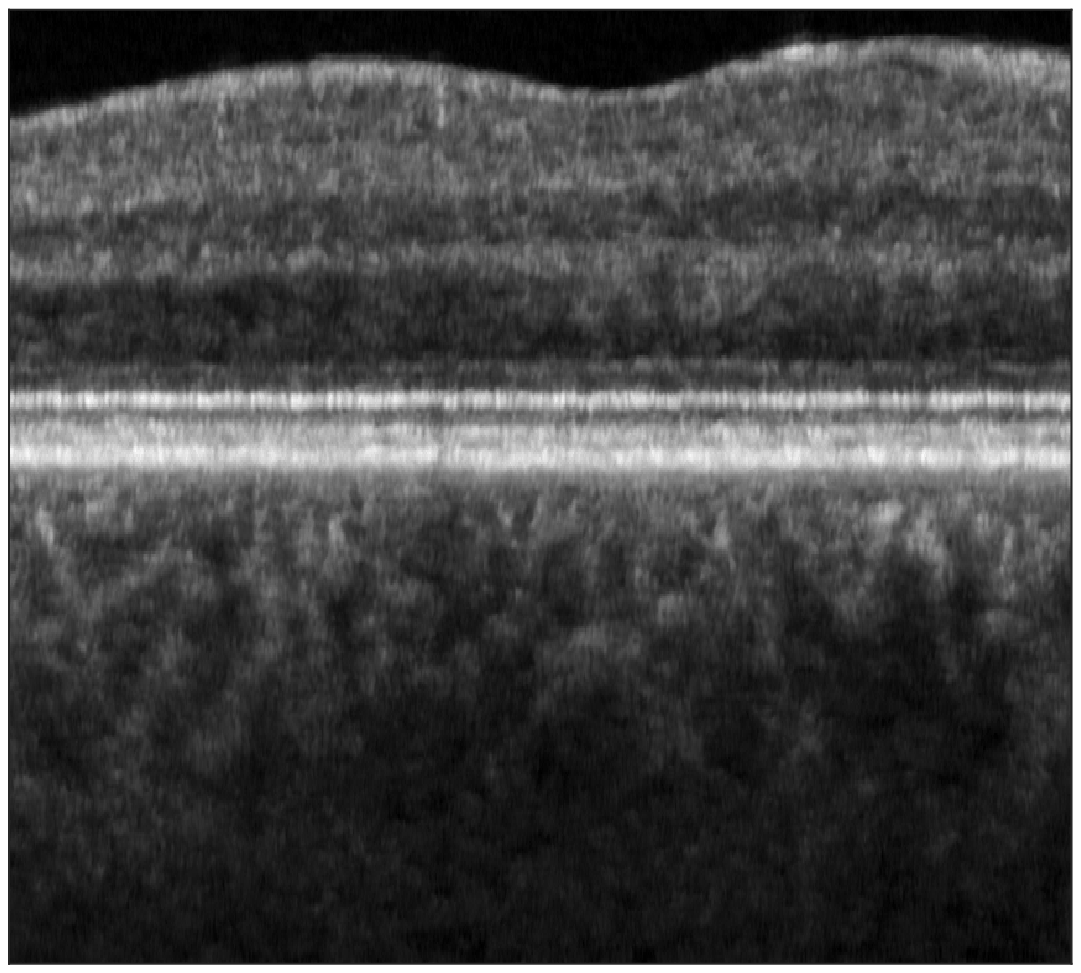}
				\caption{31.34dB, SSIM 0.85}
    \end{subfigure}%
		\hfill 
		\begin{subfigure}[t]{0.197\textwidth}
        \includegraphics[width=\linewidth]{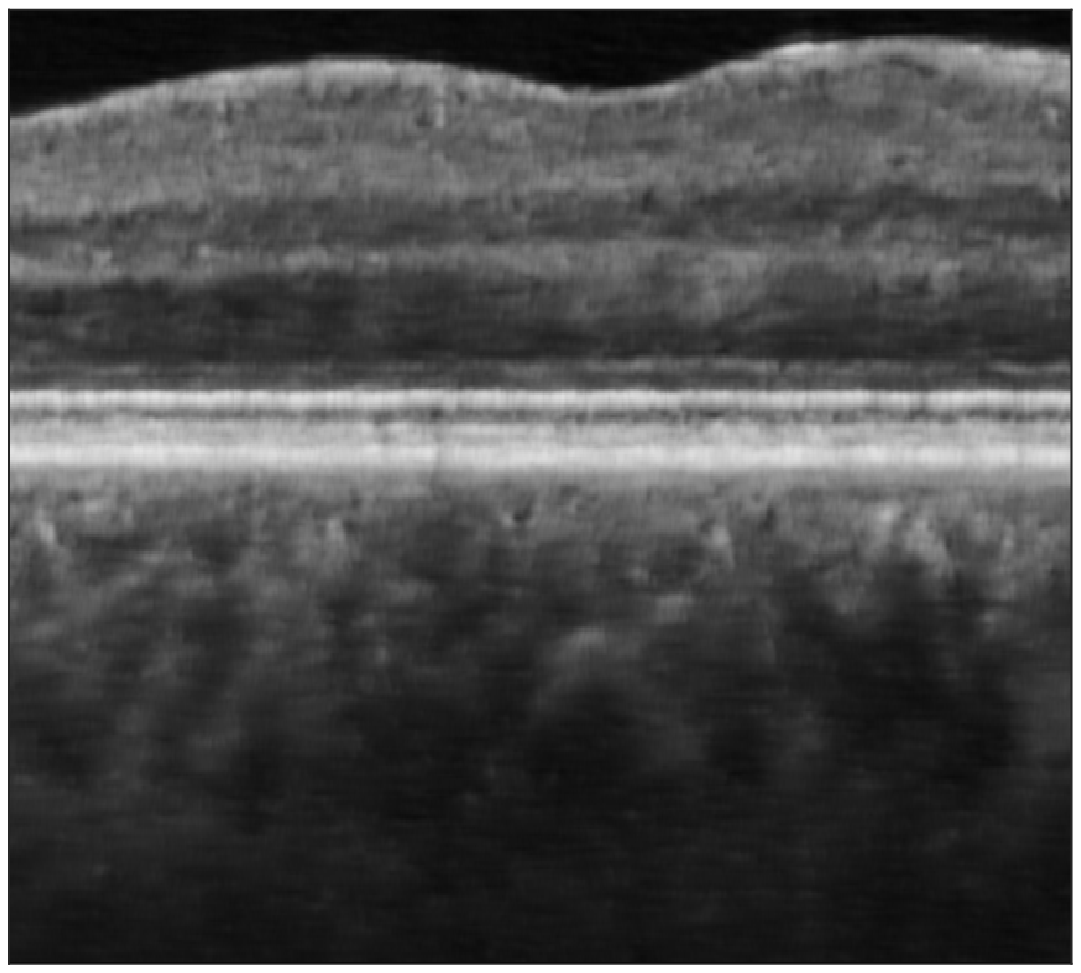}
				\caption{30.86dB, SSIM 0.87}
    \end{subfigure}%
		\hfill 
		\begin{subfigure}[t]{0.197\textwidth}
        \includegraphics[width=\linewidth]{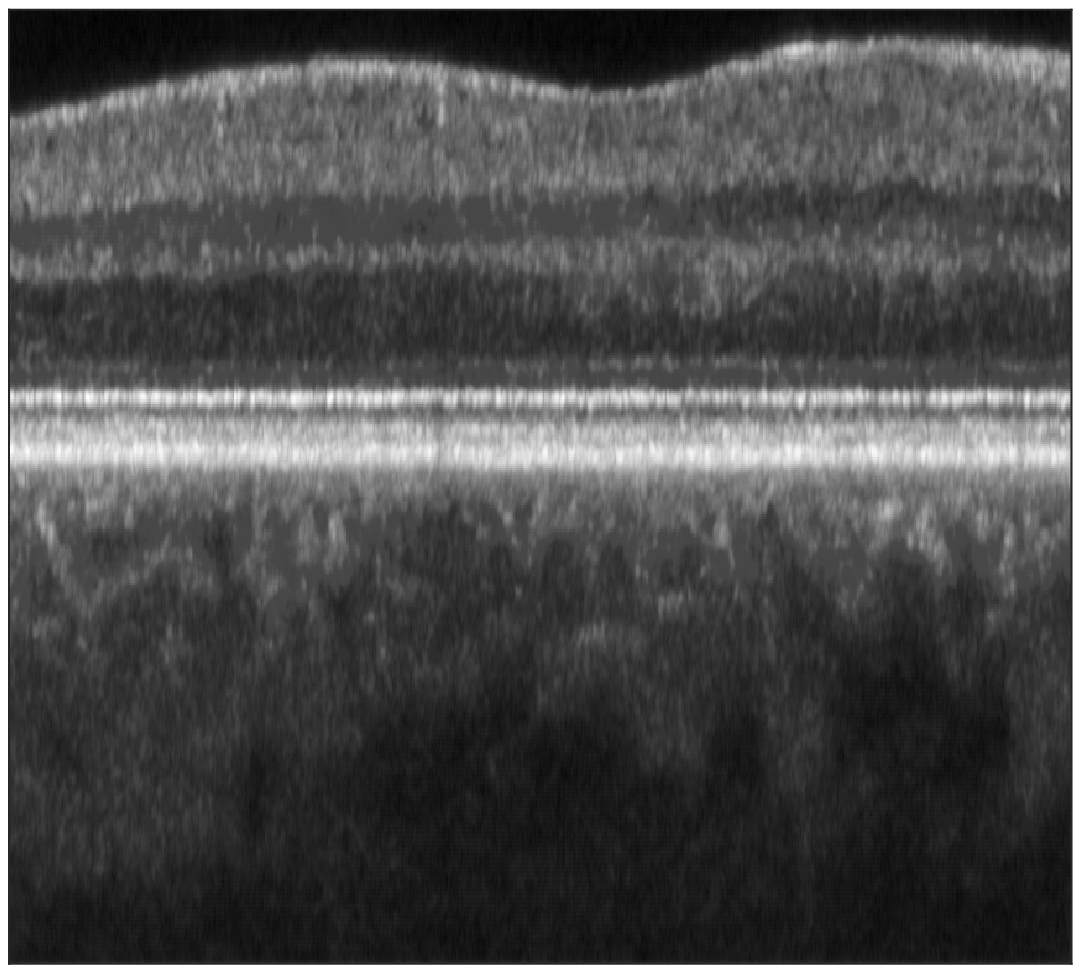}
				\caption{34.63, SSIM 0.93}
    \end{subfigure}%

\caption{Retinal data speckle suppression: (a) Cross sectional human retina in vivo, $p^t_{\mathrm{x}}=2$; (b) Despeckled (NLM) image used ground truth; (c) DRNN trained with 100 columns of retinal image $p^s_{\mathrm{x}}=p^t_{\mathrm{x}}=2$; (d) RNN-GAN trained with 100 first columns of muscle chicken and blueberry, $p^s_{\mathrm{x}}=3$; (e) U-Net trained with retinal image of size $256 \times 256$, $p^s_{\mathrm{x}}=2$. Scale bars are 200 $\mu$m.}
\label{fig2}
\end{figure*}

\begin{figure*}[t]
    \begin{subfigure}[t]{0.25\textwidth}
        \includegraphics[width=\linewidth]{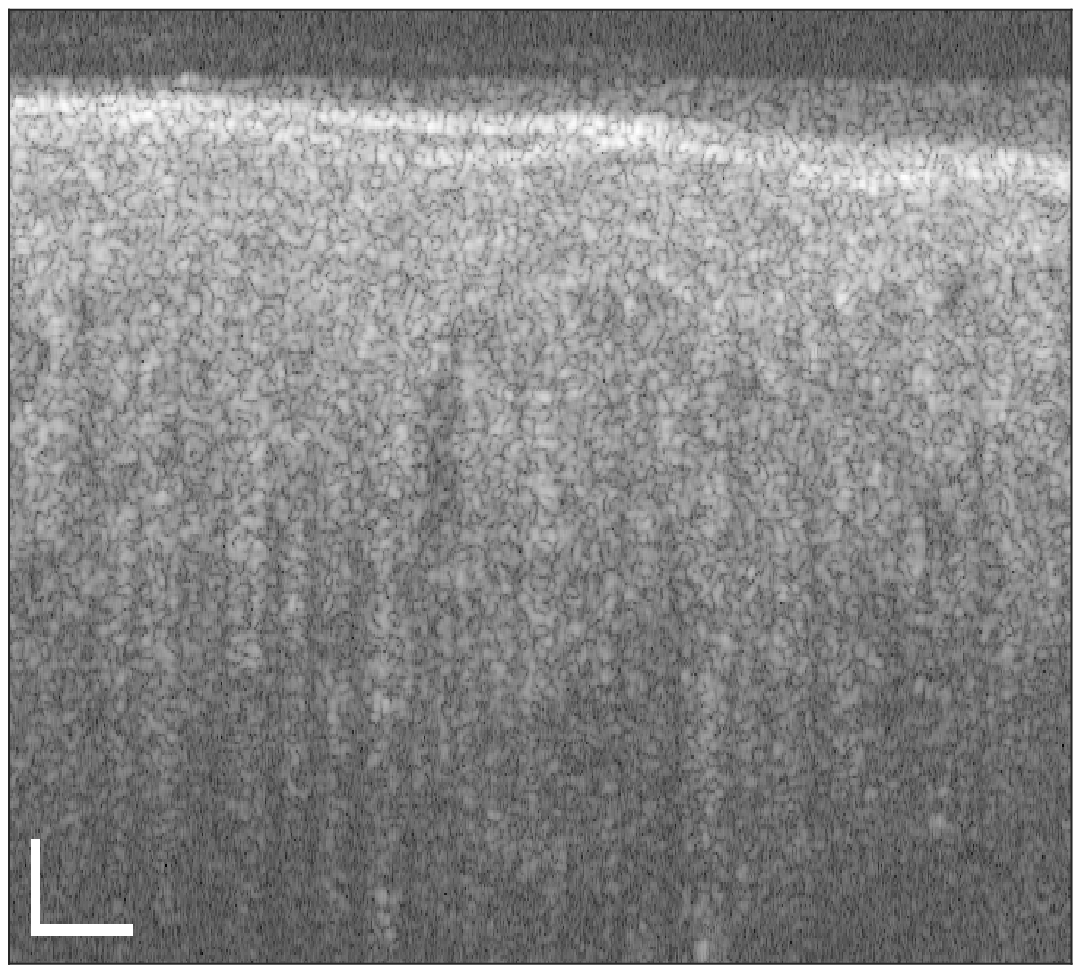}
				\caption{24.39dB, SSIM 0.21}
    \end{subfigure}%
		\hfill 
		\begin{subfigure}[t]{0.25\textwidth}
        \includegraphics[width=\linewidth]{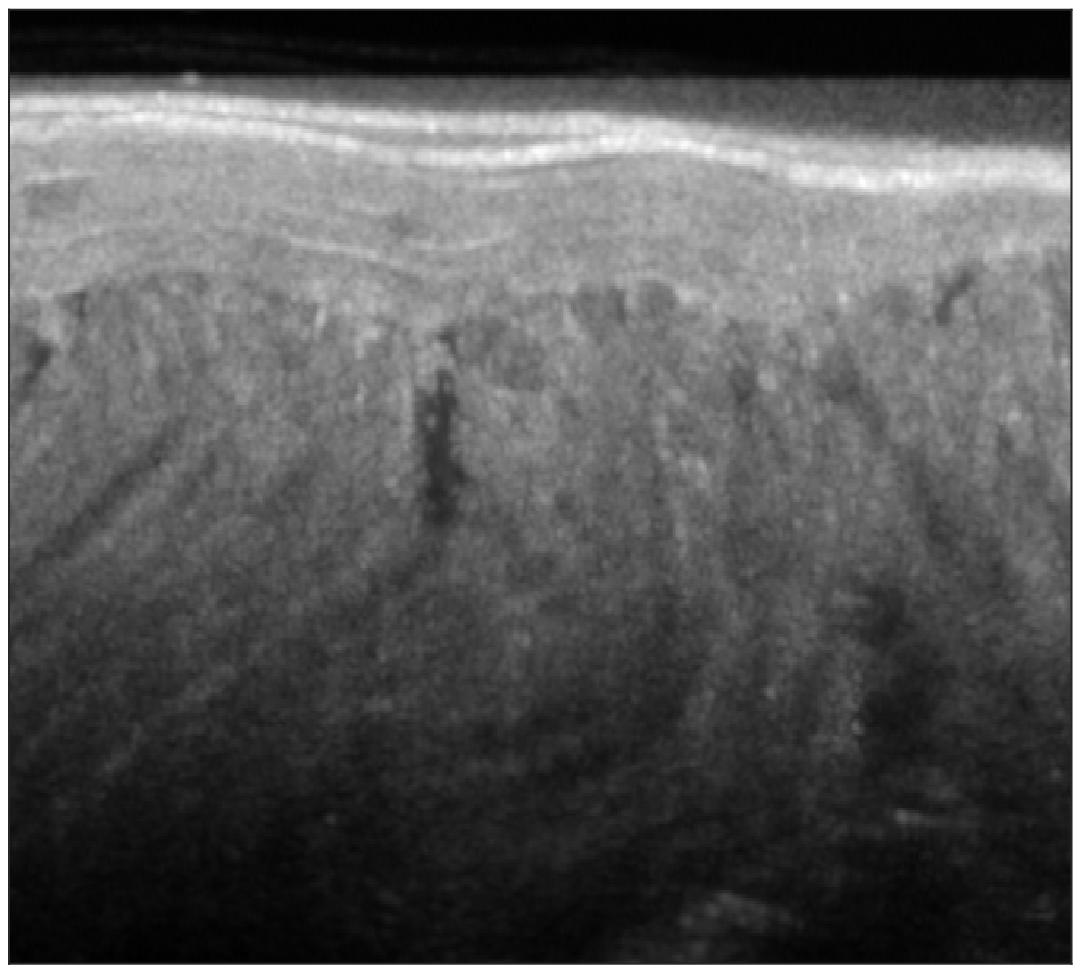}
				\caption{}
    \end{subfigure}%
		\hfill 
		\begin{subfigure}[t]{0.25\textwidth}
        \includegraphics[width=\linewidth]{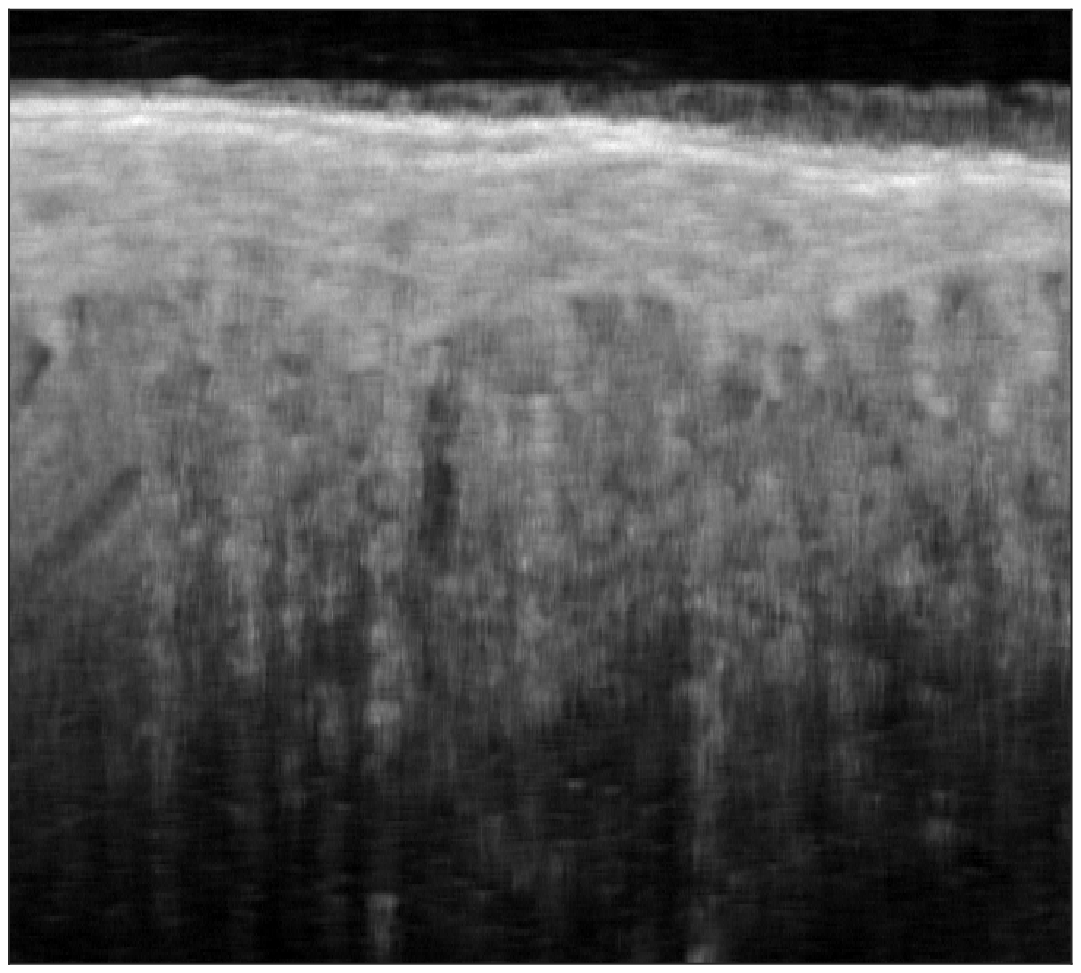}
				\caption{31.56dB, SSIM 0.68}
    \end{subfigure}%
		\hfill
		\begin{subfigure}[t]{0.25\textwidth}
        \includegraphics[width=\linewidth]{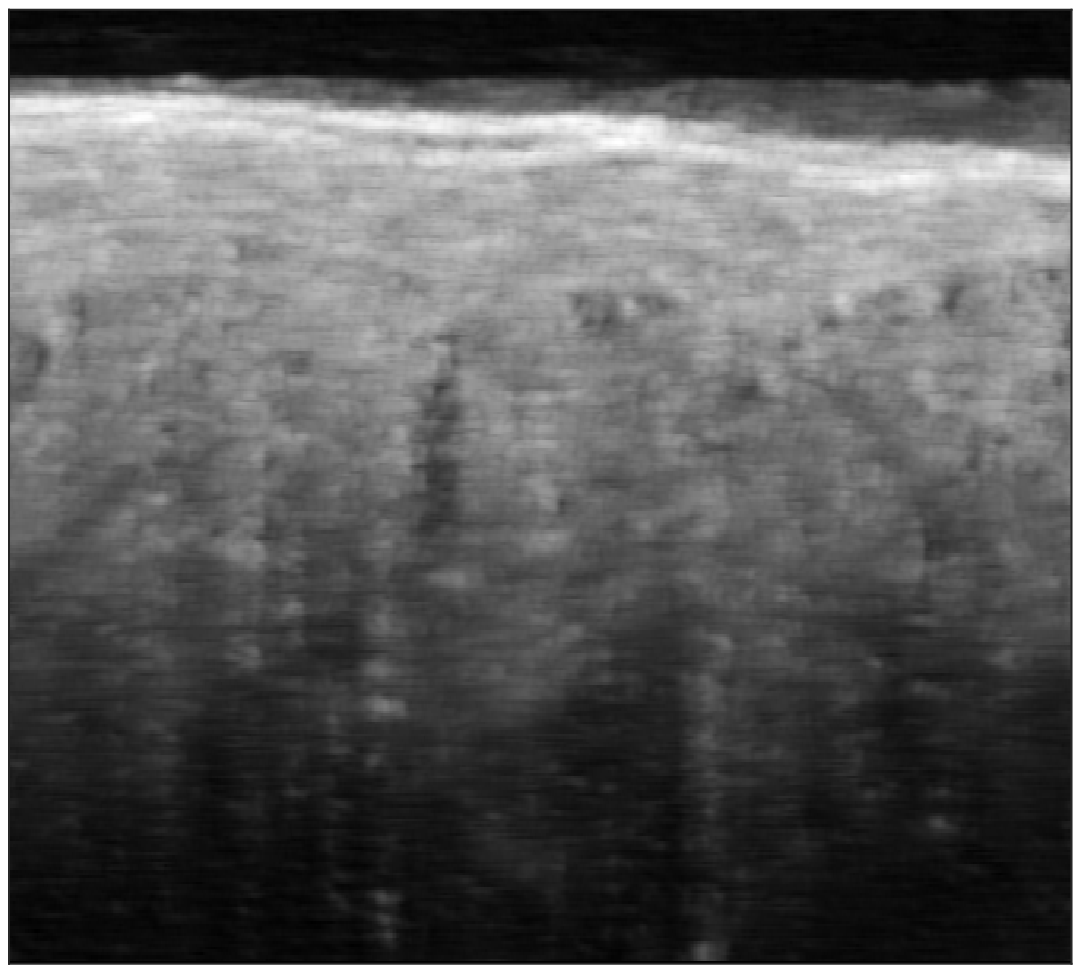}
				\caption{32.67dB SSIM 0.72}
    \end{subfigure}%
		\hfill
		\begin{subfigure}[t]{0.25\textwidth}
        \includegraphics[width=\linewidth]{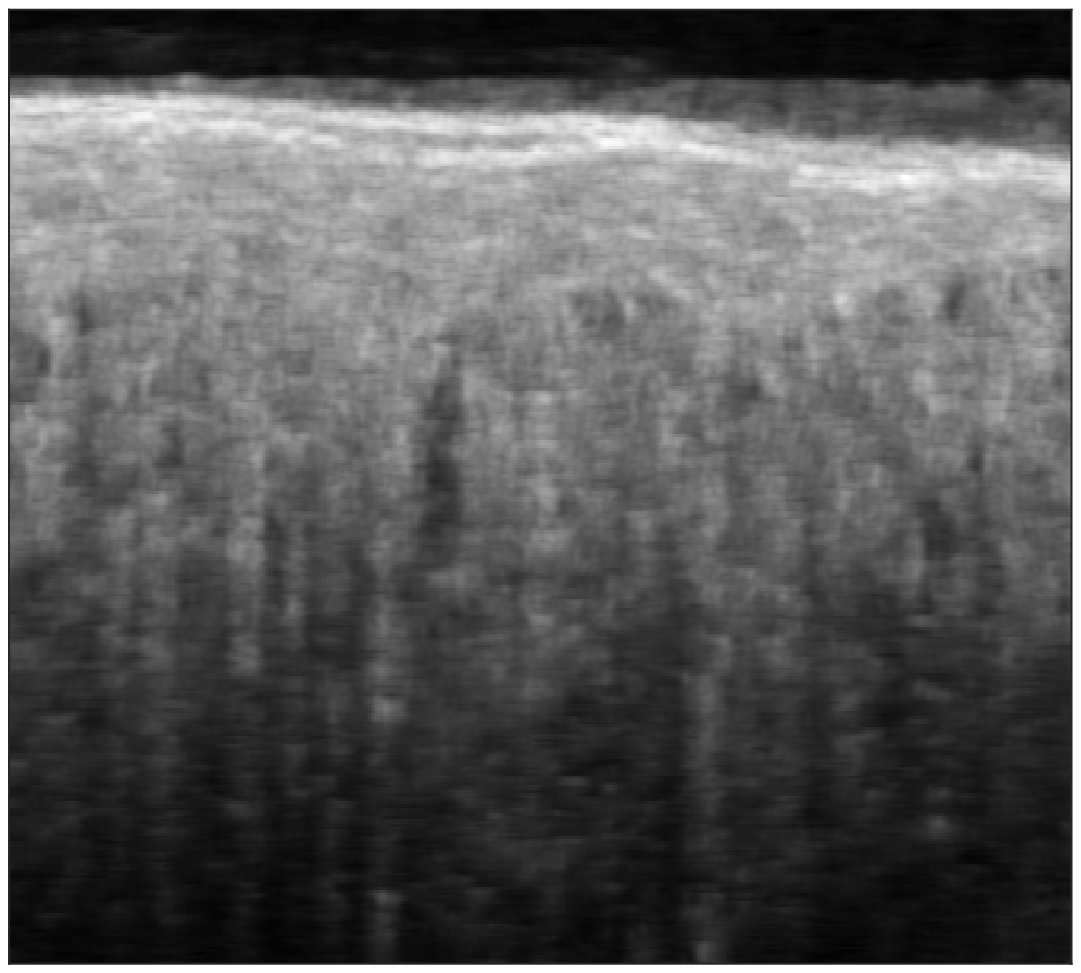}
				\caption{32.08dB SSIM 0.73}
    \end{subfigure}%
		\hfill
    \begin{subfigure}[t]{0.25\textwidth}
        \includegraphics[width=\linewidth]{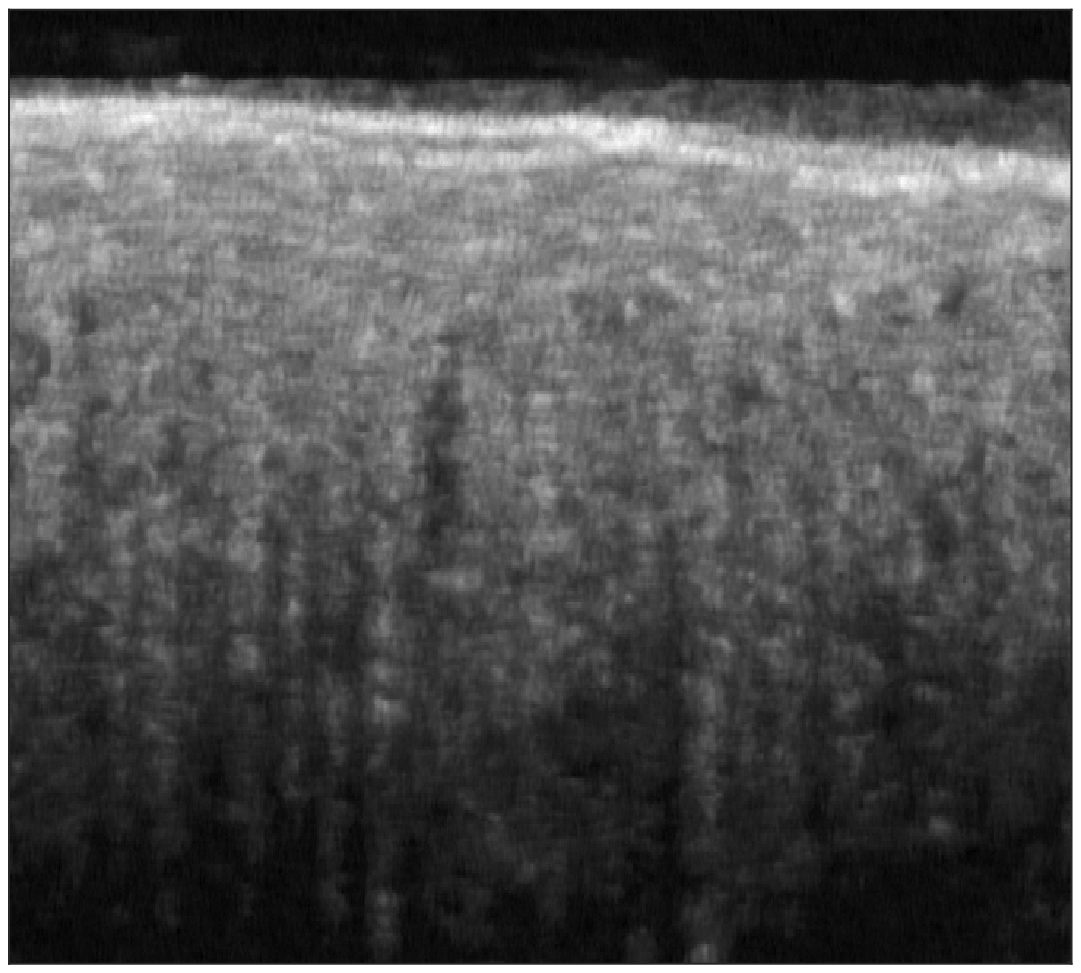}
				\caption{30.35dB, SSIM 0.62}
    \end{subfigure}%
		\begin{subfigure}[t]{0.25\textwidth}
        \includegraphics[width=\linewidth]{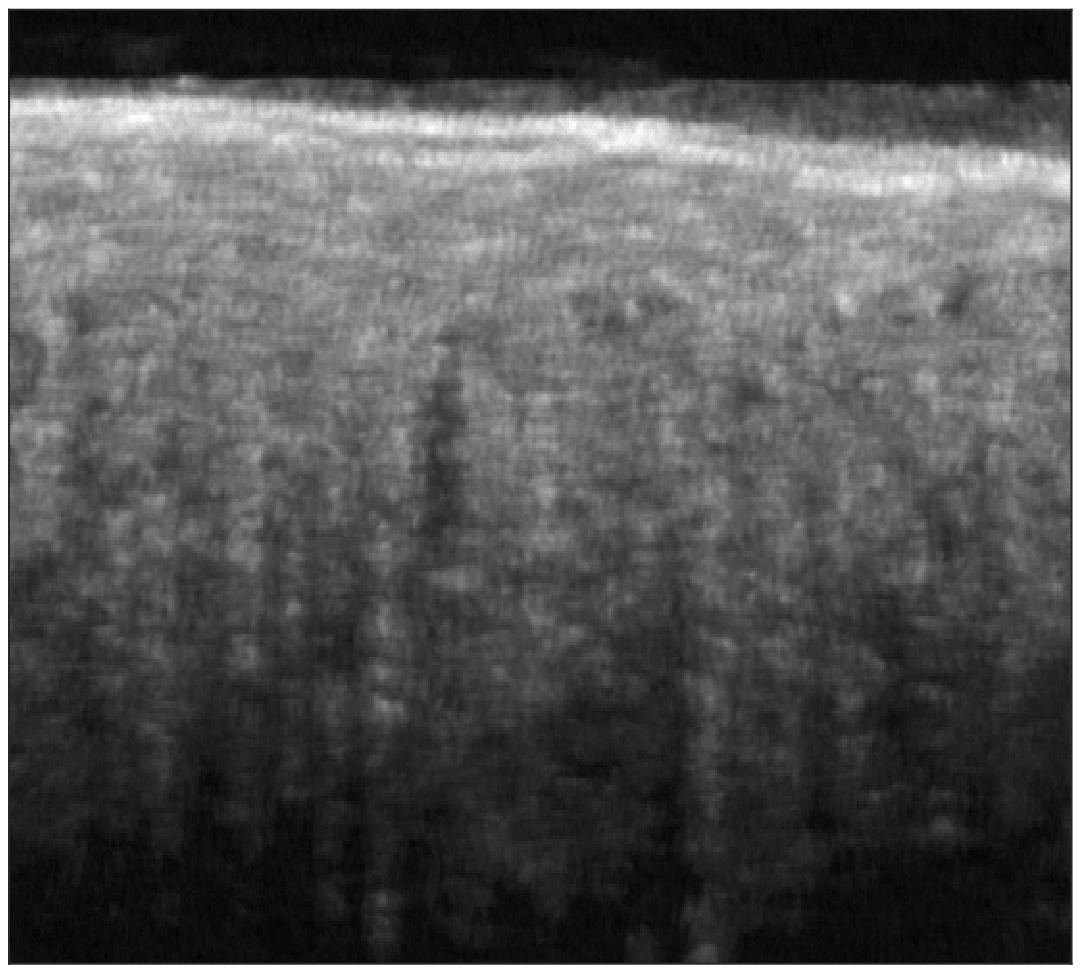}
				\caption{30.91dB, SSIM 0.64}
    \end{subfigure}%
		\begin{subfigure}[t]{0.25\textwidth}
        \includegraphics[width=\linewidth]{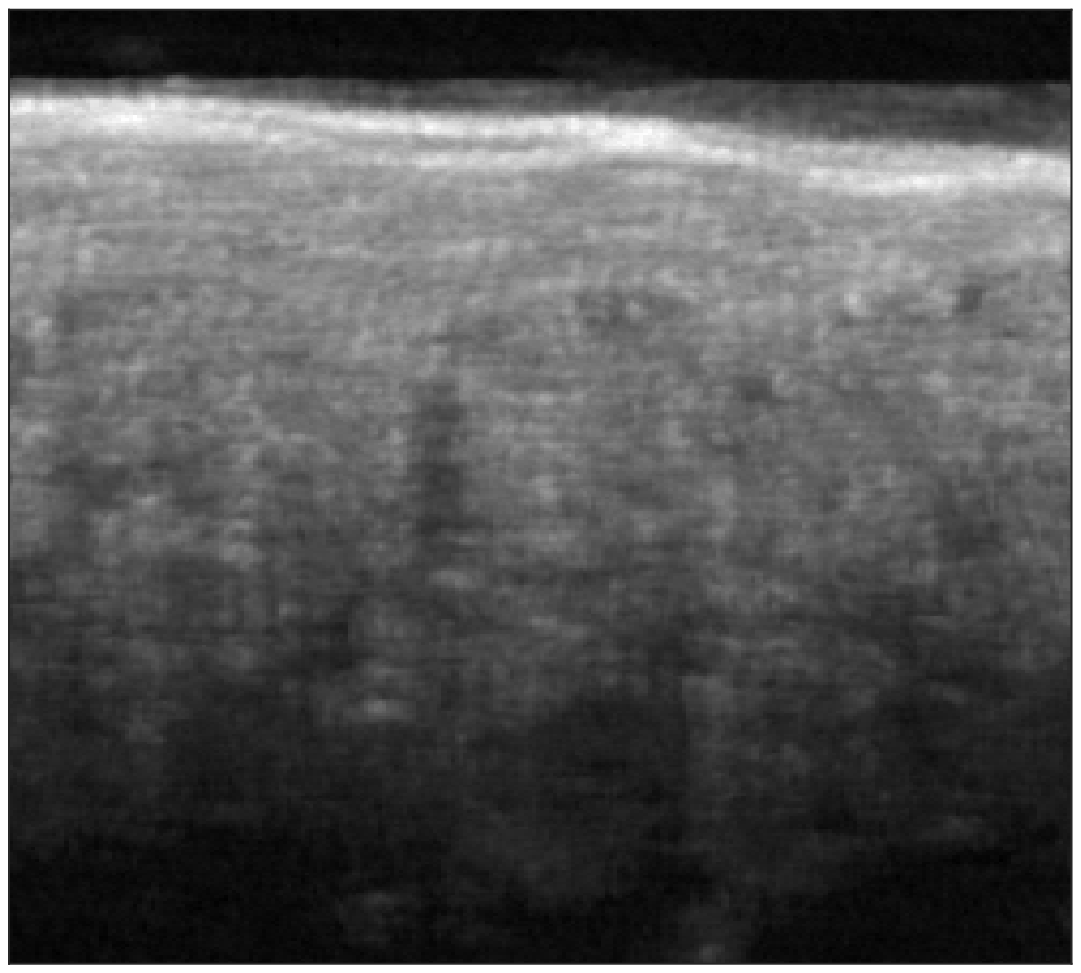}
				\caption{30.92dB, SSIM 0.66}
    \end{subfigure}%
\caption{Chicken muscle speckle suppression results: (a) Speckled acquired tomogram $p_{\mathrm{x}}=3$; 
(b) Ground truth averaged over 901 tomograms; (c) OCT-RNN trained with 100 first columns of chicken muscle; 
(d) RNN-GAN trained with 100 first columns of muscle chicken and blueberry, $p^s_{\mathrm{x}}=p^t_{\mathrm{x}}=3$;
(e) RNN-GAN trained with 200 columns of chicken decimated by a factor 8/3 in the lateral direction, $p^s_{\mathrm{x}}=1$.
System and tissue mismatch:
(f) DRNN trained with 100 columns of human retinal image, $p^s_{\mathrm{x}}=2$; 
(g) DRNN following lateral decimation of the target input by a factor of 4/3, $p^s_{\mathrm{x}}=p^t_{\mathrm{x}}=2$; 
(h) DRNN following lateral decimation of the target input by 8/3, $p^t_{\mathrm{x}}=1$.
Scale bars are 200 $\mu$m.}
\label{fig3}
\end{figure*}

\begin{figure*}[t]
    \begin{subfigure}[t]{0.199\textwidth}
        \includegraphics[width=\linewidth]{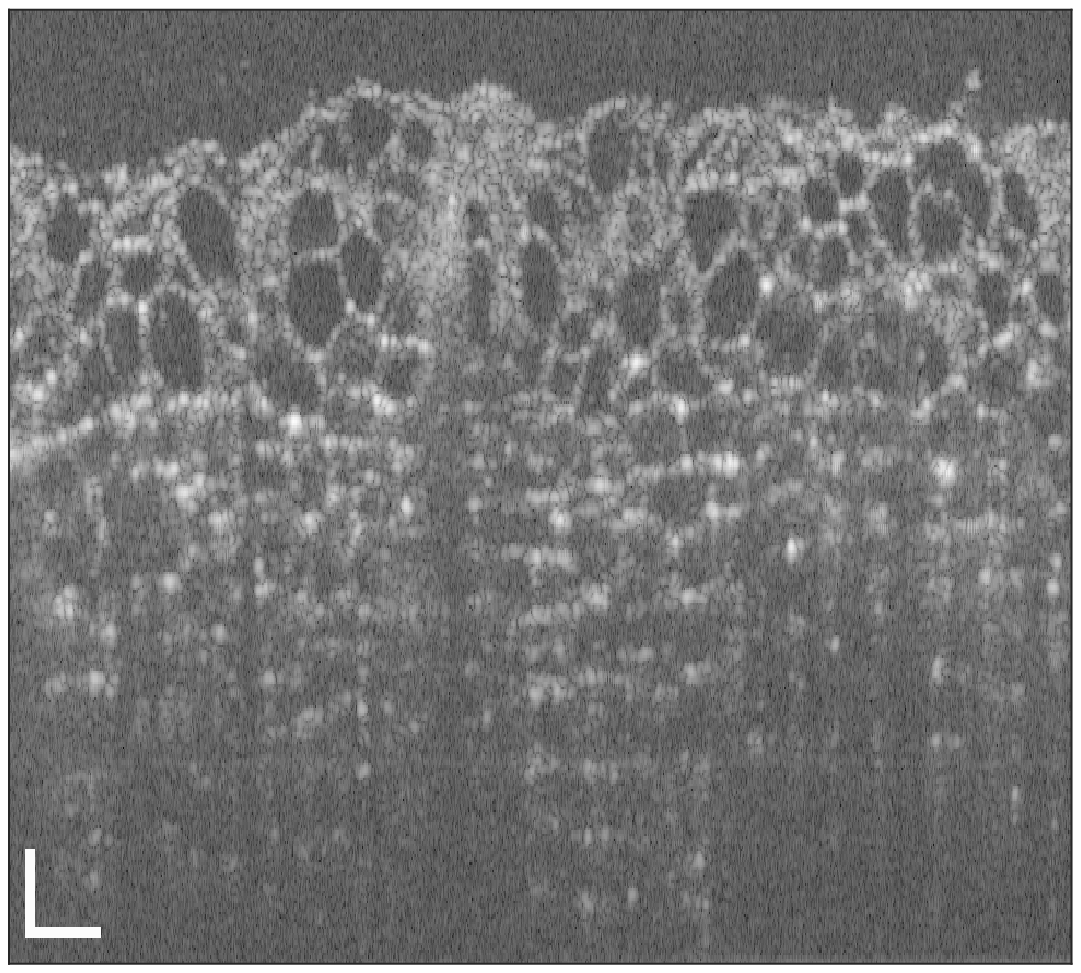}
				\caption{24.76dB, SSIM 0.35}
    \end{subfigure}%
		\hfill 
		\begin{subfigure}[t]{0.199\textwidth}
        \includegraphics[width=\linewidth]{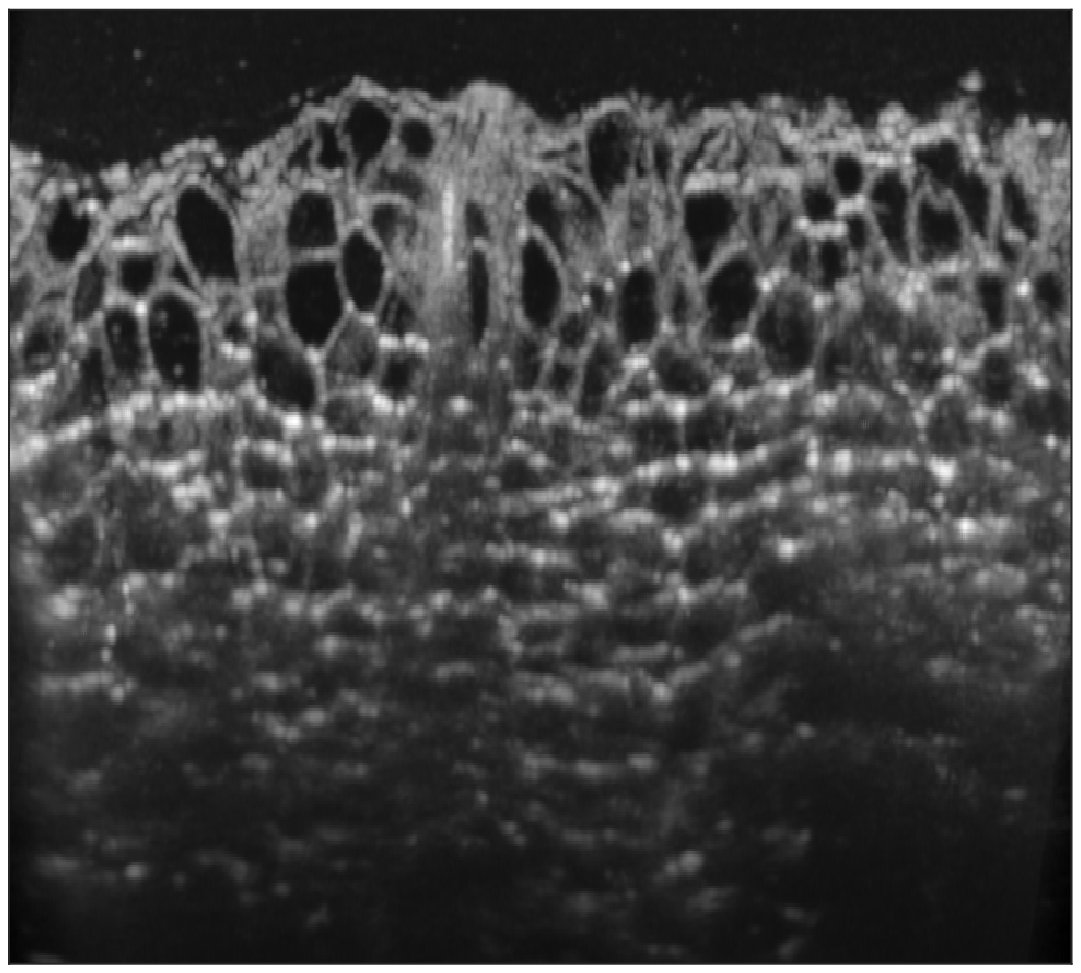}
				\caption{}
    \end{subfigure}%
		\hfill 
    \begin{subfigure}[t]{0.199\textwidth}
        \includegraphics[width=\linewidth]{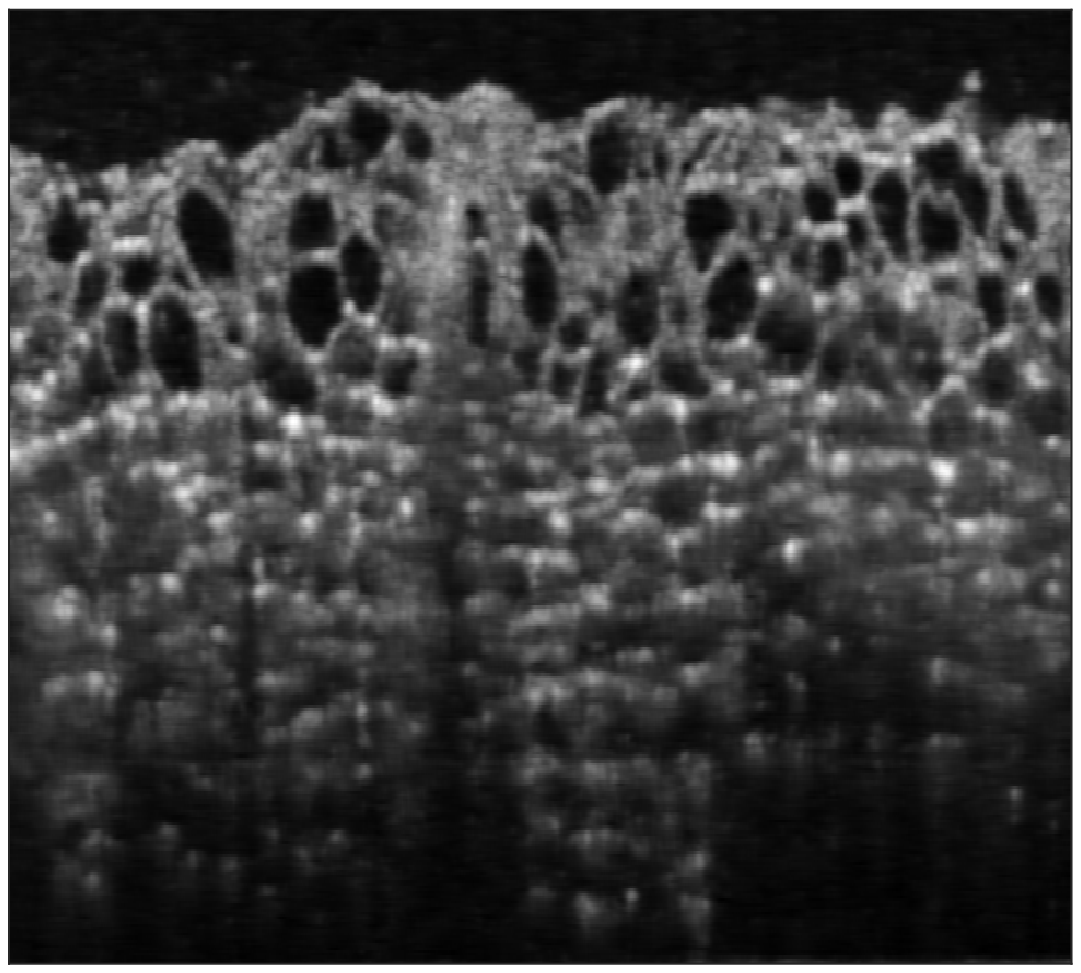}
				\caption{30.18dB, SSIM 0.81}
    \end{subfigure}%
		\hfill 
    \begin{subfigure}[t]{0.199\textwidth}
        \includegraphics[width=\linewidth]{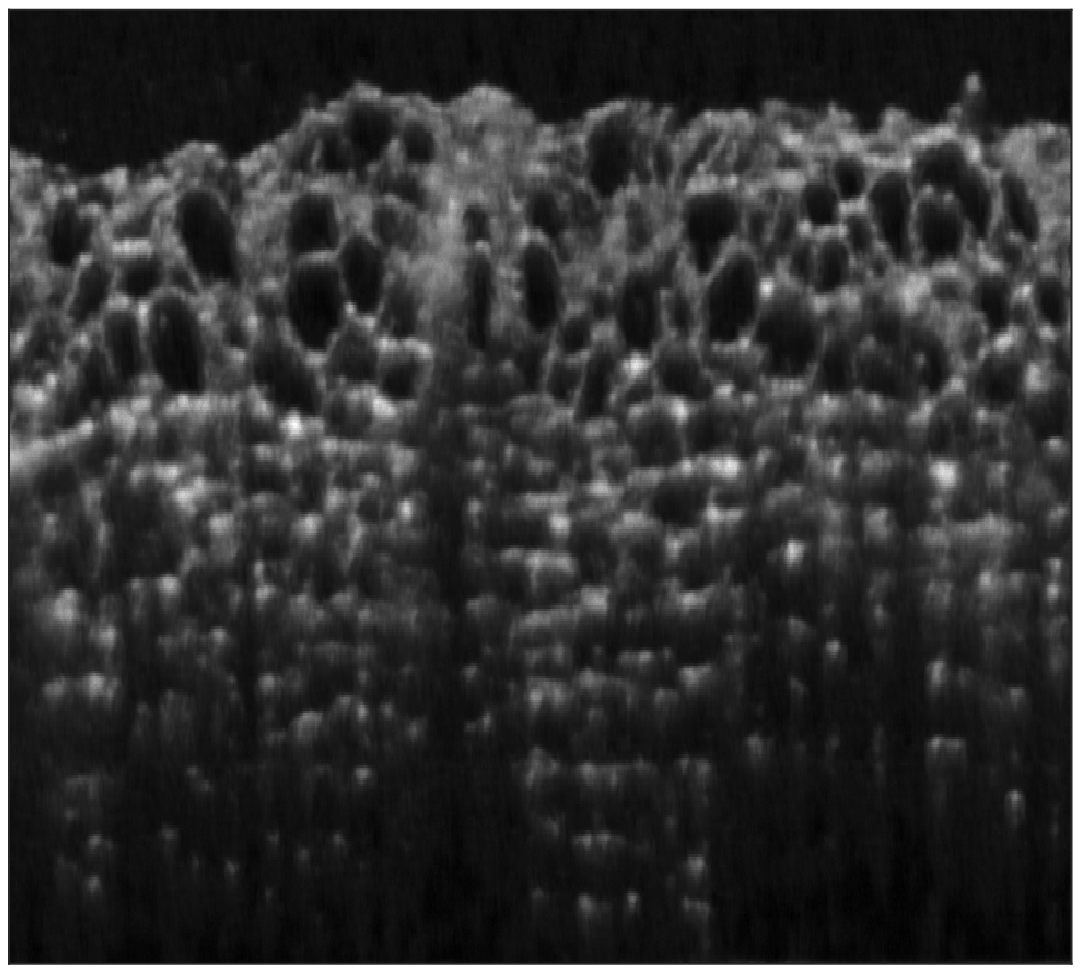}
				\caption{27.19dB, SSIM 0.72 }
    \end{subfigure}%
		\hfill 
		\begin{subfigure}[t]{0.199\textwidth}
        \includegraphics[width=\linewidth]{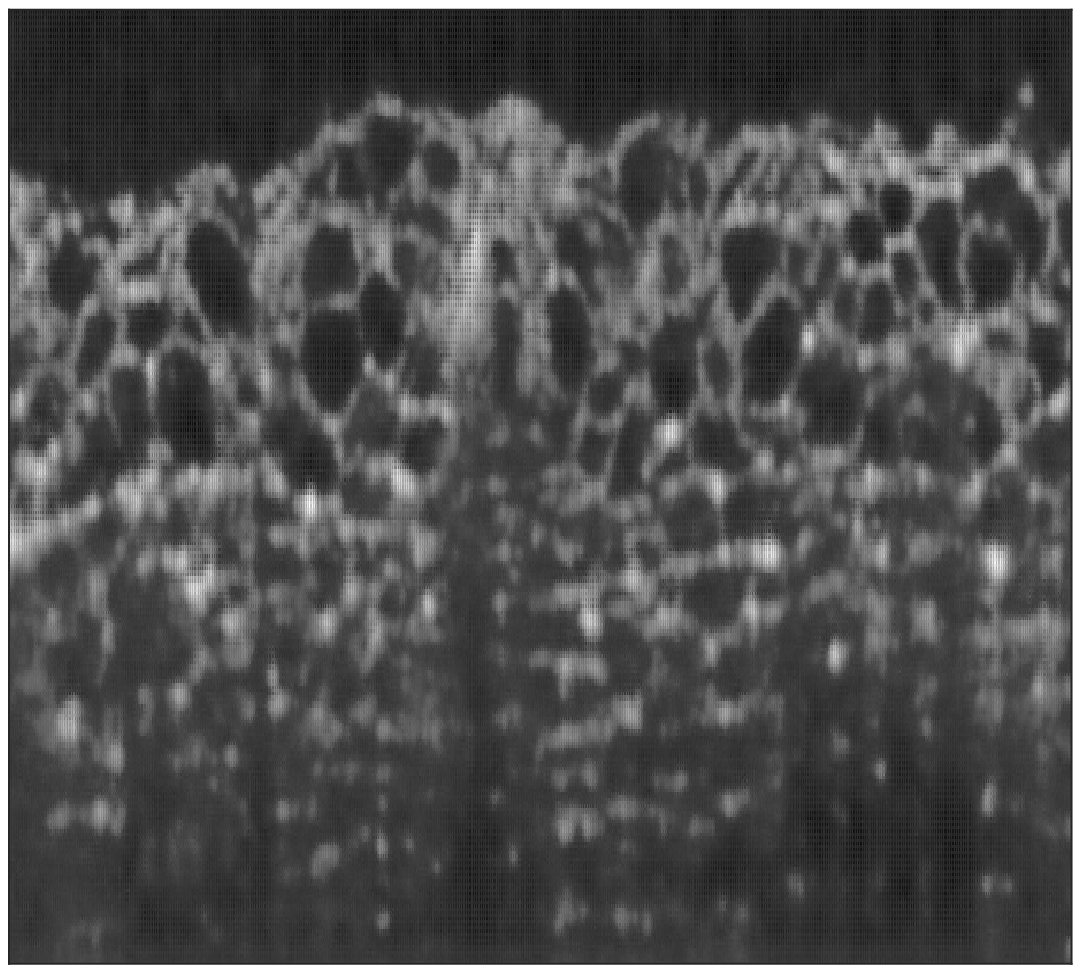}
				\caption{24.98dB, SSIM 0.53}
    \end{subfigure}%
		\hfill
\caption{Blueberry speckle suppression results: (a) Speckled acquired tomogram; (b) Despeckled via angular compounding used ground truth; (c) RNN-GAN trained with 200 last columns of blueberry, $p^s_{\mathrm{x}}=p^t_{\mathrm{x}}=3$; (d) DRNN trained with 100 columns of human retinal image, $p^s_{\mathrm{x}}=2$;  (e) U-Net trained with $256 \times 256$ chicken skin image, $p^s_{\mathrm{x}}=2$.
Scale bars are 200 $\mu$m.}
\label{fig4}
\end{figure*}

\begin{figure*}[t]
    \begin{subfigure}[t]{0.199\textwidth}
        \includegraphics[width=\linewidth]{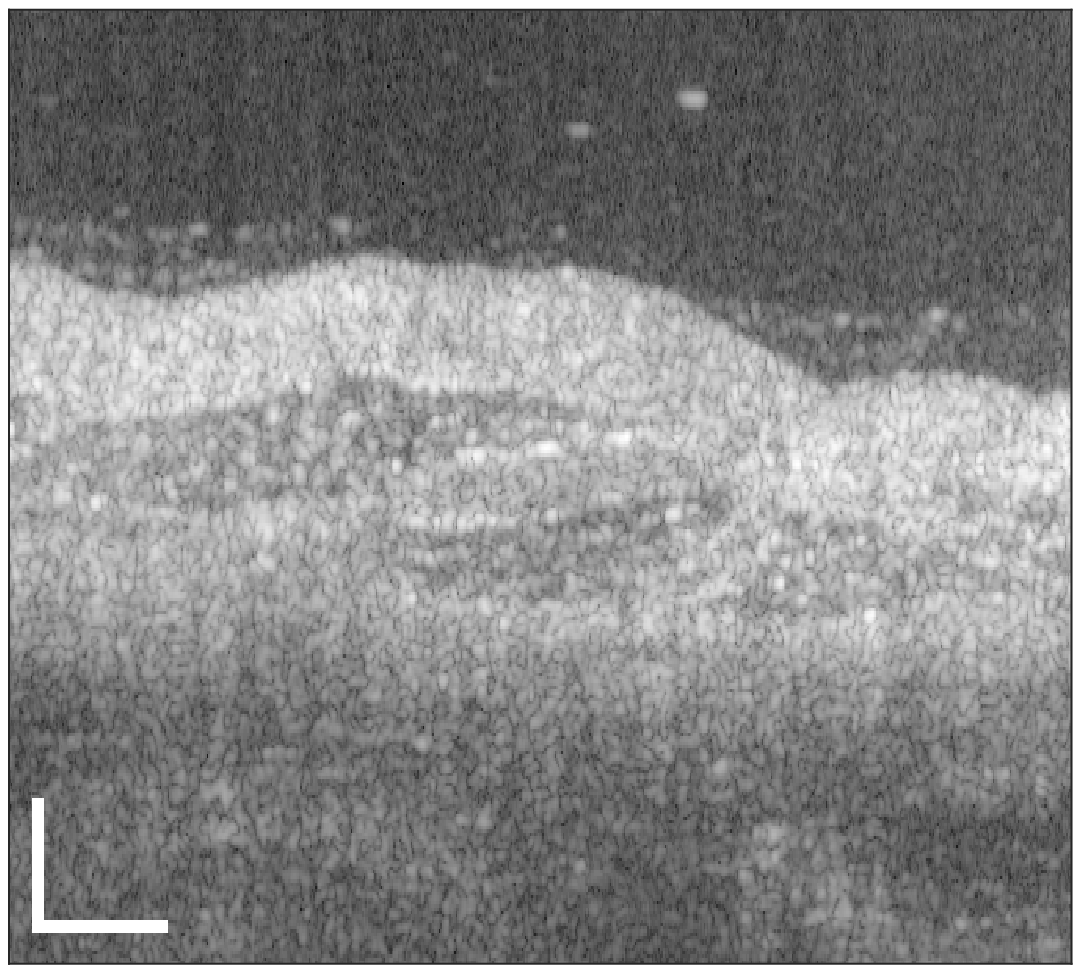}
				\caption{25.92dB, SSIM 0.37}
    \end{subfigure}%
		\hfill 
		\begin{subfigure}[t]{0.199\textwidth}
        \includegraphics[width=\linewidth]{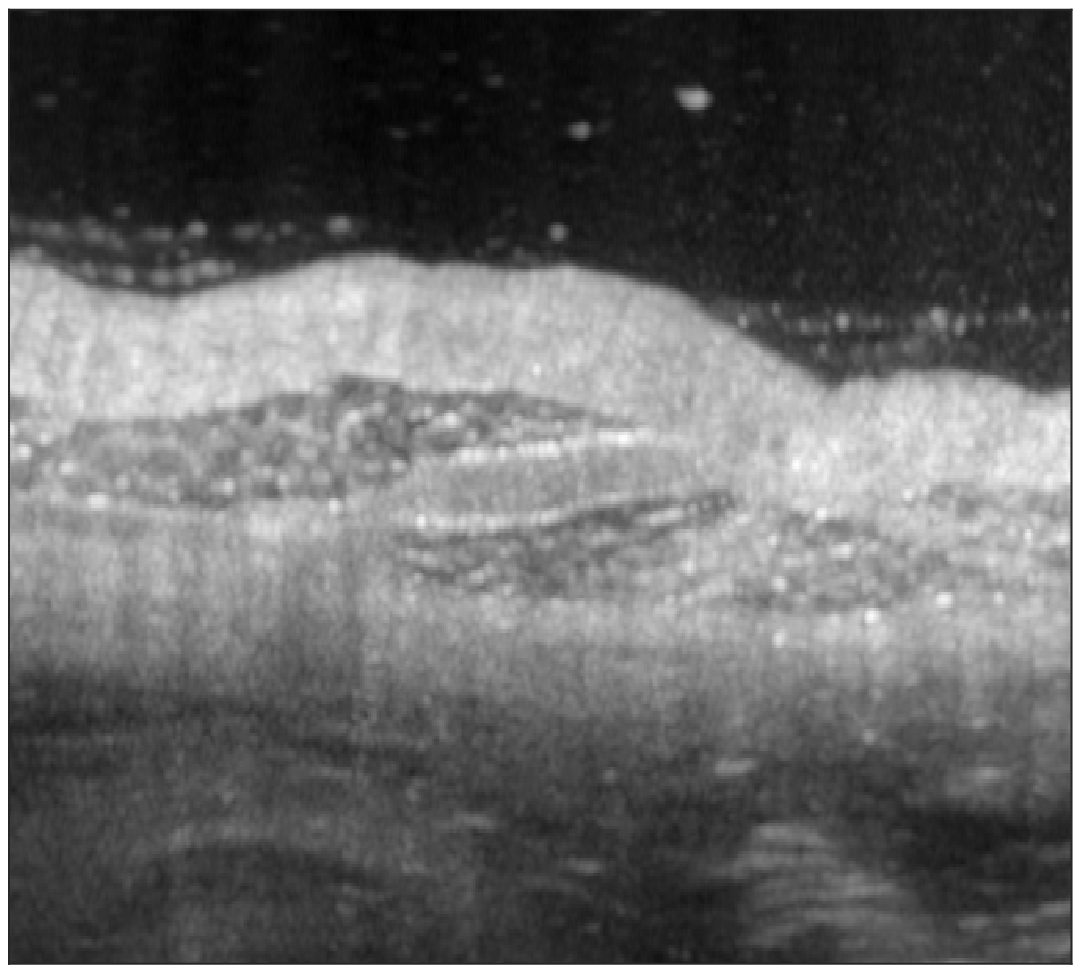}
				\caption{}
    \end{subfigure}%
		\hfill 
    \begin{subfigure}[t]{0.199\textwidth}
        \includegraphics[width=\linewidth]{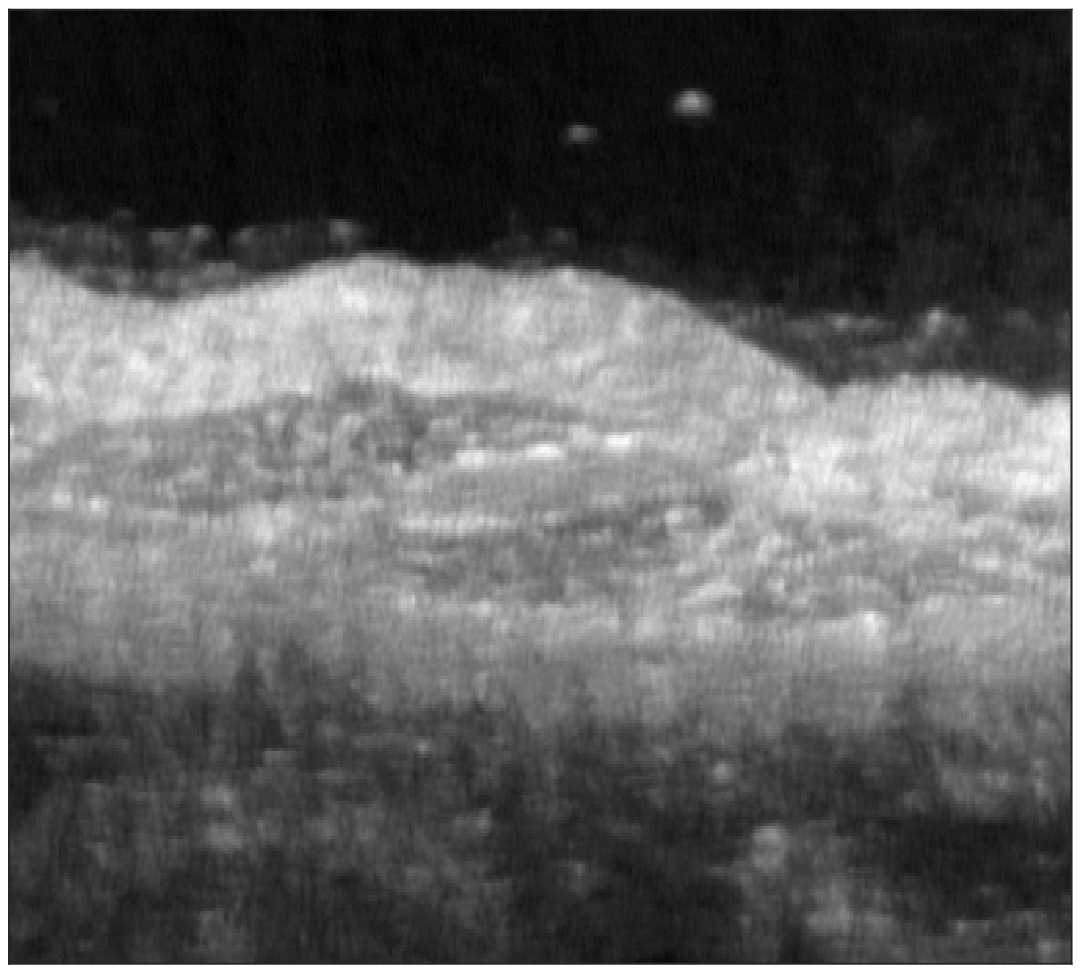}
				\caption{30.25dB, SSIM 0.71 }
    \end{subfigure}%
		\hfill 
		\begin{subfigure}[t]{0.199\textwidth}
        \includegraphics[width=\linewidth]{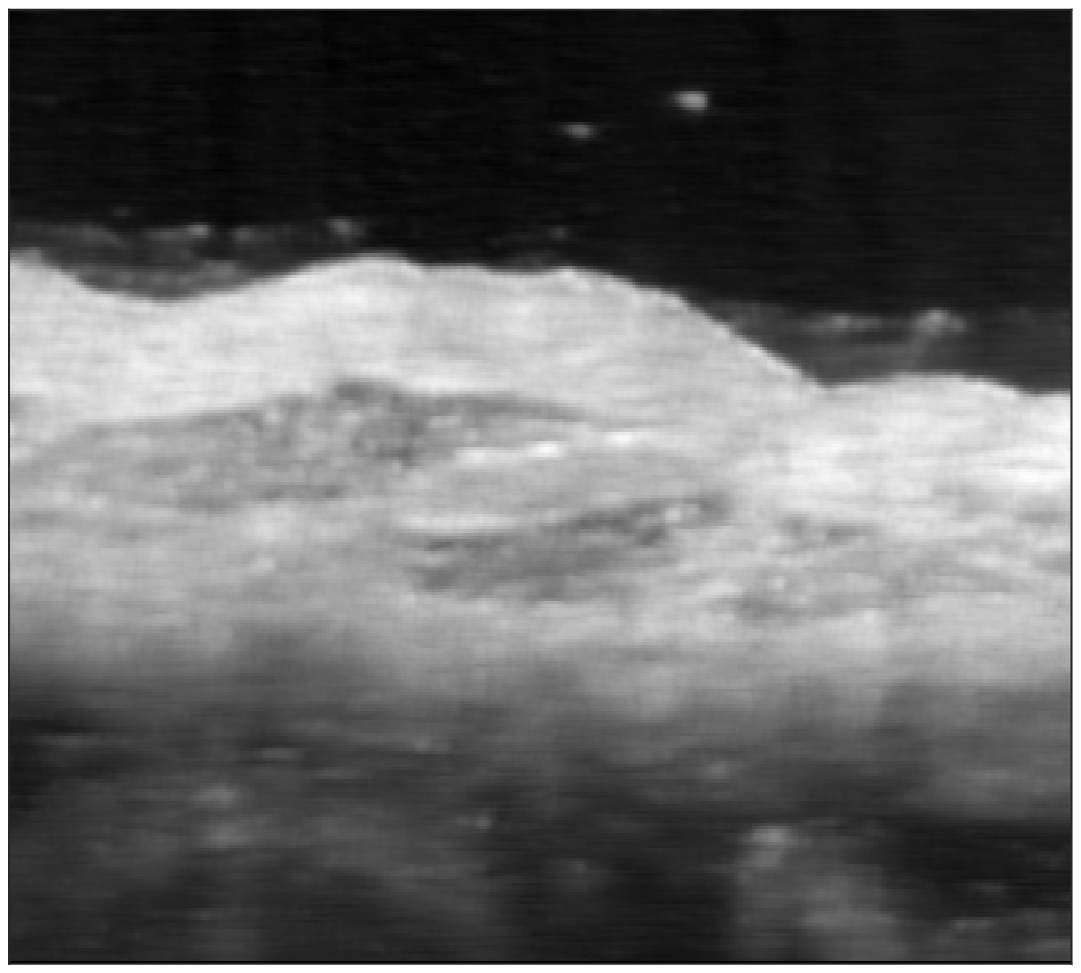}
				\caption{30.61dB , SSIM 0.78 }
    \end{subfigure}%
		\hfill
		\begin{subfigure}[t]{0.199\textwidth}
        \includegraphics[width=\linewidth]{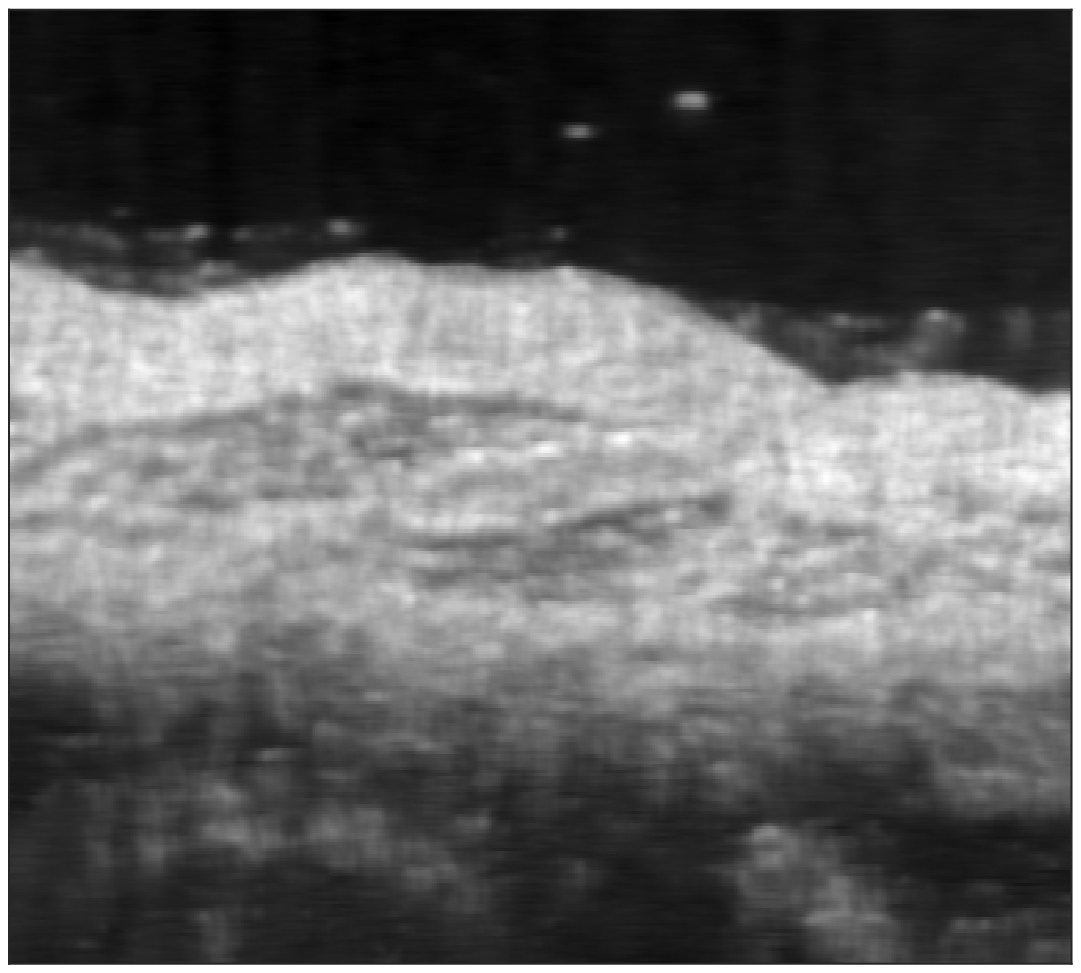}
				\caption{29.00dB, SSIM 0.75}
    \end{subfigure}%
\caption{Chicken skin speckle suppression results: (a) Speckled acquired tomogram; (b) AC ground truth, averaged over 60 tomograms; (c) DRNN trained with 100 columns of human retinal image, $p^s_{\mathrm{x}}=p^t_{\mathrm{x}}=2$; (d) RNN-GAN trained with 100 first columns of muscle chicken and blueberry, $p^s_{\mathrm{x}}=3$; (e) RNN-GAN trained with 200 last columns of blueberry, $p^s_{\mathrm{x}}=3$. Scale bars are 200 $\mu$m.}
\label{fig5}
\end{figure*}

\begin{figure*}[t]
    \begin{subfigure}[t]{0.199\textwidth}
        \includegraphics[width=\linewidth]{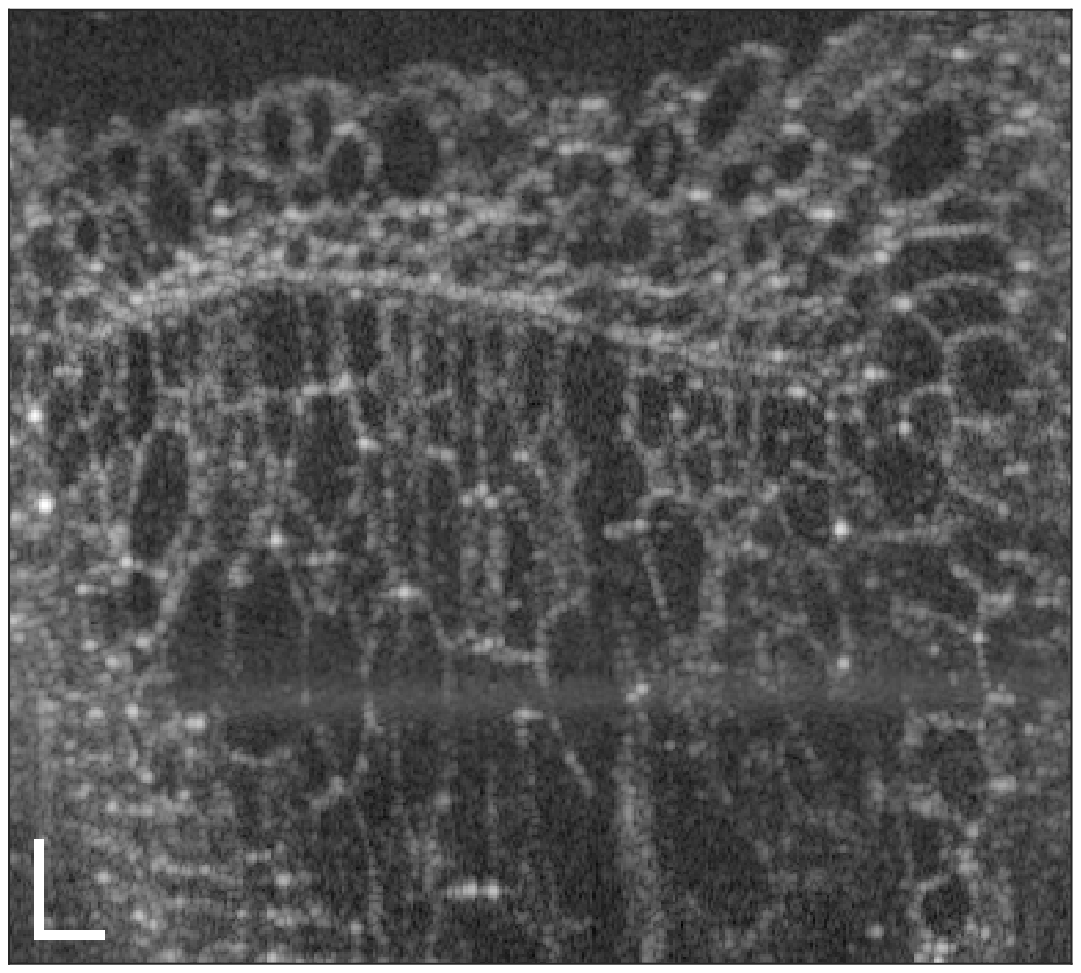}
				\caption{26.00dB, SSIM 0.60}
    \end{subfigure}%
		\hfill 
		\begin{subfigure}[t]{0.199\textwidth}
        \includegraphics[width=\linewidth]{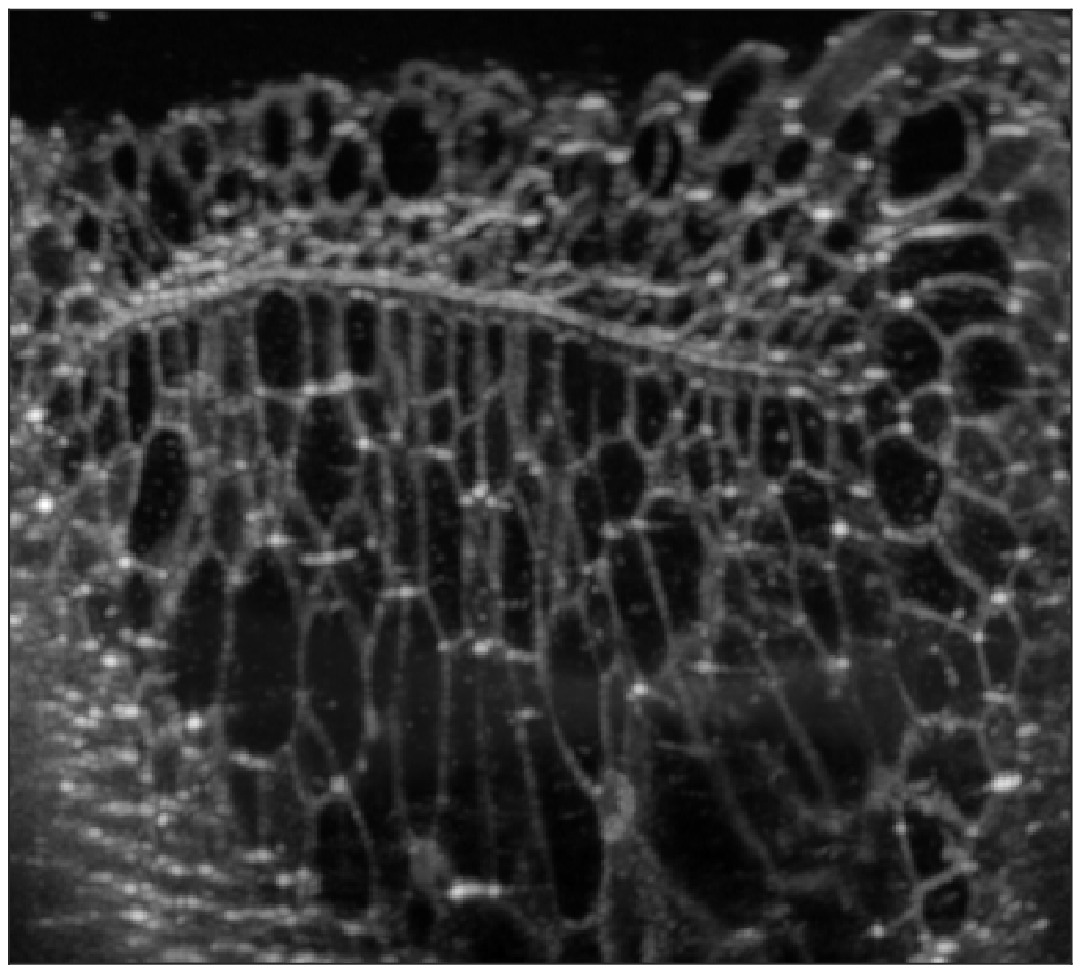}
				\caption{}
    \end{subfigure}%
		\hfill 
    \begin{subfigure}[t]{0.199\textwidth}
        \includegraphics[width=\linewidth]{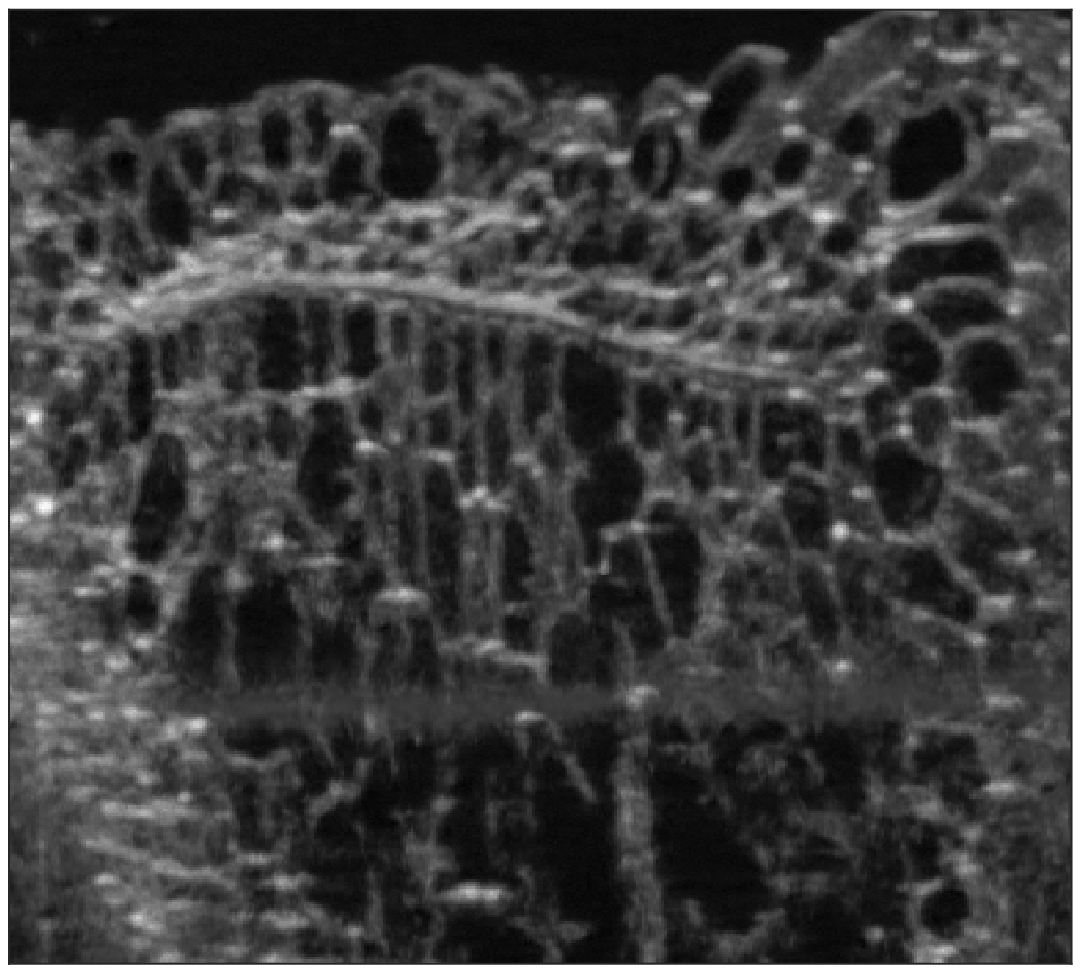}
				\caption{26.80dB, SSIM 0.70 }
    \end{subfigure}%
		\hfill 
		\begin{subfigure}[t]{0.199\textwidth}
        \includegraphics[width=\linewidth]{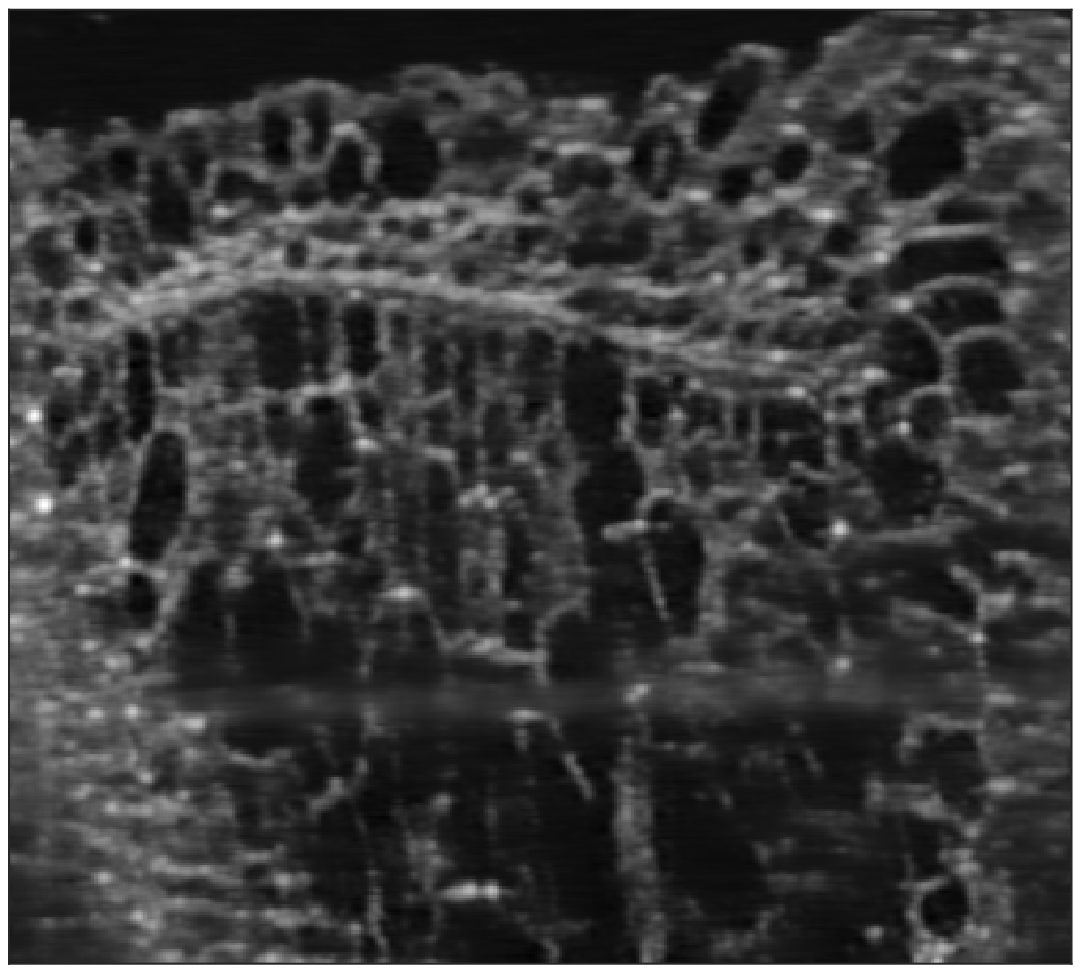}
				\caption{27.84dB, SSIM 0.75}
    \end{subfigure}%
		\hfill
		\begin{subfigure}[t]{0.199\textwidth}
        \includegraphics[width=\linewidth]{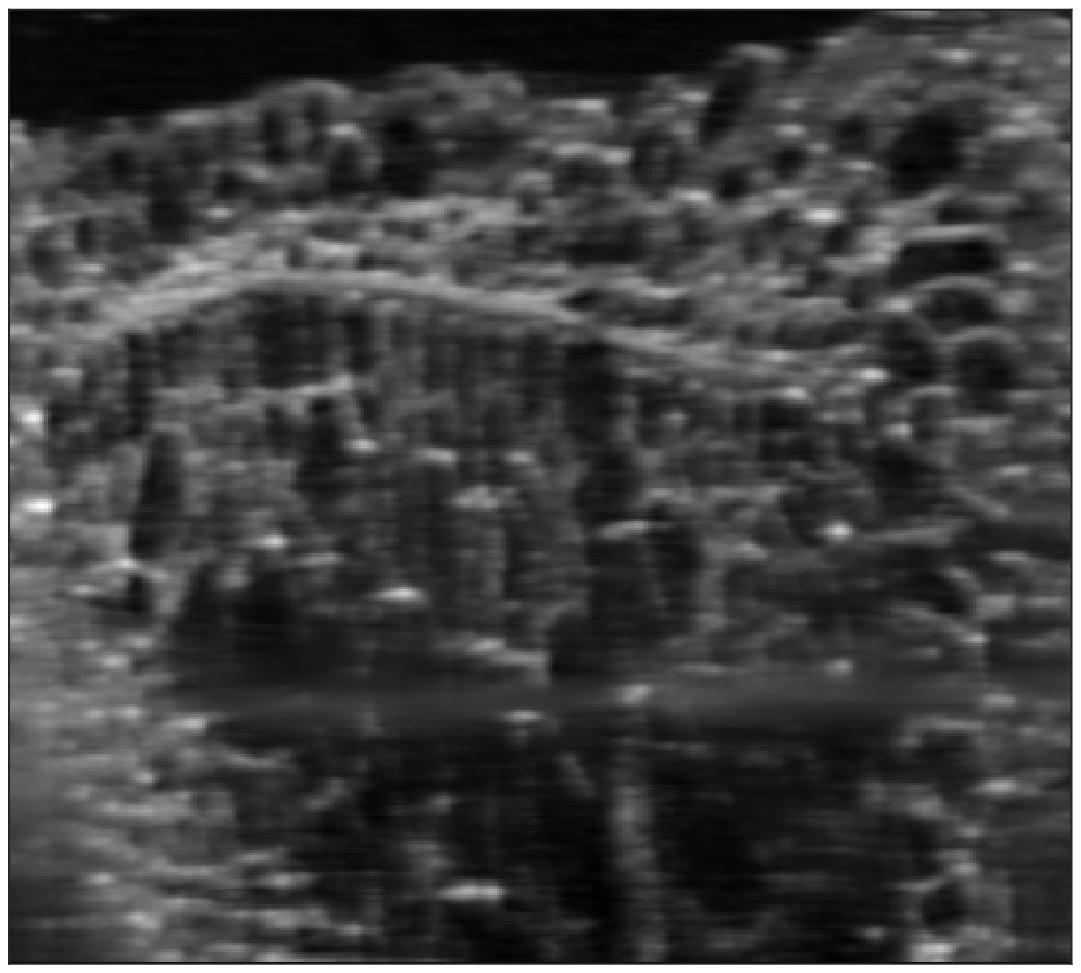}
				\caption{26.29dB, SSIM 0.64}
    \end{subfigure}%
\caption{Cucumber speckle suppression results: (a) Speckled acquired tomogram, $p^t_{\mathrm{x}}=1$; (b) 
Ground truth averaged over 301 tomograms; (c) DRNN trained with human retina image, $p^s_{\mathrm{x}}=2$; (d) RNN-GAN trained with 200 columns of blueberry image decimated in the lateral direction by a factor of 8/3, $p^s_{\mathrm{x}}=p^t_{\mathrm{x}}=1$; (e) RNN-GAN trained with 200 columns of blueberry and chicken, $p^s_{\mathrm{x}}=3$.
Scale bars are 200 $\mu$m. }
\label{fig6}
\end{figure*}

\begin{figure*}[ht]
    \begin{subfigure}[t]{0.247\textwidth}
        \includegraphics[width=\linewidth]{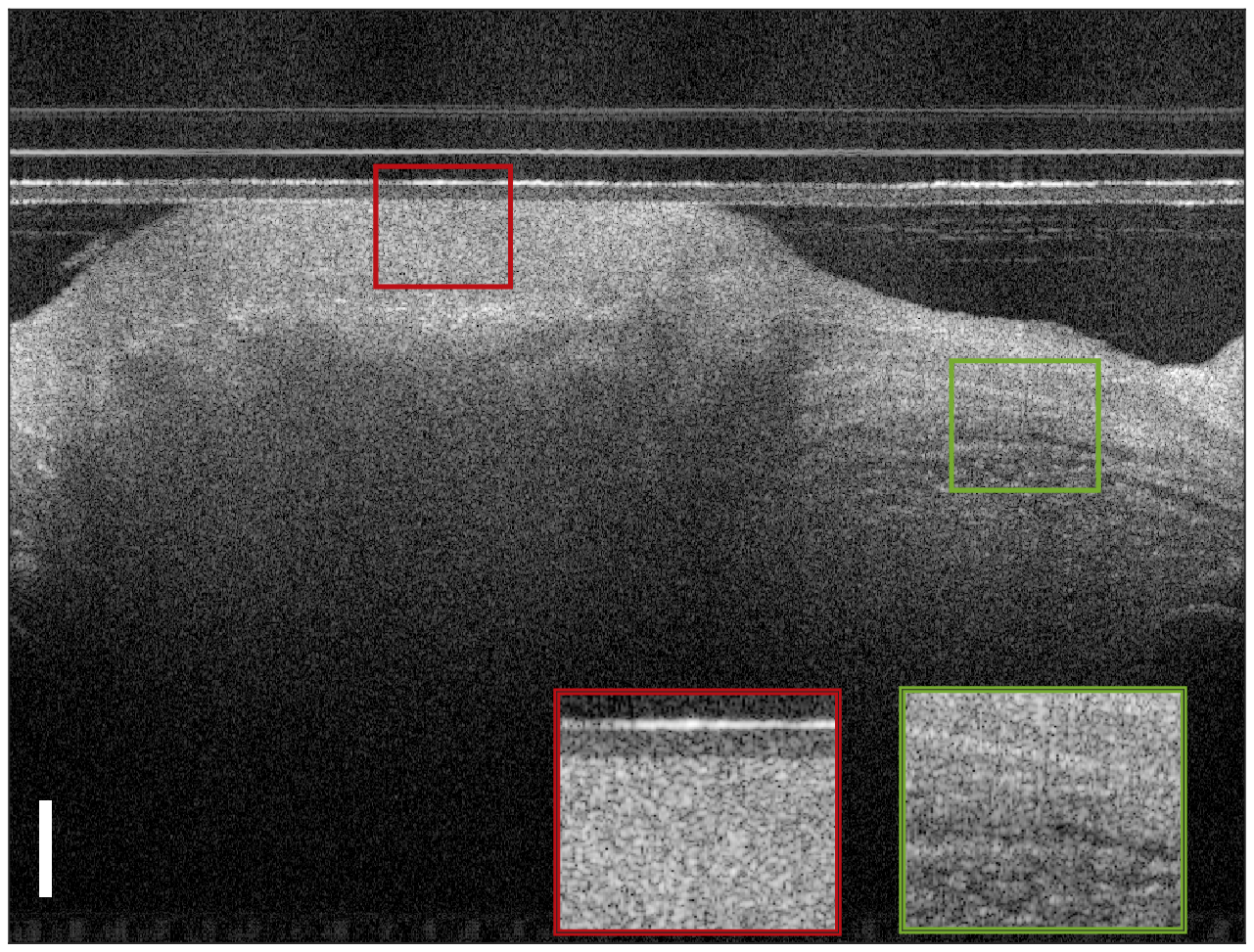}
				\caption{}
    \end{subfigure}%
		\hfill 
		\begin{subfigure}[t]{0.247\textwidth}
        \includegraphics[width=\linewidth]{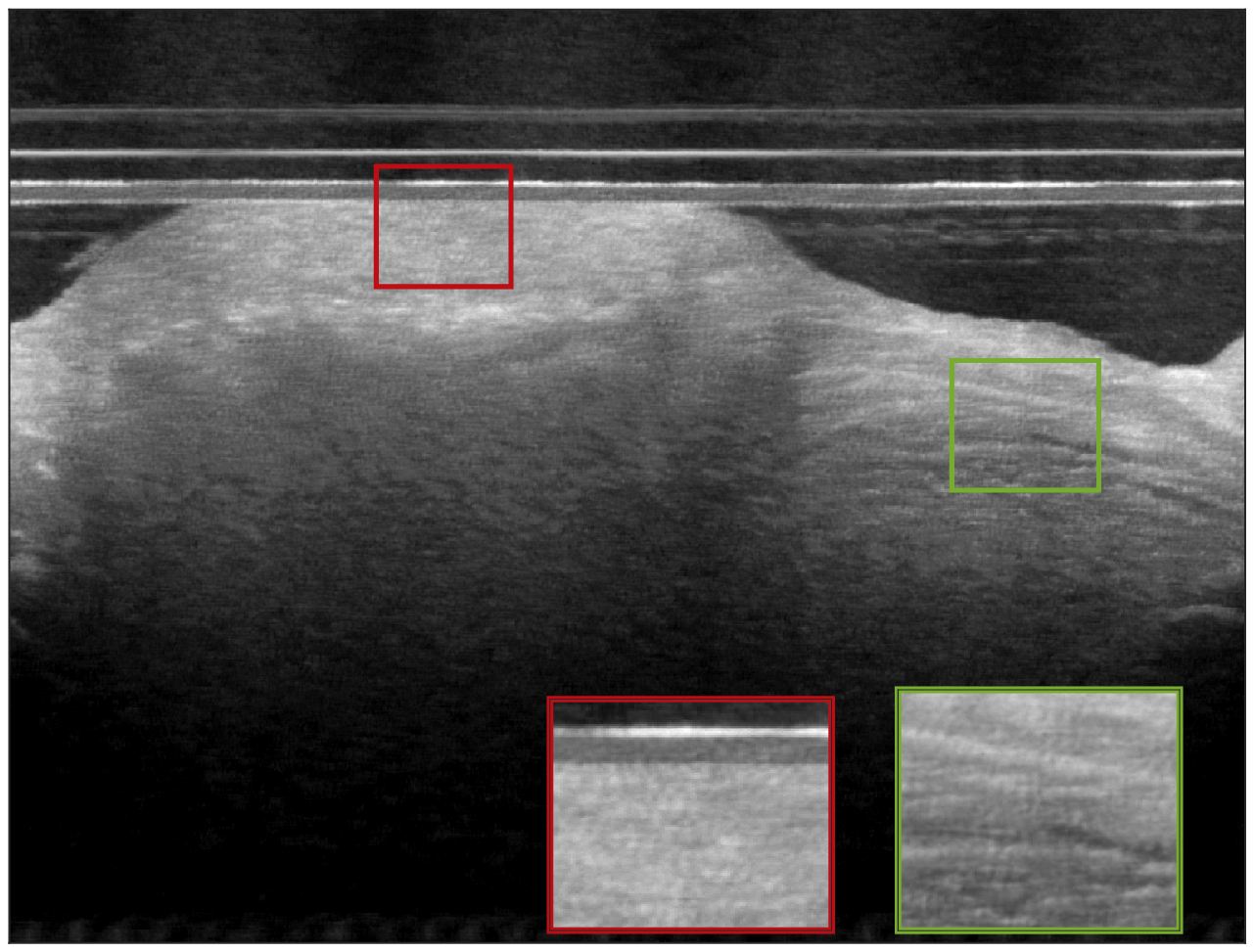}
				\caption{}
    \end{subfigure}%
		\hfill 
		\begin{subfigure}[t]{0.247\textwidth}
        \includegraphics[width=\linewidth]{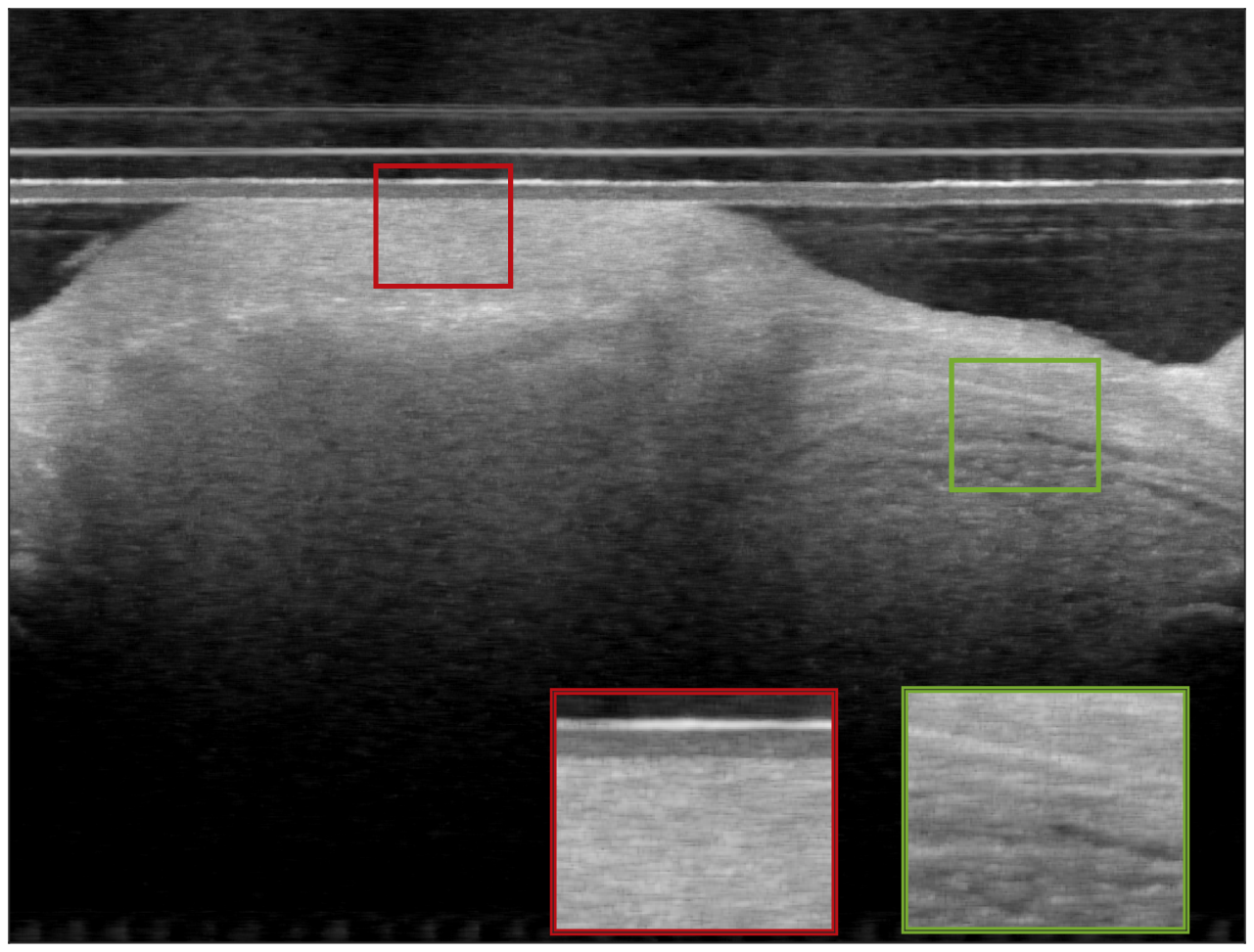}
				\caption{}
    \end{subfigure}%
		\hfill 
    \begin{subfigure}[t]{0.247\textwidth}
        \includegraphics[width=\linewidth]{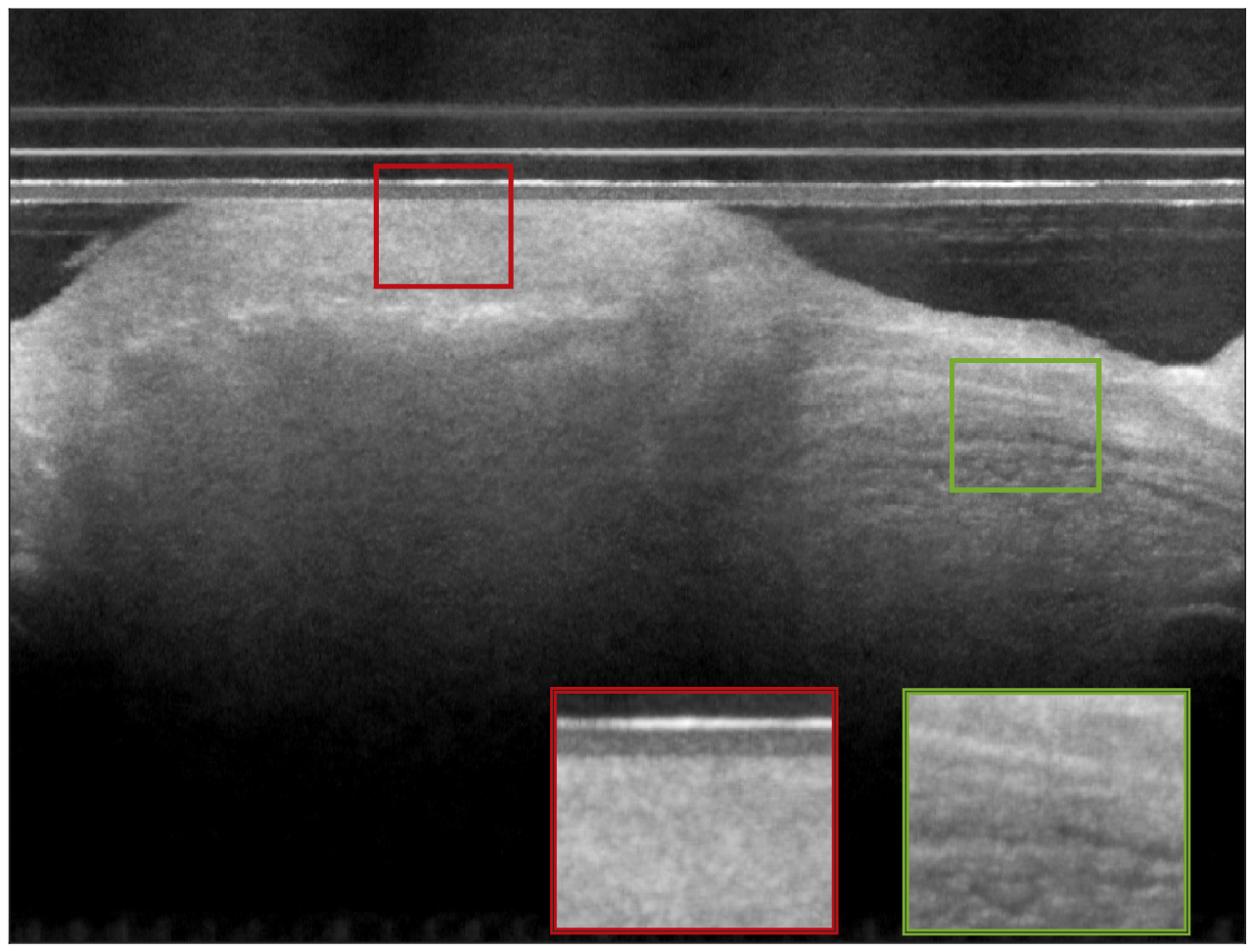}
				\caption{}
    \end{subfigure}%
		\hfill 
\caption{Cardiovascular-I speckle suppression results (in Cartesian coordinates): (a) Speckled acquired tomogram, $p^t_{\mathrm{x}}=2$; (b) DRNN trained with 100 columns of human retinal image, $p^s_{\mathrm{x}}=2$; (c) RNN-GAN trained with 200 columns of blueberry and chicken images, $p^s_{\mathrm{x}}=3$; (d) U-Net trained with retinal data image of size $448 \times 256$, $p^s_{\mathrm{x}}=2$;. Scale bar is 500 $\mu$m. }
\label{fig7}
\end{figure*}

\begin{figure*}[ht]
    \begin{subfigure}[t]{0.16\textwidth}
        \includegraphics[width=\linewidth]{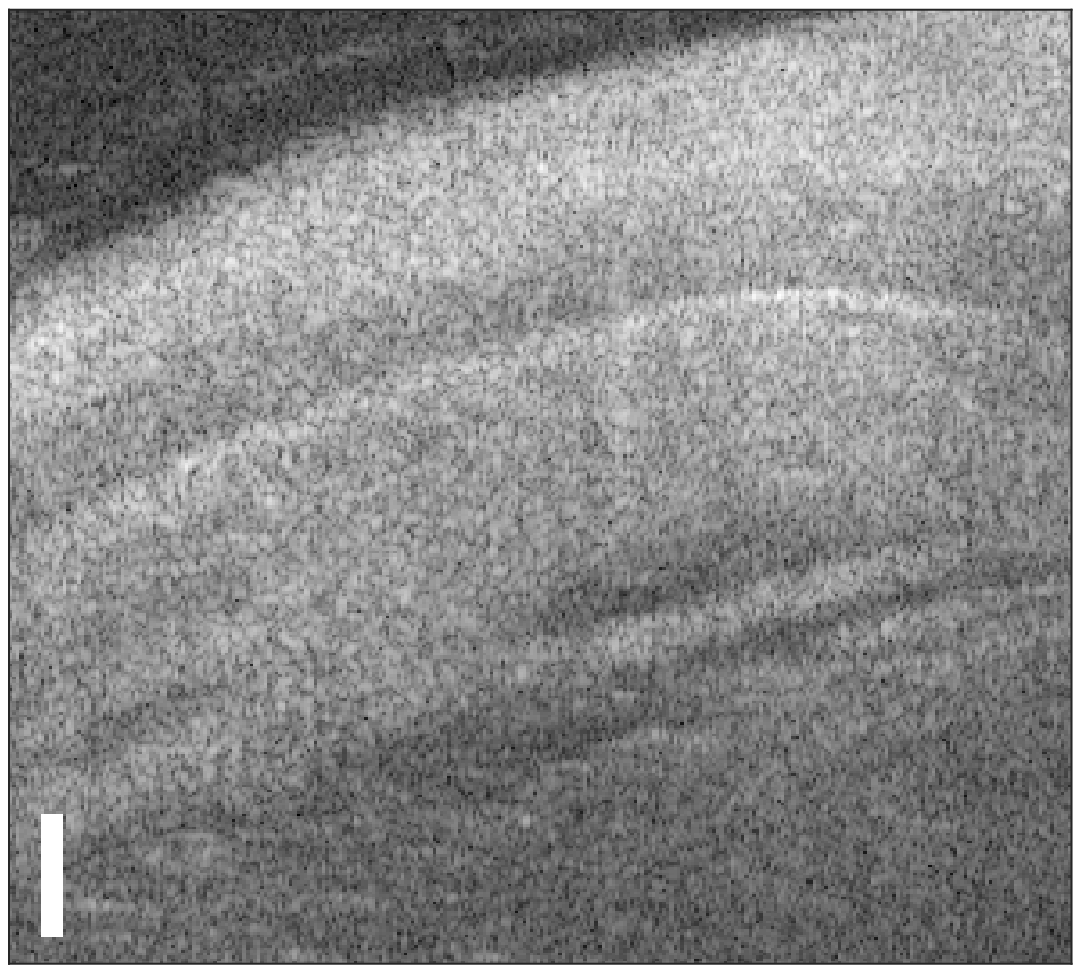}
				\caption{}
    \end{subfigure}%
		\hfill 
		\begin{subfigure}[t]{0.16\textwidth}
        \includegraphics[width=\linewidth]{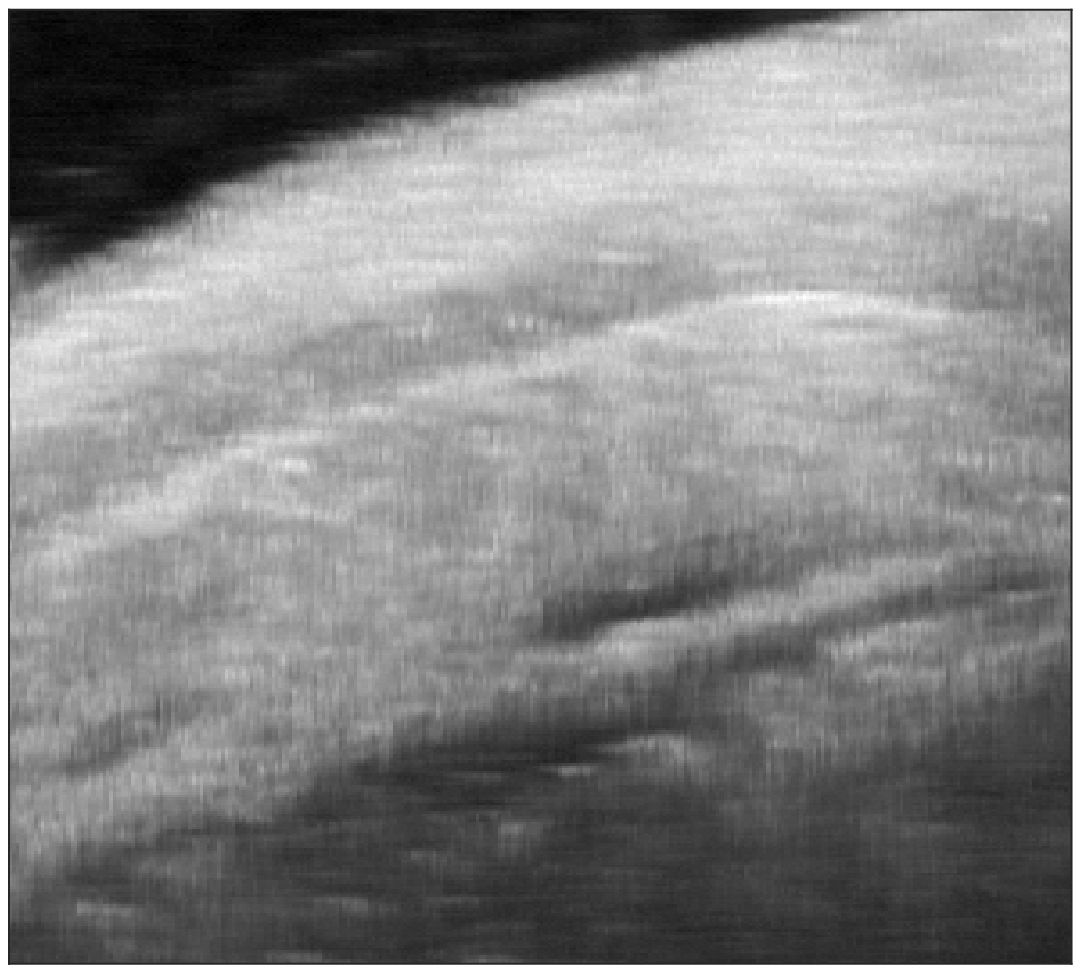}
				\caption{}
    \end{subfigure}%
		\hfill 
		\begin{subfigure}[t]{0.16\textwidth}
        \includegraphics[width=\linewidth]{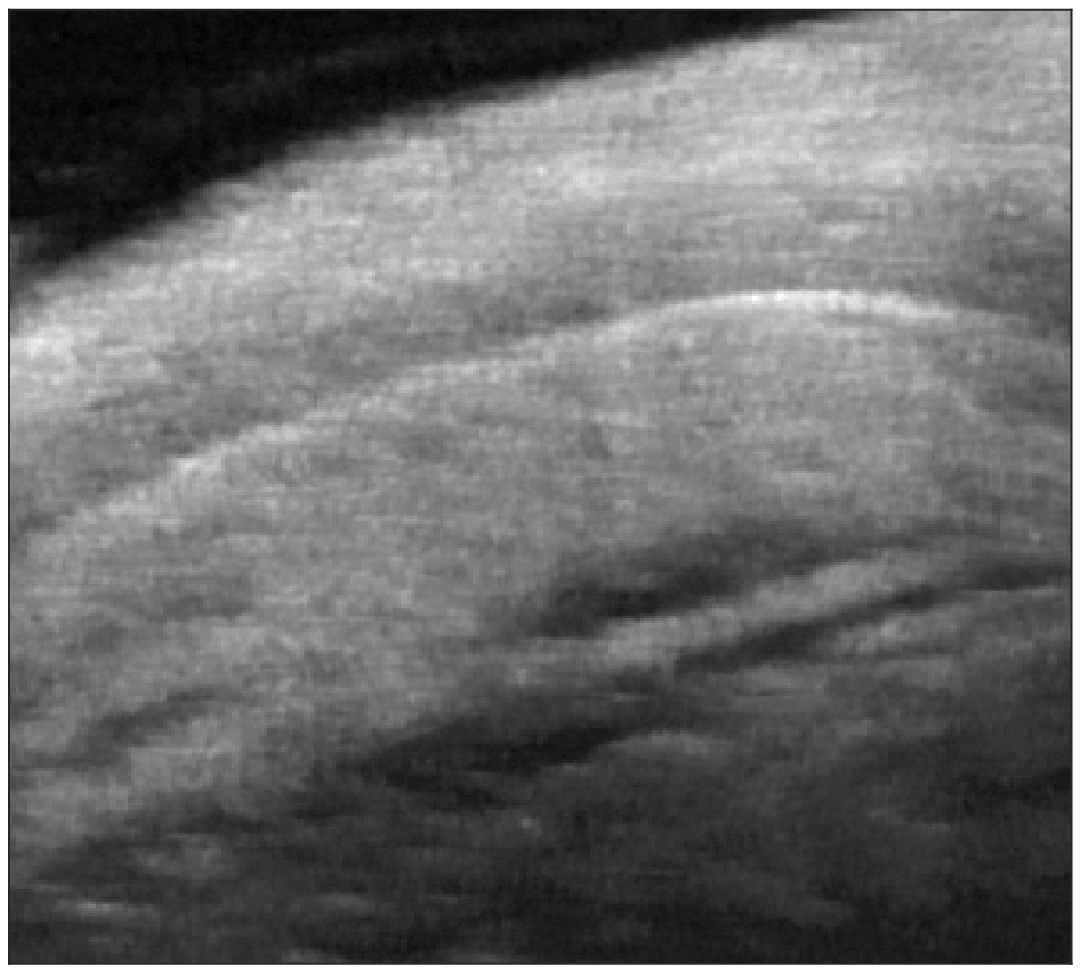}
				\caption{}
    \end{subfigure}%
		\hfill 
		\begin{subfigure}[t]{0.16\textwidth}
        \includegraphics[width=\linewidth]{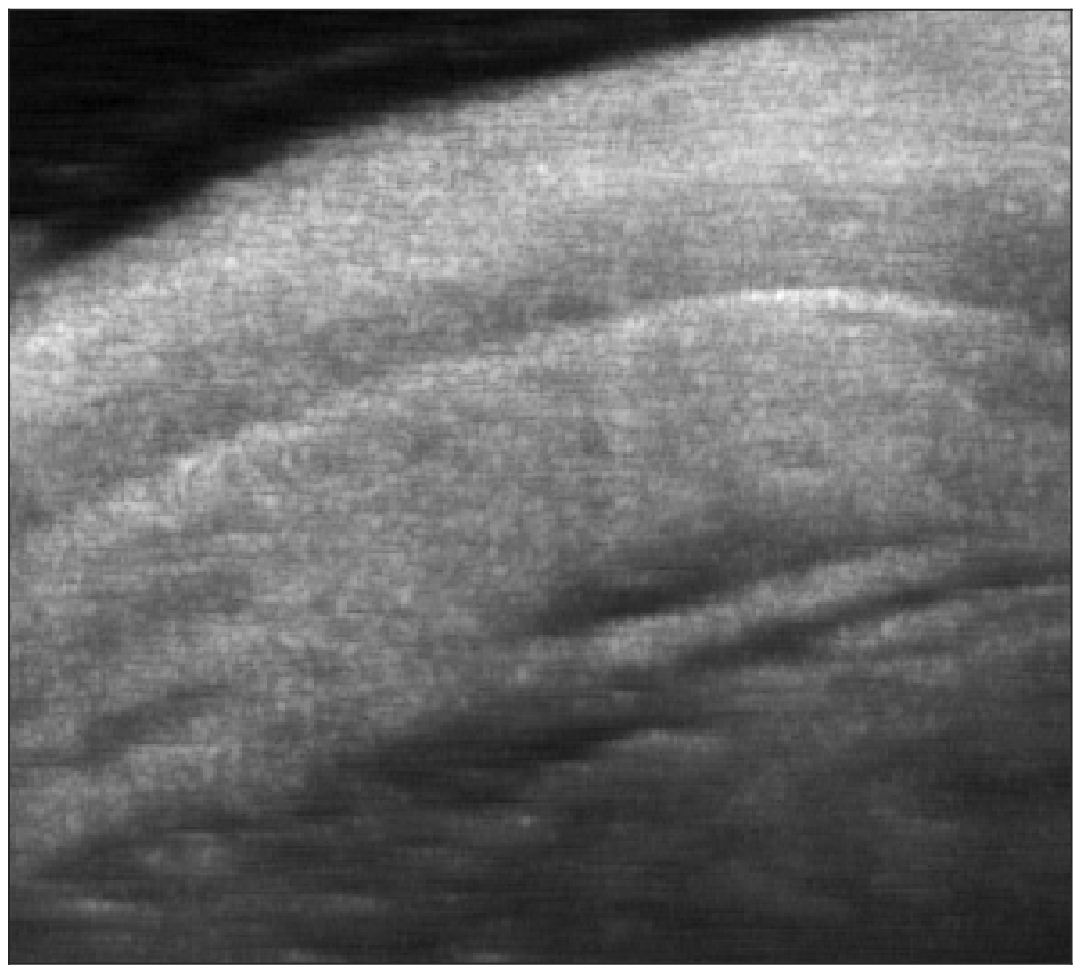}
				\caption{}
    \end{subfigure}%
		\hfill 
		\begin{subfigure}[t]{0.16\textwidth}
        \includegraphics[width=\linewidth]{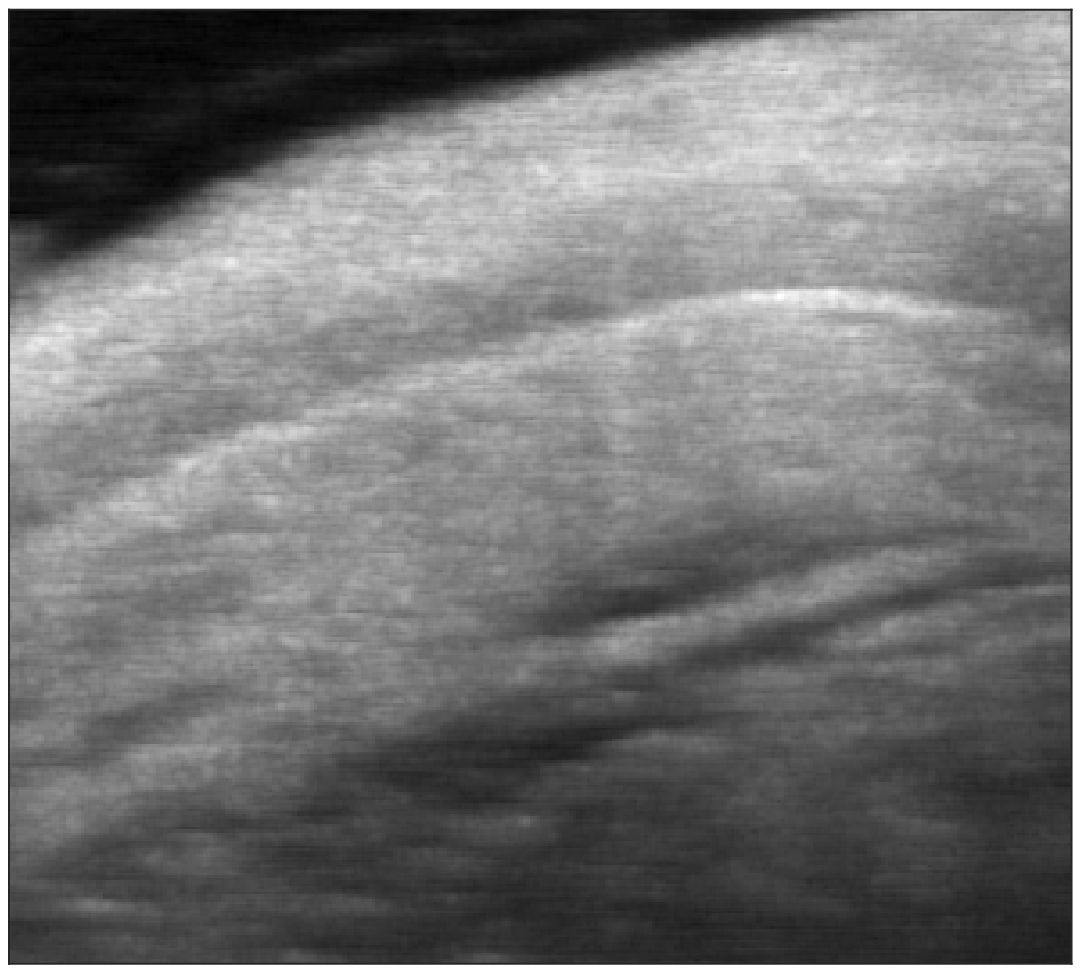}
				\caption{}
    \end{subfigure}%
		\hfill 
		\begin{subfigure}[t]{0.16\textwidth}
        \includegraphics[width=\linewidth]{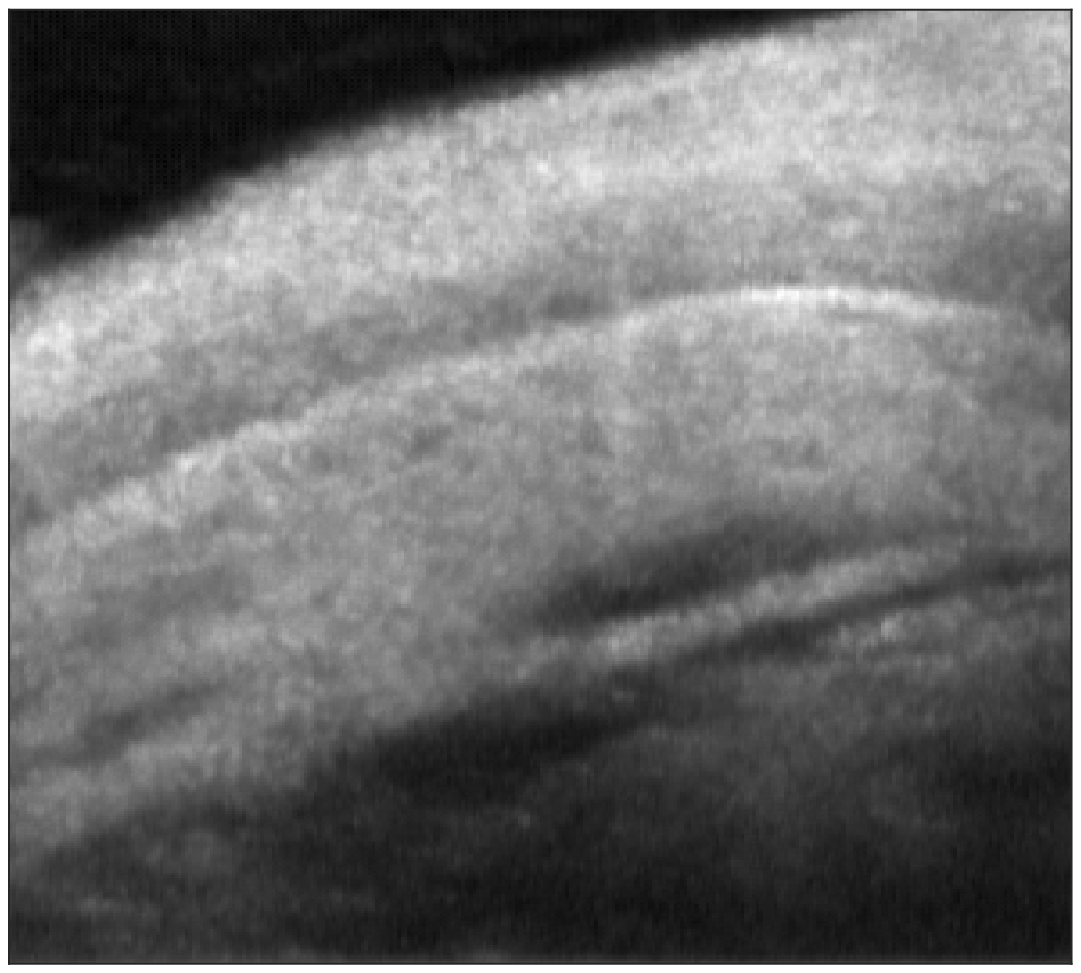}
				\caption{}
    \end{subfigure}%
		\hfill 
\caption{Cardiovascular-II speckle suppression results (in Cartesian coordinates): (a) Cropped speckled acquired tomogram of size $371\times 311$, $p^t_{\mathrm{x}}=1$; (b) OCT-RNN trained with 100 first columns of muscle chicken, $p^s_{\mathrm{x}}=3$; (c) DRNN trained with 100 columns of human retinal image, $p^s_{\mathrm{x}}=2$;  
(d) RNN-GAN trained with decimated retinal data, $p^s_{\mathrm{x}}=1$;
(e) RNN-GAN trained with interpolated retinal data, $p^s_{\mathrm{x}}=3$;
(f) U-Net trained with retinal data image of size $448 \times 256$, $p^s_{\mathrm{x}}=2$. Scale bar is 200 $\mu$m.}
\label{fig8}
\end{figure*}

%
Figs.~\ref{fig1}-\ref{fig8} depict the obtained despeckled predictions for ex-vivo samples, as well as for in vivo retinal data and intravascular OCT images, employing 4 methods: RNN, DRNN, RNN-GAN and U-Net \cite{ronneberger:2015}. Please zoom-in on screen to see the differences. The U-Net has about $8.2 \times 10^6$ parameters (8 times the RNN number of parameters), and it trains on patches of size $64 \times 64$.
Visually observing the results in different scenarios, overall, the proposed approach efficiently suppresses speckle, while preserving and enhancing visual detailed structure.

To test the DRNN's performance in different domains, we trained it with 100 columns of acquired in vivo human retinal cross section presented in Fig.~\ref{fig1}(a). 
Fig.~\ref{fig1}(b) presents the ground truth obtained as described in \cite{cuartas:2018}.
As can be observed the DRNN approach generalizes well, both with matching tissues and imaging systems, as well as in cases of tissue and system mismatch. The DRNN produces good visual quality and efficiently suppresses speckle, even without preprocssing domain adaptation. As we theoretically established, applying the source trained system to a target with lower lateral sampling-resolution ratio indeed smooths the result, whereas a target input with higher lateral sampling-resolution ratio results in detailed structure with minor speckle residuals. Visually observing the other methods' results leads to similar conclusions. 

We quantitatively evaluated the proposed approaches by comparing the peak signal to noise ratio (PSNR) and structural similarity index (SSIM) of their results with respect to the images assumed as ground truth.
Table \ref{Table2} compares the average PSNR and SSIM score for the above 4 methods with matching system and tissue. As can be seen, significant increase in PSNR and SSIM scores is achieved for all methods. RNN-GAN and U-Net have the highest scores in most cases. The U-Net usually yields the highest scores, yet, as can be observed in Fig.~\ref{fig4}(e) it can produce unexpected visible artifacts in some cases. The U-Net has more capacity, therefore, it tends to memorize the training image better, but generalize worse.
Note that PSNR and SSIM scores not always reliably represent perceptual quality or desired features of the images \cite{Blau:2018}. Also, keep in mind that AC despeckled images are a result of averaging of numerous images, whereas our system's predictions rely solely on a single observation, therefore reconstructions are notably more loyal to the single observed speckled image. Furthermore, although AC images are referred to as ground truth, they may suffer from inaccuracies related to the stage tilting and its processing.

Table \ref{Table3} provides quantitative scores for the proposed domain adaptation approach for various pairs of source and target, differing in acquisition system and tissue type, for RNN-GAN and U-Net. Notably, both approaches result in significant increase in PSNR and SSIM scores.  
Note that the images differ not only in their sampling-resolution ratio, but also by the nature of the ground truth used for training. Namely, AC images have different texture and visual appearance than NLM. Regardless of PSNR and SSIM scores the learning system often tends to adopt the visual characteristics of the source data. This tendency may also be perceived as an advantage in the absence of ground truth, as can be seen in Fig.~\ref{fig1}(g). The observed speckled image may originate in many plausible reconstructions with varying textures and ﬁne details, and different semantic information \cite{bahat:2020}. The above results somewhat offer a user-dependent degree of freedom. Unfortunately, in our experiments domain randomization strategy \cite{BILLOT:2023} failed to generalize well.


\begin{table*}[ht] 
\caption{Average PSNR / SSIM obtained for different methods and datasets with training and testing matching acquisition systems and tissue-types. Average scores are over 100 tomograms of size $256 \times 256$.}
\label{Table2}
\begin{center}
{\small
\begin{tabular}{|m{6.4em}|m{5.3em}|m{5.8em}|m{5em}|m{6em}|m{5em}|}
    \hline 
	  \textbf{Dataset}			& \textbf{input} & \textbf{RNN-OCT} & \textbf{DRNN} & \textbf{RNN-GAN} & 	\textbf{U-Net}   
										\\ \hline \hline 
		\textbf{Retina}			  & 24.87 / 0.46	 & 33.60/ 0.87    & 30.46 / 0.82  & 32.24 / 0.86  & 33.66 / 0.89   \\ \hline
		\textbf{Chicken}  		& 24.29 / 0.29	 & 27.97 / 0.61 	& 29.41 / 0.63  & 30.81 / 0.74  & 32.50 / 0.77   \\ \hline
		\textbf{Blueberry} 		& 24.98 / 0.48 	 & 27.15 / 0.63   & 27.57 / 0.69  & 28.18 / 0.76	& 28.09 / 0.76  	\\ \hline
		\textbf{Chicken Skin} &	26.12 / 0.44	 & 29.64 / 0.71  	& 29.59 / 0.69  & 30.49 / 0.78 	& 30.26 / 0.77   \\ \hline
		\textbf{Cucumber}			& 25.91 / 0.59	 & 27.31 / 0.73		& 28.69 / 0.73  & 28.85 / 0.79  & 28.52 / 0.81   \\ \hline
\end{tabular}  
}
\end{center}
\vspace{-5pt}
\end{table*}

\begin{table}[h] 
\caption{Domain-aware PSNR / SSIM obtained for different methods and datasets, with training and testing acquisition systems and tissue-types mismatch, with preprocessing adapting sampling-resolution ratio.
$^*$ denotes cases where domain adaption was not applied.}
\label{Table3}
\begin{center}
{\footnotesize
\begin{tabular}{|m{5.5em}|m{5.5em}|m{5.5em}|m{5.5em}|}
    \hline
	  \textbf{Target Data}  & \textbf{Source Data} & \textbf{RNN-GAN} & \textbf{U-Net}  \\ \hline
		Retina			   & 	Blueberry $\&$ Chicken 	 & 30.86 /0.87      &  31.80 / 0.85    			\\ \hline
		Chicken  		   & 	Chicken Skin       			 & 31.33 / 0.69 		&  31.08 / 0.68				 	\\ \hline
		Chicken 		   &  Retina 		         			 & 29.08 / 0.63    	&  31.89 / 0.70				  \\ \hline
		Blueberry 		 &  Retina 		         			 & 27.78 / 0.69    	&  28.33 / 0.77		      \\ \hline
	  Chicken Skin   &	Blueberry $\&$ Chicken	 & 30.68 / 0.76     &  30.43 / 0.71        	\\ \hline
		Chicken Skin   &	Retina						     	 & 31.51 / 0.76$^*$ &  30.88 / 0.77$^*$         \\ \hline
		Cucumber			 & 	Blueberry		       			 & 27.84 / 0.75  		&  27.61 / 0.72$^*$    		\\ \hline
		Cucumber			 & 	Retina       			 			 & 28.71 / 0.77$^*$			&  28.11 / 0.73            	  			\\ \hline
\end{tabular}  
}
\end{center}
\vspace{-18pt}
\end{table}

Training of the proposed model is extremely fast. 
The number of epochs for the first training content-loss stage is 5-12 epochs, depending on analysis patch size, batch size and training image size. Adversarial loss training stage takes about 10-30 epochs. The total time of training is \textit{\textbf{5-25 seconds} on a laptop GPU}. Training without adversarial stages normally takes about 12 seconds. 
As a rule of thumb, training for too long can cause over-fitting, and blurry images.
Training times were measured on a standard laptop workstation equipped with a 12th Gen Intel(R) Core(TM) i7-12800H 2.40 GHz with 32.0 GB RAM, NVIDIA RTX A2000 8GB Laptop GPU. Training can also be performed on a CPU in a few minutes. Inference time is 110.5 ms.
U-Net training is usually longer and takes about 5.76 minutes (for 16 epochs). 
As far as we know, our results are the state-of-the-art in terms of optimized real-time training with minimal available training data. Our code will be available upon acceptance.


\section{Discussion $\&$ Conclusions}
In this work, we analyzed the critical challenge of supervised learning domain adaptation for the task of OCT speckle suppression, with diverse imaging systems, given limited ground-truth data. 
We focused on a RNN patch-based approach that is both flexible and efficient in terms of patch-size and number of parameters, and less prone to low-frequency bias. We further designed a suitable adversarial loss training stage with dramatically reduced complexity. We also demonstrated the applicability of our proposed point of view to a U-Net. Future research can potentially investigate other architectures.
For the challenge of domain adaptation, we illuminate a rather simplified point-of-view, that can facilitate efficient deployment both for research and industrial purposes. 

Our results challenge the assumption that training data sets must consist of a large representation of the entire input data probability distribution, thus evading the curse of dimensionality. The proposed few-shot learning framework is potentially related to the information theoretical asymptotic equipartition property (\cite{Pereg:2022}, Section 2). 
Our results can inspire the design of novel few-shot learning systems for medical imaging, and can also be of interest to the wider deep learning and signal processing community.
The models are formulated for two-dimensional (2D) signals, but can be easily be adapted to other data dimensions. 
Future work can investigate the applicability of the proposed approach to other tasks and to cross-modality domain adaptation. 

\vspace{-10pt}
\section*{Appendix A}

\textbf{Proof of Theorem 2}

Let us assume source input-output pairs $\{\mathbf{y}^s,\mathbf{x}^s\}\in \mathbb{R}^{M \times N}$, with distribution $\{\mathbf{y}^s,\mathbf{x}^s\} \sim P_{\mathcal{S}}(\mathbf{y}^s,\mathbf{x}^s)$ in the source domain $\mathcal{S}$, and samples $\{\mathbf{y}^t,\mathbf{x}^t\} \in \mathbb{R}^{M \times N}$ in the target domain $\mathcal{T}$ such that $\{\mathbf{y}^t,\mathbf{x}^t\} \sim P_{\mathcal{T}}(\mathbf{y}^t,\mathbf{x}^t)$.
The predictor is trained and generalizes well, such that 
{\small
\begin{equation}\label{A.0}
\mathcal{L}(\mathcal{F}_\mathcal{S}) = E_{(x,y) \sim P_{\mathcal{S}}} \big\| \mathcal{F}_\mathcal{S} \big(y^s[m,n]\big)-x^s[m,n]\big\|_2 <\varepsilon,
\end{equation}
}
where $E_{(x,y) \sim P_{X,Y}}(\cdot)$ denotes the expectation over $P_{X,Y}$, $\|\cdot\|_2$ denotes the $\ell_2$ norm, $\varepsilon << \min \|\mathbf{x}\|_2 $.
As mentioned above, in the OCT speckle suppression problem, the coherent speckled input is 
{\small
\begin{equation}\label{A.1}
y^s[m,n]=\big|f^s[m,n]*\alpha^s[m,n]\big|^2.
\end{equation}
}
While the corresponding incoherent output is formulated as
{\small
\begin{equation}\label{A.2}
x^s[m,n]=\big|f^s[m,n]\big|^2*\big|\alpha^s[m,n]\big|^2.
\end{equation}
}
Similarly, the target input is
{
\begin{equation}\label{A.4}
y^t[m,n]=\big|f^t[m,n]*\alpha^t[m,n]\big|^2,
\end{equation}
and the desired despeckled image in the target domain is
\begin{equation}\label{A.3}
x^t[m,n]=\big|f^t[m,n]\big|^2*\big|\alpha^t[m,n]\big|^2.
\end{equation}
}We omit the scaled logarithm for simplicity.
Applying the trained source system to the source data, we can assume
$\forall y^s[m,n] \in \mathbb{R}^{M \times N} \exists x^s[m,n] : x^s[m,n] \approx \mathcal{F}_\mathcal{S}(y^s[m,n])$, therefore applying the trained system to the target output we have,
{\small
\begin{equation}\label{A.5}
\mathcal{F}_\mathcal{S}\big(y^t[m,n]\big)=x^s[m,n]\Big|_{y^s[m,n]=y^t[m,n]} 
\end{equation}
}
In other words, the system's output is its prediction for 
{\footnotesize
\begin{equation}\label{A.6}
y^s[m,n]=\big|f^s[m,n]*\alpha^s[m,n]\big|^2=\big|f^t[m,n]*\alpha^t[m,n]\big|^2=y^t[m,n]
\end{equation}
}where {\small $\alpha^t[m,n]=\alpha_{\mathrm{z}}^t[m]*\alpha_{\mathrm{x}}^t[n]$} and {\small $\alpha^s[m,n]=\alpha_{\mathrm{z}}^s[m]*\alpha_{\mathrm{x}}^s[n]$} are separable PSFs.
The predictor does not "know'' that its input does not belong to the same domain, and it would output the corresponding prediction in the source domain. For the sake of the proof's outline, let us assume the target and source domains share an equivalent sampling resolution ratio in the axial direction, that is {\small $\alpha^s_{\mathrm{z}}[m]=\alpha^t_{\mathrm{z}}[m]$}. Let us examine two possible scenarios. (1) If $p_\mathrm{x}^s<p_\mathrm{x}^t$, there exist $\alpha^{s \rightarrow t}_{\mathrm{x}}[n]$ such that $\alpha_{\mathrm{x}}^t[n] = \alpha_{\mathrm{x}}^s[n]* \alpha_{\mathrm{x}}^{s \rightarrow t}[n]$.
Then (\ref{A.6}) is satisfied if
{\footnotesize
\begin{flalign*}
\nonumber
\big|f^t[m,n]*\alpha^t[m,n]\big| & = \big|f^t[m,n]*\alpha^{s \rightarrow t}[m,n]*\alpha^s[m,n]\big| 
\\
& = \big|f^s[m,n]*\alpha^s[m,n]\big|.
\end{flalign*}
}
Therefore,
{\footnotesize
\begin{flalign}\label{A.10}
\nonumber
\mathcal{F}_\mathcal{S}\big(y^t[m,n]\big) & =x^s[m,n]\Big|_{y^t[m,n]=y^s[m,n]} \\
& = \big|f^t[m,n]*\alpha^{s \rightarrow t}[m,n]\big|^2*\big|\alpha^s[m,n]\big|^2.
\end{flalign}
}In other words, the output resolution is determined by the source resolution $\alpha^s[m,n]$.
The tomogram component $|f^t[m,n]*\alpha^{s \rightarrow t}[m,n]|$ is a lower resolution version of the desired target tomogram $|f^t[m,n]|$. 
(2) If $p_\mathrm{x}^s>p_\mathrm{x}^t$, there exist $\alpha_{\mathrm{x}}^{t \rightarrow s}[n]$ such that {\small $\alpha^{s}_{\mathrm{x}}[n]=\alpha^t_{\mathrm{x}}[n]* \alpha_{\mathrm{x}}^{t \rightarrow s}[n]$}.
Then (\ref{A.6}) is satisfied if
{\footnotesize
\begin{flalign*}
\nonumber
\big|f^t[m,n]*\alpha^t[m,n]\big| &= \big|f^s[m,n]*\alpha^s[m,n]\big|
\\
& = \big|f^s[m,n]*\alpha^{t \rightarrow s}[m,n]*\alpha^{t}[m,n]\big|
\end{flalign*}
}
Thus,
\vspace{-5pt}
{\small
\begin{flalign}\label{A.12}
\nonumber
&\mathcal{F}_\mathcal{S}\big(y^t[m,n]\big) =x^s[m,n]\Big|_{y^s[m,n]=y^t[m,n]}\\
&= \big|f^s[m,n]\big|^2*\big|\alpha^s[m,n]\big|^2
\Big|_{f^t[m,n]=f^s[m,n]*\alpha^{t \rightarrow s}[m,n]}.
\end{flalign}
}In other words, the resolution of the output will be determined by the source resolution $\alpha^s[m,n]$.
Furthermore, since the solution obeys $f^t[m,n]=f^s[m,n]*\alpha^{t \rightarrow s}[m,n]$, the tomogram component $|f^s[m,n]|$ is in a sense a speckled super-resolved version of the target tomogram $|f^t[m,n]|$.  
Note that (\ref{A.12}) is true under the assumption that $\forall f^t[m,n] \exists f^s[m,n] : f^t[m,n]=f^s[m,n]*\alpha^{t \rightarrow s}[m,n]$. In other words $f^s[m,n]$ is a super-resolved version of $f^t[m,n]$, which is safe to assume since the resolution of $f[m,n]$ is well above the size of point-scatters particles. 
Specifically, for a Gaussian PSF we know the convolution of two Gaussian with mean $\mu_1, \mu_2$ and variance $\sigma^2_1$, $\sigma^2_2$ is a Gaussian with mean $\mu=\mu_1+\mu_2$ and variance $\sigma^2 =\sigma^2_1+\sigma^2_2$.
The same proof applies to the cases where $p_\mathrm{z}^s \lessgtr p_\mathrm{z}^t$ interchangeably, or a combination of $p_\mathrm{z}, p_\mathrm{x}$ due to separability. 
For the general case, we can write
\begin{equation}\label{A.13}
\alpha^s[m,n]*\alpha^{s \rightarrow t}[m,n]=\alpha^t[m,n]*\alpha^{t \rightarrow s}[m,n].
\end{equation}
Therefore,
{\small
\begin{flalign}\label{A.14}
\nonumber
& \mathcal{F}_\mathcal{S}\big(y^t[m,n]\big) =x^s[m,n]\Big|_{y^s[m,n]=y^t[m,n]} =\\
& \big|f^s[m,n]\big|^2*\big|\alpha^s[m,n]\big|^2
\Big|_{f^t[m,n]*\alpha^{s \rightarrow t}[m,n]=f^s[m,n]*\alpha^{t \rightarrow s}[m,n]}.
\end{flalign}
}
For example, when $p_\mathrm{x}^s<p_\mathrm{x}^t$ and $p_\mathrm{z}^s>p_\mathrm{z}^t$, $\hat{x}^t=\mathcal{F}_{\mathcal{S}}\big\{y^t\big\}$, we have,
{\scriptsize
\begin{equation*}
\mathcal{F}_{\mathcal{S}}\big(y^t\big)=
\big|f^s[m,n]\big|^2*\big|\alpha^s[m,n]\big|^2
\Big|_{f^t[m,n]*\alpha_{\mathrm{x}}^{s \rightarrow t}[n]= f^s[m,n]*\alpha_{\mathrm{z}}^{t \rightarrow s}[m]}.
\end{equation*}
}
\qed
\vspace{-10pt}
\section*{Acknowledgment}
The author thanks Sebasti\'an Ruiz-Lopera, N\'estor Uribe-Patarroyo and Martin Villiger (Wellman Center for Photomedicine) for providing the OCT data. This work was supported in part by the Zuckerman STEM Leadership Program.

\vspace{-5pt}
\bibliography{dpereg_RNN_OCT}

\end{document}